\documentclass[aps,prd,twocolumn,superscriptaddress,groupedaddress,amsmath,amssymb,nofootinbib,preprintnumbers,letterpaper]{revtex4-2} 
\usepackage{graphicx}
\usepackage{xcolor}
\usepackage{dcolumn}
\usepackage{bm}
\usepackage{subcaption} 
\usepackage{enumitem} 
\usepackage{xspace} 
\usepackage{hyperref}
\usepackage{multirow}
\usepackage{mathptmx}
\usepackage{placeins} 

\usepackage[Symbol]{upgreek}
\usepackage{caption}
\usepackage{float}
\floatstyle{plaintop}
\restylefloat{table}

\hypersetup{
    colorlinks=true,
    linkcolor=cyan,
    filecolor=magenta,      
    urlcolor=cyan,
    citecolor=violet,
}
\usepackage[nameinlink, capitalize]{cleveref}

\makeatletter
\g@addto@macro\bfseries{\boldmath}
\makeatother

\newcommand{\xmax}{\ensuremath{X_{\mathrm{max}}}\xspace}
\newcommand{\xmaxdnn}{\ensuremath{X_{\mathrm{max,DNN}}}\xspace}
\newcommand{\xmaxfd}{\ensuremath{X_{\mathrm{max,FD}}}\xspace}
\newcommand{\xmaxsigma}{\ensuremath{\sigma(X_{\mathrm{max}})}\xspace}
\newcommand{\xmaxmu}{\ensuremath{\langle X_{\mathrm{max}}\rangle}\xspace}
\newcommand{\lnamu}{\ensuremath{\langle \ln\mathrm{A}\rangle}\xspace}

\newcommand{\gcm}{\ensuremath{\mathrm{g\,cm^{-2}}}\xspace}
\newcommand{\gcmd}{\ensuremath{\mathrm{g\,cm^{-2}\,\mathrm{decade}^{-1}}}\xspace}
\newcommand{\grad}{\ensuremath{^{\circ}}\xspace}

\def\etal{\textit{et~al.}\xspace}

\newcommand{\sibyllc}{Sibyll2.3c\xspace}
\newcommand{\sibylld}{Sibyll2.3d\xspace} 
\newcommand{\qgs}{QGSJetII-04\xspace}
\newcommand{\epos}{EPOS-LHC\xspace}
\newcommand{\sibstar}{Sibyll$^{\bigstar}$\xspace}

\begin{document}

\title{Measurement of the Depth of Maximum of Air-Shower Profiles with energies between $\mathbf{10^{18.5}}$ and $\mathbf{10^{20}}$ eV using the Surface Detector of the Pierre Auger Observatory and Deep Learning}

\author{
A.~Abdul Halim$^{13}$,
P.~Abreu$^{71}$,
M.~Aglietta$^{53,51}$,
I.~Allekotte$^{1}$,
K.~Almeida Cheminant$^{79,78,69}$,
A.~Almela$^{7,12}$,
R.~Aloisio$^{44,45}$,
J.~Alvarez-Mu\~niz$^{77}$,
J.~Ammerman Yebra$^{77}$,
G.A.~Anastasi$^{57,46}$
% ...
}
\affiliation{}
\collaboration{The Pierre Auger Collaboration}
\email{spokespersons@auger.org}
\author{\phantom{1}}
\affiliation{
\begin{description}[labelsep=0.2em,align=right,labelwidth=0.7em,labelindent=0em,leftmargin=2em,noitemsep,before={\renewcommand\makelabel[1]{##1 }}]
\item[$^{1}$] Centro At\'omico Bariloche and Instituto Balseiro (CNEA-UNCuyo-CONICET), San Carlos de Bariloche, Argentina
\item[$^{2}$] Departamento de F\'\i{}sica and Departamento de Ciencias de la Atm\'osfera y los Oc\'eanos, FCEyN, Universidad de Buenos Aires and CONICET, Buenos Aires, Argentina
\item[$^{3}$] IFLP, Universidad Nacional de La Plata and CONICET, La Plata, Argentina
\item[$^{4}$] Instituto de Astronom\'\i{}a y F\'\i{}sica del Espacio (IAFE, CONICET-UBA), Buenos Aires, Argentina
\item[$^{5}$] Instituto de F\'\i{}sica de Rosario (IFIR) -- CONICET/U.N.R.\ and Facultad de Ciencias Bioqu\'\i{}micas y Farmac\'euticas U.N.R., Rosario, Argentina
\item[$^{6}$] Instituto de Tecnolog\'\i{}as en Detecci\'on y Astropart\'\i{}culas (CNEA, CONICET, UNSAM), and Universidad Tecnol\'ogica Nacional -- Facultad Regional Mendoza (CONICET/CNEA), Mendoza, Argentina
\item[$^{7}$] Instituto de Tecnolog\'\i{}as en Detecci\'on y Astropart\'\i{}culas (CNEA, CONICET, UNSAM), Buenos Aires, Argentina
\item[$^{8}$] International Center of Advanced Studies and Instituto de Ciencias F\'\i{}sicas, ECyT-UNSAM and CONICET, Campus Miguelete -- San Mart\'\i{}n, Buenos Aires, Argentina
\item[$^{9}$] Laboratorio Atm\'osfera -- Departamento de Investigaciones en L\'aseres y sus Aplicaciones -- UNIDEF (CITEDEF-CONICET), Argentina
\item[$^{10}$] Observatorio Pierre Auger, Malarg\"ue, Argentina
\item[$^{11}$] Observatorio Pierre Auger and Comisi\'on Nacional de Energ\'\i{}a At\'omica, Malarg\"ue, Argentina
\item[$^{12}$] Universidad Tecnol\'ogica Nacional -- Facultad Regional Buenos Aires, Buenos Aires, Argentina
\item[$^{13}$] University of Adelaide, Adelaide, S.A., Australia
\item[$^{14}$] Universit\'e Libre de Bruxelles (ULB), Brussels, Belgium
\item[$^{15}$] Vrije Universiteit Brussels, Brussels, Belgium
\item[$^{16}$] Centro Brasileiro de Pesquisas Fisicas, Rio de Janeiro, RJ, Brazil
\item[$^{17}$] Centro Federal de Educa\c{c}\~ao Tecnol\'ogica Celso Suckow da Fonseca, Petropolis, Brazil
\item[$^{18}$] Instituto Federal de Educa\c{c}\~ao, Ci\^encia e Tecnologia do Rio de Janeiro (IFRJ), Brazil
\item[$^{19}$] Universidade de S\~ao Paulo, Escola de Engenharia de Lorena, Lorena, SP, Brazil
\item[$^{20}$] Universidade de S\~ao Paulo, Instituto de F\'\i{}sica de S\~ao Carlos, S\~ao Carlos, SP, Brazil
\item[$^{21}$] Universidade de S\~ao Paulo, Instituto de F\'\i{}sica, S\~ao Paulo, SP, Brazil
\item[$^{22}$] Universidade Estadual de Campinas (UNICAMP), IFGW, Campinas, SP, Brazil
\item[$^{23}$] Universidade Estadual de Feira de Santana, Feira de Santana, Brazil
\item[$^{24}$] Universidade Federal de Campina Grande, Centro de Ciencias e Tecnologia, Campina Grande, Brazil
\item[$^{25}$] Universidade Federal do ABC, Santo Andr\'e, SP, Brazil
\item[$^{26}$] Universidade Federal do Paran\'a, Setor Palotina, Palotina, Brazil
\item[$^{27}$] Universidade Federal do Rio de Janeiro, Instituto de F\'\i{}sica, Rio de Janeiro, RJ, Brazil
\item[$^{28}$] Universidad de Medell\'\i{}n, Medell\'\i{}n, Colombia
\item[$^{29}$] Universidad Industrial de Santander, Bucaramanga, Colombia
\item[$^{30}$] Charles University, Faculty of Mathematics and Physics, Institute of Particle and Nuclear Physics, Prague, Czech Republic
\item[$^{31}$] Institute of Physics of the Czech Academy of Sciences, Prague, Czech Republic
\item[$^{32}$] Palacky University, Olomouc, Czech Republic
\item[$^{33}$] CNRS/IN2P3, IJCLab, Universit\'e Paris-Saclay, Orsay, France
\item[$^{34}$] Laboratoire de Physique Nucl\'eaire et de Hautes Energies (LPNHE), Sorbonne Universit\'e, Universit\'e de Paris, CNRS-IN2P3, Paris, France
\item[$^{35}$] Univ.\ Grenoble Alpes, CNRS, Grenoble Institute of Engineering Univ.\ Grenoble Alpes, LPSC-IN2P3, 38000 Grenoble, France
\item[$^{36}$] Universit\'e Paris-Saclay, CNRS/IN2P3, IJCLab, Orsay, France
\item[$^{37}$] Bergische Universit\"at Wuppertal, Department of Physics, Wuppertal, Germany
\item[$^{38}$] Karlsruhe Institute of Technology (KIT), Institute for Experimental Particle Physics, Karlsruhe, Germany
\item[$^{39}$] Karlsruhe Institute of Technology (KIT), Institut f\"ur Prozessdatenverarbeitung und Elektronik, Karlsruhe, Germany
\item[$^{40}$] Karlsruhe Institute of Technology (KIT), Institute for Astroparticle Physics, Karlsruhe, Germany
\item[$^{41}$] RWTH Aachen University, III.\ Physikalisches Institut A, Aachen, Germany
\item[$^{42}$] Universit\"at Hamburg, II.\ Institut f\"ur Theoretische Physik, Hamburg, Germany
\item[$^{43}$] Universit\"at Siegen, Department Physik -- Experimentelle Teilchenphysik, Siegen, Germany
\item[$^{44}$] Gran Sasso Science Institute, L'Aquila, Italy
\item[$^{45}$] INFN Laboratori Nazionali del Gran Sasso, Assergi (L'Aquila), Italy
\item[$^{46}$] INFN, Sezione di Catania, Catania, Italy
\item[$^{47}$] INFN, Sezione di Lecce, Lecce, Italy
\item[$^{48}$] INFN, Sezione di Milano, Milano, Italy
\item[$^{49}$] INFN, Sezione di Napoli, Napoli, Italy
\item[$^{50}$] INFN, Sezione di Roma ``Tor Vergata'', Roma, Italy
\item[$^{51}$] INFN, Sezione di Torino, Torino, Italy
\item[$^{52}$] Istituto di Astrofisica Spaziale e Fisica Cosmica di Palermo (INAF), Palermo, Italy
\item[$^{53}$] Osservatorio Astrofisico di Torino (INAF), Torino, Italy
\item[$^{54}$] Politecnico di Milano, Dipartimento di Scienze e Tecnologie Aerospaziali , Milano, Italy
\item[$^{55}$] Universit\`a del Salento, Dipartimento di Matematica e Fisica ``E.\ De Giorgi'', Lecce, Italy
\item[$^{56}$] Universit\`a dell'Aquila, Dipartimento di Scienze Fisiche e Chimiche, L'Aquila, Italy
\item[$^{57}$] Universit\`a di Catania, Dipartimento di Fisica e Astronomia ``Ettore Majorana``, Catania, Italy
\item[$^{58}$] Universit\`a di Milano, Dipartimento di Fisica, Milano, Italy
\item[$^{59}$] Universit\`a di Napoli ``Federico II'', Dipartimento di Fisica ``Ettore Pancini'', Napoli, Italy
\item[$^{60}$] Universit\`a di Palermo, Dipartimento di Fisica e Chimica ''E.\ Segr\`e'', Palermo, Italy
\item[$^{61}$] Universit\`a di Roma ``Tor Vergata'', Dipartimento di Fisica, Roma, Italy
\item[$^{62}$] Universit\`a Torino, Dipartimento di Fisica, Torino, Italy
\item[$^{63}$] Benem\'erita Universidad Aut\'onoma de Puebla, Puebla, M\'exico
\item[$^{64}$] Unidad Profesional Interdisciplinaria en Ingenier\'\i{}a y Tecnolog\'\i{}as Avanzadas del Instituto Polit\'ecnico Nacional (UPIITA-IPN), M\'exico, D.F., M\'exico
\item[$^{65}$] Universidad Aut\'onoma de Chiapas, Tuxtla Guti\'errez, Chiapas, M\'exico
\item[$^{66}$] Universidad Michoacana de San Nicol\'as de Hidalgo, Morelia, Michoac\'an, M\'exico
\item[$^{67}$] Universidad Nacional Aut\'onoma de M\'exico, M\'exico, D.F., M\'exico
\item[$^{68}$] Universidad Nacional de San Agustin de Arequipa, Facultad de Ciencias Naturales y Formales, Arequipa, Peru
\item[$^{69}$] Institute of Nuclear Physics PAN, Krakow, Poland
\item[$^{70}$] University of \L{}\'od\'z, Faculty of High-Energy Astrophysics,\L{}\'od\'z, Poland
\item[$^{71}$] Laborat\'orio de Instrumenta\c{c}\~ao e F\'\i{}sica Experimental de Part\'\i{}culas -- LIP and Instituto Superior T\'ecnico -- IST, Universidade de Lisboa -- UL, Lisboa, Portugal
\item[$^{72}$] ``Horia Hulubei'' National Institute for Physics and Nuclear Engineering, Bucharest-Magurele, Romania
\item[$^{73}$] Institute of Space Science, Bucharest-Magurele, Romania
\item[$^{74}$] Center for Astrophysics and Cosmology (CAC), University of Nova Gorica, Nova Gorica, Slovenia
\item[$^{75}$] Experimental Particle Physics Department, J.\ Stefan Institute, Ljubljana, Slovenia
\item[$^{76}$] Universidad de Granada and C.A.F.P.E., Granada, Spain
\item[$^{77}$] Instituto Galego de F\'\i{}sica de Altas Enerx\'\i{}as (IGFAE), Universidade de Santiago de Compostela, Santiago de Compostela, Spain
\item[$^{78}$] IMAPP, Radboud University Nijmegen, Nijmegen, The Netherlands
\item[$^{79}$] Nationaal Instituut voor Kernfysica en Hoge Energie Fysica (NIKHEF), Science Park, Amsterdam, The Netherlands
\item[$^{80}$] Stichting Astronomisch Onderzoek in Nederland (ASTRON), Dwingeloo, The Netherlands
\item[$^{81}$] Universiteit van Amsterdam, Faculty of Science, Amsterdam, The Netherlands
\item[$^{82}$] Case Western Reserve University, Cleveland, OH, USA
\item[$^{83}$] Colorado School of Mines, Golden, CO, USA
\item[$^{84}$] Department of Physics and Astronomy, Lehman College, City University of New York, Bronx, NY, USA
\item[$^{85}$] Michigan Technological University, Houghton, MI, USA
\item[$^{86}$] New York University, New York, NY, USA
\item[$^{87}$] University of Chicago, Enrico Fermi Institute, Chicago, IL, USA
\item[$^{88}$] University of Delaware, Department of Physics and Astronomy, Bartol Research Institute, Newark, DE, USA
\item[$^{89}$] University of Wisconsin-Madison, Department of Physics and WIPAC, Madison, WI, USA
\item[] -----
\item[$^{a}$] Max-Planck-Institut f\"ur Radioastronomie, Bonn, Germany
\item[$^{b}$] also at Kapteyn Institute, University of Groningen, Groningen, The Netherlands
\item[$^{c}$] School of Physics and Astronomy, University of Leeds, Leeds, United Kingdom
\item[$^{d}$] Fermi National Accelerator Laboratory, Fermilab, Batavia, IL, USA
\item[$^{e}$] Pennsylvania State University, University Park, PA, USA
\item[$^{f}$] Colorado State University, Fort Collins, CO, USA
\item[$^{g}$] Louisiana State University, Baton Rouge, LA, USA
\item[$^{h}$] now at Agenzia Spaziale Italiana (ASI).\ Via del Politecnico 00133, Roma, Italy
\item[$^{i}$] now at ECAP, FAU Erlangen-N\"urnberg, Erlangen, Germany
\item[$^{j}$] now at Graduate School of Science, Osaka Metropolitan University, Osaka, Japan
\item[$^{k}$] Institut universitaire de France (IUF), France
\end{description}
}

\begin{abstract}
We report an investigation of the mass composition of cosmic rays with energies from 3 to 100~EeV (1~EeV=$10^{18}$~eV) using the distributions of the depth of shower maximum \xmax. 
The analysis relies on ${\sim}50,000$ events recorded by the Surface Detector of the Pierre Auger Observatory and a deep-learning-based reconstruction algorithm.
Above energies of 5~EeV, the data set offers a 10-fold increase in statistics with respect to fluorescence measurements at the Observatory. After cross-calibration using the Fluorescence Detector, this enables the first measurement of the evolution of the mean and the standard deviation of the \xmax distributions up to 100 EeV.
Our findings are threefold: 
\begin{enumerate}[rightmargin=6em, label=(\roman*)]
    \item The evolution of the mean logarithmic mass towards a heavier composition with increasing energy can be confirmed and is extended to 100~EeV.
    \item The evolution of the fluctuations of \xmax towards a heavier and purer composition with increasing energy can be confirmed with high statistics. We report a rather heavy composition and small fluctuations in \xmax at the highest energies.
    \item We find indications for a characteristic structure beyond a constant change in the mean logarithmic mass, featuring three breaks that are observed in proximity to the ankle, instep, and suppression features in the energy spectrum.
\end{enumerate}
\end{abstract}

\maketitle

\hspace{10em}
\section{\label{sec:intro}Introduction}
To understand the physics of ultra-high-energy cosmic rays (UHECRs), including their origin, the measurement of their mass composition is of fundamental importance.
On the one hand, an event-by-event determination of mass enables estimations of the particle charges, which are valuable when performing arrival-direction analyses in the presence of magnetic fields.
On the other hand, it provides insights into whether the observed flux suppression at the end of the cosmic-ray spectrum~\cite{Hires2008, Auger_PRL_spectrum_2008, TA_energy_spectrum, the_pierre_auger_collaboration_features_2020} is a signature of the interaction of the particles with the cosmic microwave background~\cite{greisen_end_1966, zatsepin_upper_1966}, a consequence of a limit of the maximum injection energy of the cosmic accelerators~\cite{Peters1961PrimaryCR, Allard_2008}, or a combination of both~\cite{combined_fit_auger, combined_fit_eleonora}.
Whereas for the former, due to photodisintegration during the propagation, a change in the composition is expected that scales with the energy per nucleon ($E/A$), for the latter, the so-called Peters cycle, a change in composition scaling with rigidity ($E/Z$) is expected.

Due to the rapid decrease in particle flux at ultra-high energies, modern cosmic-ray observatories perform indirect measurements of the rare particles by detecting generated air showers instead.
The influence of the primary mass on the shower development can be characterized mainly by the number of muons and the atmospheric depth of the shower maximum \xmax at which the shower reaches its maximum size.
At a given primary energy, with increasing primary mass, the number of induced sub-showers increases, and the energy per nucleon reduces, leading to an \xmax higher up in the atmosphere and decreasing shower fluctuations.
The increase in the number of sub-showers additionally causes an increase in the number of produced muons.
Since the current generation of hadronic interaction models cannot describe the muon component in full detail~\cite{Aab_2015_muons, Aab_2016_test_models, muons_amiga, the_pierre_auger_collaboration_measurement_muon_fluctuations}, currently the most precise composition studies at ultra-high energies rely on measurements of \xmax.

\begin{figure*}[t!]
    \begin{centering}
        \begin{subfigure}[b]{0.475\textwidth}
            \begin{centering}
                \includegraphics[height=7.5cm]{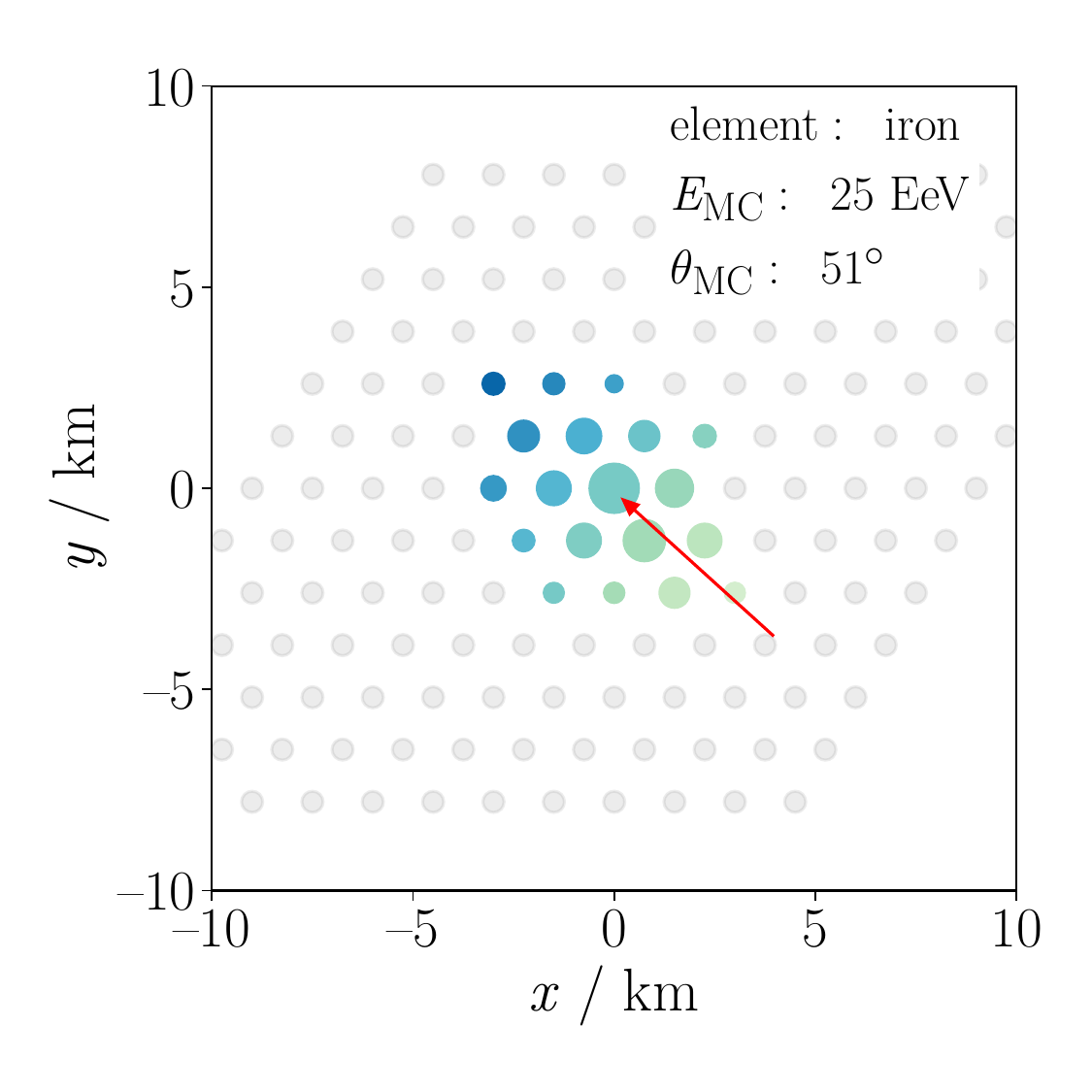}
                \subcaption{}
                \label{fig:footprint}
            \end{centering}
        \end{subfigure}
        \begin{subfigure}[b]{0.475\textwidth}
            \begin{centering}
                \includegraphics[height=7.5cm]{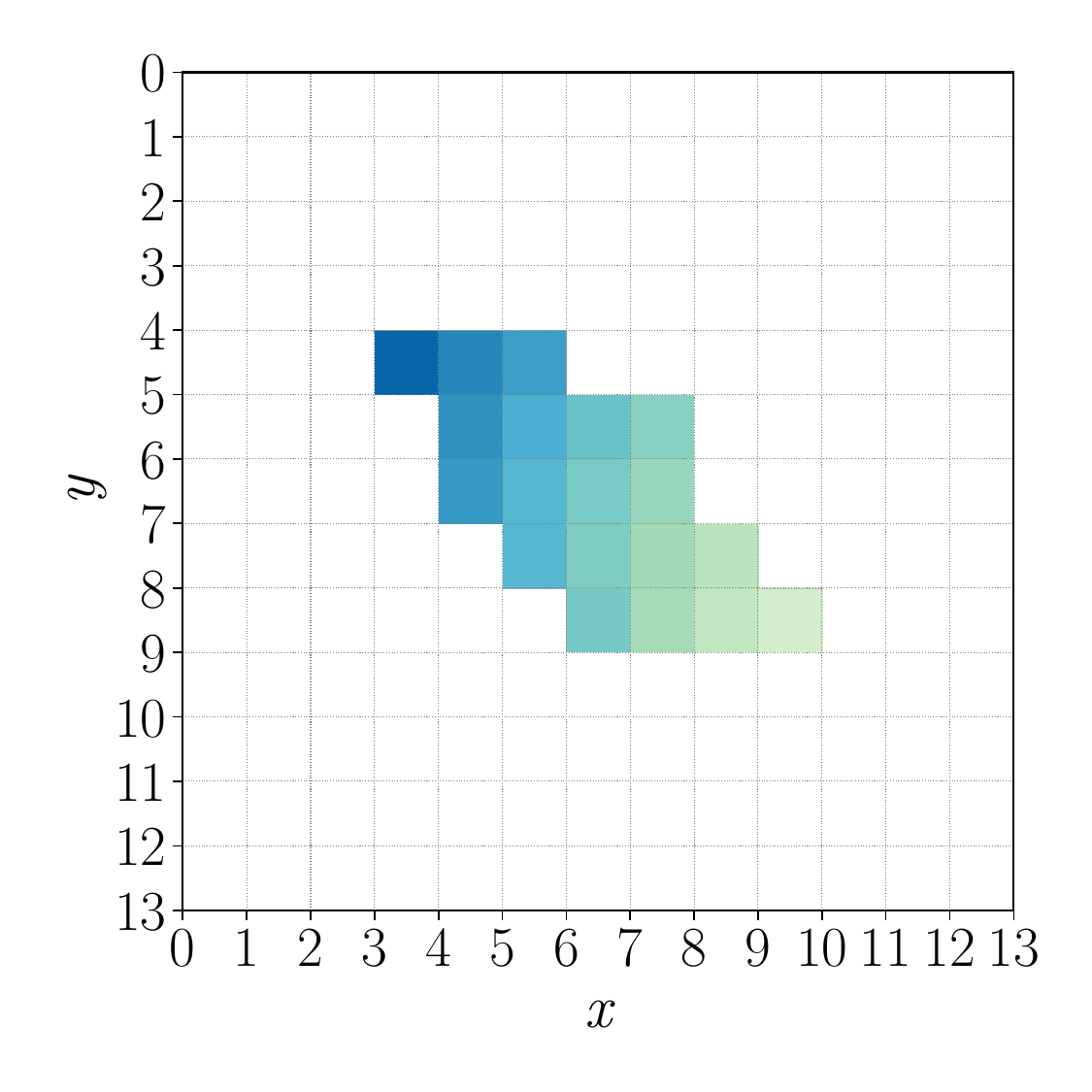}
                \subcaption{}
                \label{fig:footprint_processed}
            \end{centering}
        \end{subfigure}
    \end{centering}
    \caption{Simulated footprint of a cosmic-ray event measured by the SD. (a) Cutout of $13\times 13$ stations of an SD event containing the induced signal pattern on the triangular grid. The marker size indicates the logarithm of the total measured signal, the color denotes the arrival time (green for early, blue for late), the arrow marks the projection of the shower axis on the ground, and its tip denotes the shower core. (b) Representation of the event on a Cartesian grid after pre-processing as a cutout with dimensions $13\times 13$ as used for the DNN after axial indexing. The color indicates the arrival time of the shower front at each station.}
    \label{fig:footprint_data}
\end{figure*}

Studying the energy evolution of \xmaxmu, the mean of the distribution, enables us to directly examine the evolution of \lnamu, the mean logarithmic mass.
The evolution of the fluctuations \xmaxsigma, i.e., the standard deviation of the distribution, provides additional insights into the composition and its mixing~\cite{Kampert_unger_2012, interpretation_auger_jcap}. 

In the past two decades, significant progress in our understanding of UHECRs has been made, largely attributed to the establishment of the Pierre Auger Observatory~\cite{auger_nim} and the Telescope Array Project~\cite{ta_kawai}.
The Pierre Auger Observatory is the world's largest cosmic ray detector and is composed of a Surface Detector (SD) and a Fluorescence Detector (FD).
By observing the longitudinal shower profile of extensive air showers, the FD telescopes of the Observatory not only enable the precise determination of the shower energy but also provide an accurate determination of \xmax~\cite{abraham_pierre_auger_collaboration_measurement_2010, aab_pierre_auger_collaboration_depth_2014}.
Currently, the most precise mass composition studies rely on these fluorescence observations.
However, the operation of fluorescence telescopes is confined to dark and moonless nights, resulting in a duty cycle of around 15\%.
Additionally, for an unbiased \xmax data set, further cuts have to be applied.
In contrast, the duty cycle of the SD is close to 100\%, enabling composition studies with high statistics.

Recently, several methods have been developed to infer mass-sensitive information using the SD. Using the risetime of signals in the water-Cherenkov detectors, the evolution of the average mass composition as a function of energy can be studied with good precision~\cite{aab_pierre_auger_collaboration_inferences_2017}.
Furthermore, the phenomenological approach of shower universality~\cite{univ}, based on a decomposition of the measured detector signals into the different shower components, has shown first promising results in the reconstruction of \lnamu.
To determine the fluctuations in \xmax, i.e., to measure \xmaxsigma, yet more precise, event-by-event measurements are needed.
Measuring \xmaxsigma is particularly important as its interpretation does not depend strongly on hadronic interaction models, as a considerable part of the fluctuations depends on the mean free path of the first interaction and, thus, the cross-section at the highest energies.

With the advent of deep learning, new possibilities have emerged for designing learning algorithms, i.e., deep neural networks (DNNs), to analyze high-dimensional and complex data in computational sciences~\cite{lecun2015deep, Goodfellow-et-al-2016} as well as in physics~\cite{dlfpr}.
Trained on large simulation libraries, these algorithms are capable of recognizing small patterns~\cite{zeiler2013visualizing}, like complex mesh structures, skin texture, or dog faces, to which conventional methods were previously not sensitive.
This recent progress provided improved reconstruction algorithms in astroparticle physics, e.g., imaging air Cherenkov telescopes~\cite{Shilon_2019}, gravitational wave detection~\cite{Cuoco_2020}, neutrino~\cite{abbasi_convolutional_2021, km3net_dl, gal_plane_2023} and cosmic-ray observatories~\cite{erdmann_deep_2018}, including the reconstruction of \xmax~\cite{xmax_wcd, glombitza_icrc_23} and other air shower properties~\cite{pao_muon_dnn, glombitza_icrc_19}.
So far, the potential of deep-learning-based methods for improved reconstruction in astroparticle physics has been demonstrated, but the application to measured data, including a comprehensive study of systematic uncertainties and associated new insights, is limited.
This work aims to close this gap and shows a successful application to measured data starting at the raw detector signals.

In this article and the accompanying Letter~\cite{wcd_dnn_prl}, we report on the first investigation of the UHECR mass composition based on the first and second moment of the \xmax distributions from 3 to 100~EeV using the SD.
The data set, reconstructed using a novel deep-learning-based reconstruction method, offers an increase in statistics with respect to analyses based on fluorescence observations, which amounts to a factor of ten above 5~EeV.

By cross-calibrating the developed algorithm using hybrid events --- events that feature SD and FD reconstruction --- we find an excellent agreement with previous analyses.
The new measurement of \xmaxmu and \xmaxsigma up to the highest energies is subject to minor systematic uncertainties and avoids the large statistical uncertainties present in previous work. As a result, it offers new insights into the composition at ultra-high energies.

\section{\label{sec:pao}The Pierre Auger Observatory}
The Pierre Auger Observatory, fully commissioned in 2008, is located in the Pampa Amarilla in Argentina at an altitude of ${\sim}1500\,$m, which corresponds to about $875~\gcm$ of atmospheric overburden.
The SD of the Observatory~\cite{auger_nim} comprises an array of 1660 water-Cherenkov detectors (WCDs) placed on a triangular grid with a spacing of $1500\,$m and covering an area of about $3000\,\mathrm{km}^2$.
Each WCD is composed of a sealed liner with a diameter of 3.6~m and a height of 1.2~m filled with $12,000$~liters of ultra-pure water.
Three 9-inch photomultiplier tubes (PMTs) look downward through transparent windows into the water volume to record the Cherenkov light of relativistic charged particles penetrating the walls.
The signal measured by each PMT is digitized by a 40~MHz Flash Analog-to-Digital Converter, corresponding to a bin width of $25$~ns in time.
Due to the limited available bandwidth, the time-dependent signals, i.e., \emph{signal traces}, are only collected if a signal was measured in at least three WCD stations in temporal and spatial coincidence.
In addition, the current parameters for calibrating the signals into units of VEM~\cite{Bertou_2006} (vertical equivalent muon) --- defined as the average signal of a single muon induced when passing the detector vertically through the center of the tank --- are sent.
These are updated every minute and provide reliable signal sizes even during strongly varying operation conditions.
This in situ calibration, together with solar-powered electronics and a battery, offers a duty cycle of the SD close to $100\%$.

The SD array is overlooked by 27 telescopes located at four different sites at the borders of the Observatory.
Three sites host six, and one hosts nine Schmidt telescopes, each composed of a $13~\mathrm{m}^2$ mirror and a 440-pixel camera to observe the longitudinal shower development using the isotropically-emitted fluorescence light.
At the Coihueco site, three High Elevation Auger Telescopes are used to detect low-energy (down to $10^{17}$~eV) showers.
To ensure the most precise observations, the atmospheric conditions are monitored using probing beams of two laser facilities placed close to the center of the array.
For more details on the design and operation of the Observatory we refer to Ref.~\cite{auger_nim}.

\subsection{Surface Detector data}
\begin{figure*}[t!]
    \begin{centering}
        \begin{subfigure}{0.489\textwidth}
            \begin{centering}
                \includegraphics[width=0.99\textwidth]{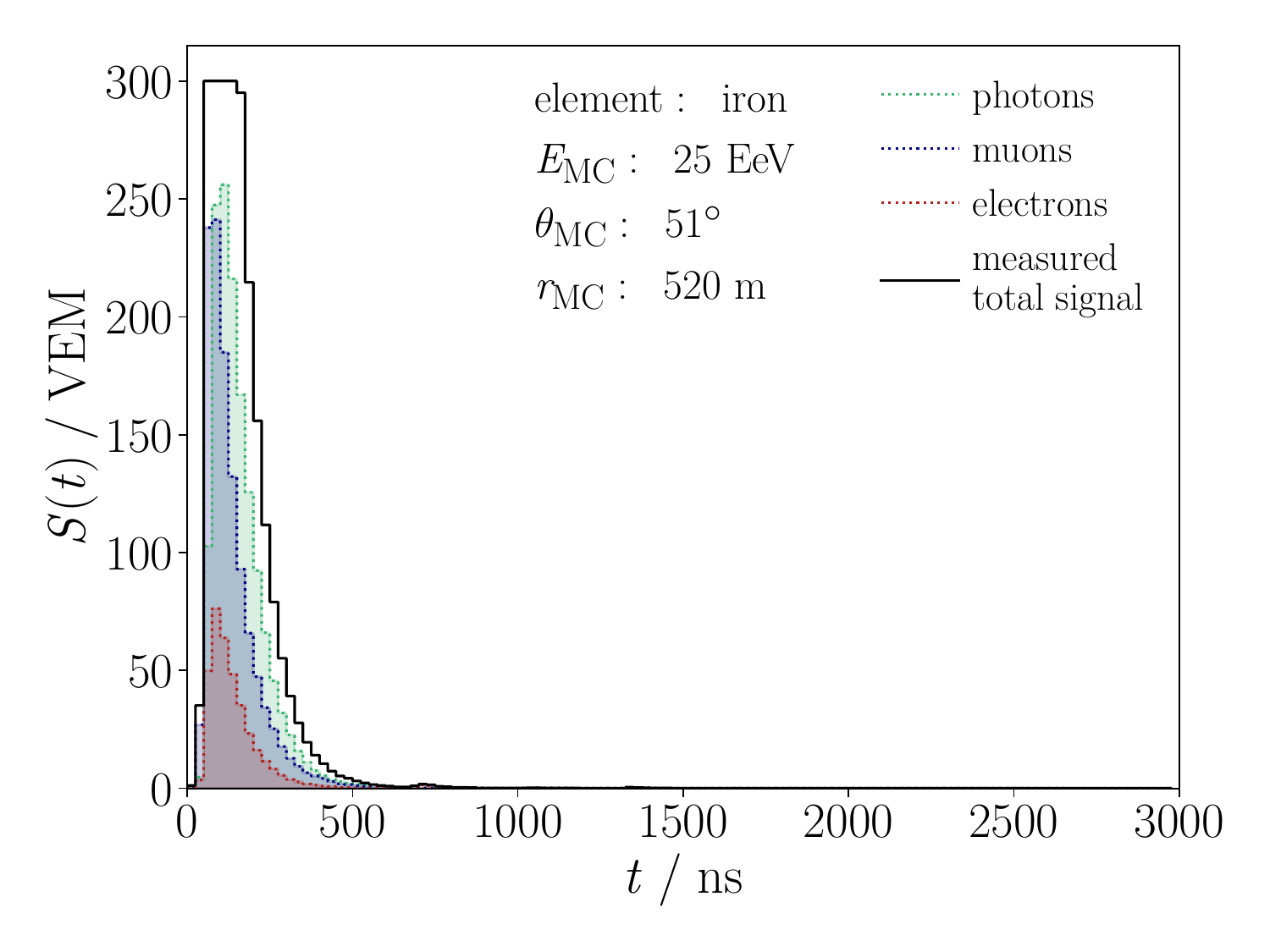}
                \subcaption{}
                \label{fig:trace_raw_1}
            \end{centering}
        \end{subfigure}
        \begin{subfigure}{0.489\textwidth}
            \begin{centering}
                \includegraphics[width=0.99\textwidth]{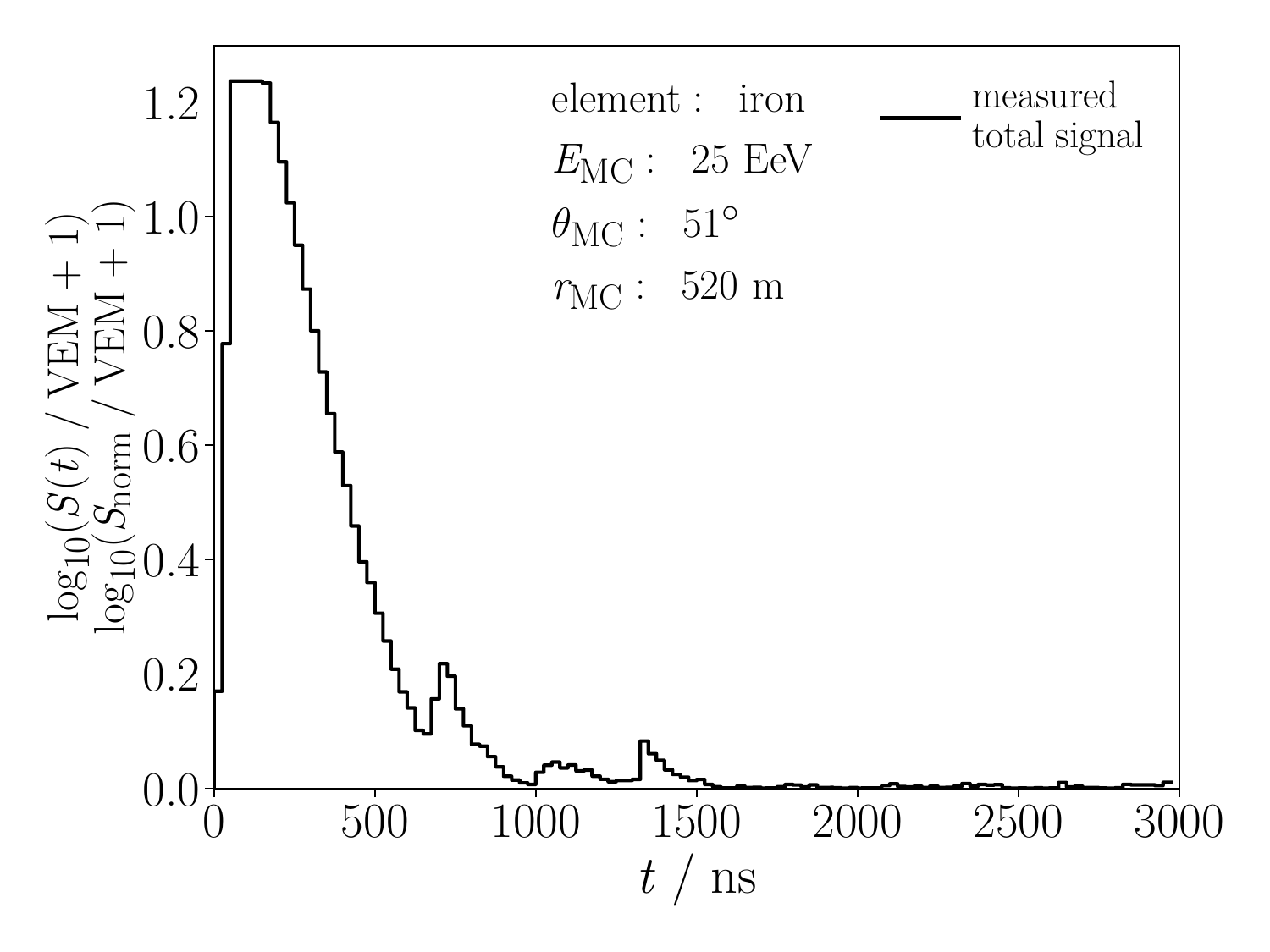}
                \subcaption{}
                \label{fig:trace_processed_1}
            \end{centering}
        \end{subfigure}

        \begin{subfigure}{0.489\textwidth}
            \begin{centering}
                \includegraphics[width=0.99\textwidth]{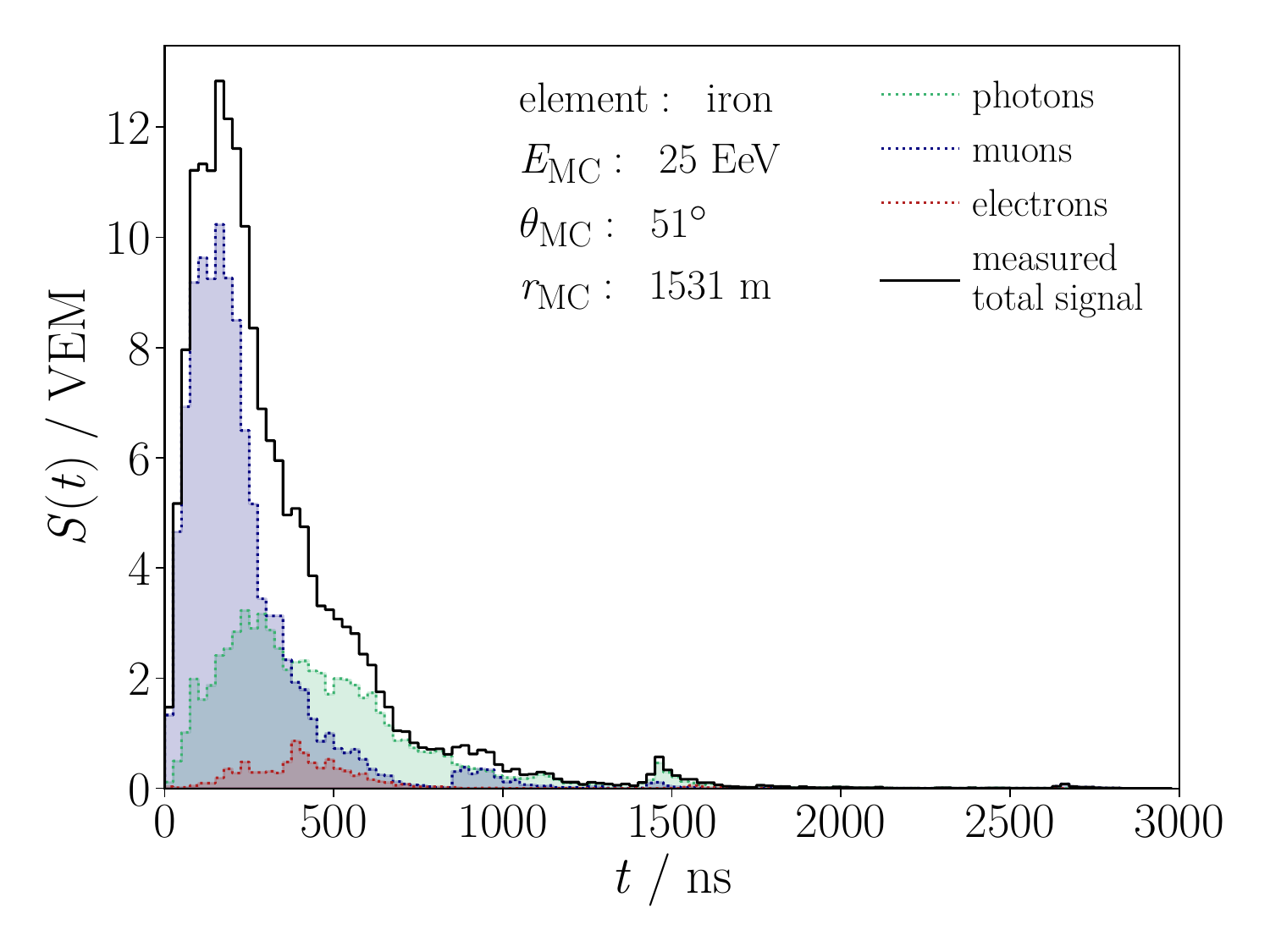}
                \subcaption{}
                \label{fig:trace_raw_2}
            \end{centering}
        \end{subfigure}
        \begin{subfigure}{0.489\textwidth}
            \begin{centering}
                \includegraphics[width=0.99\textwidth]{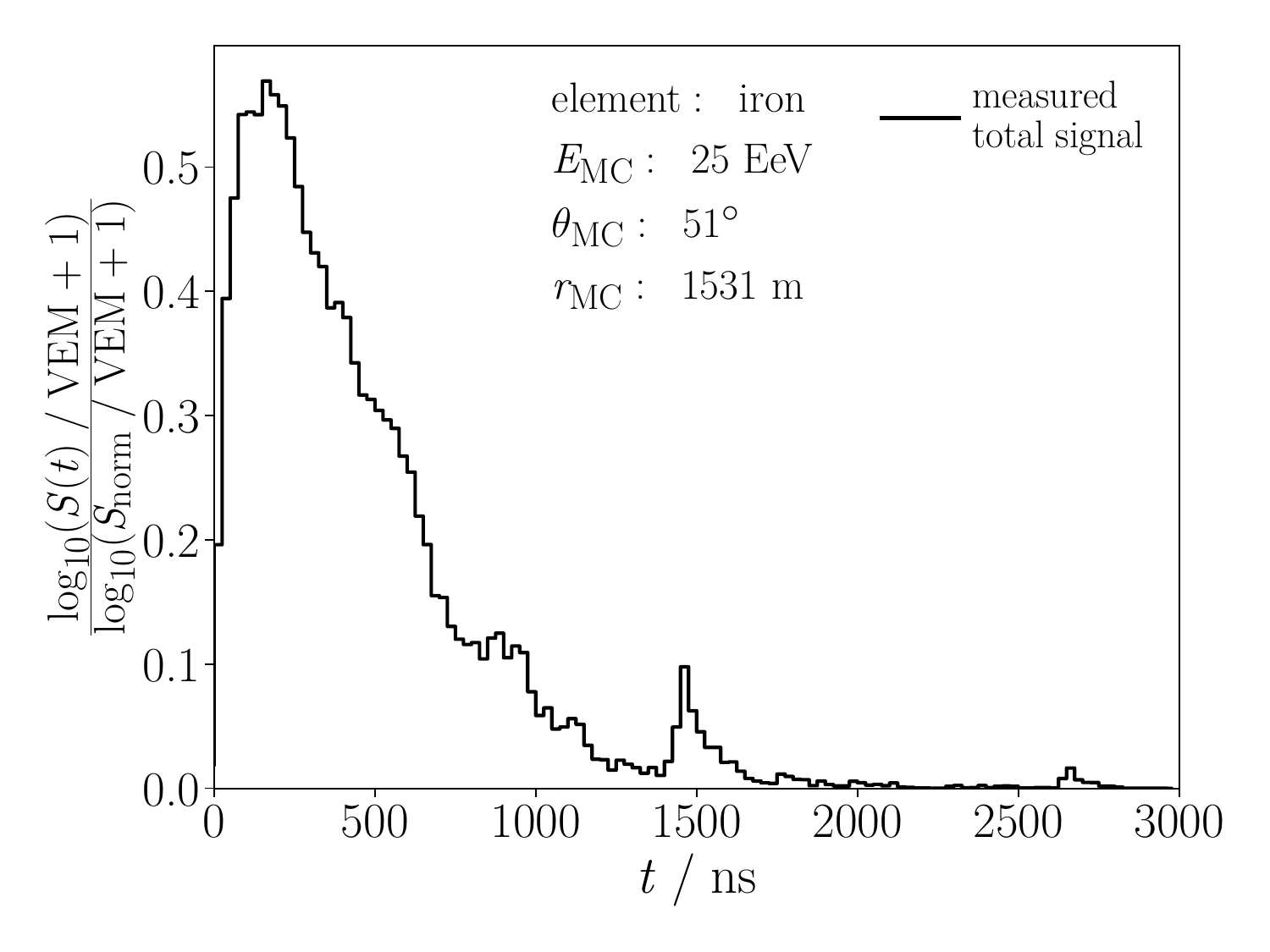}
                \subcaption{}
                \label{fig:trace_processed_2}
            \end{centering}
        \end{subfigure}

        \begin{subfigure}{0.489\textwidth}
            \begin{centering}
                \includegraphics[width=0.99\textwidth]{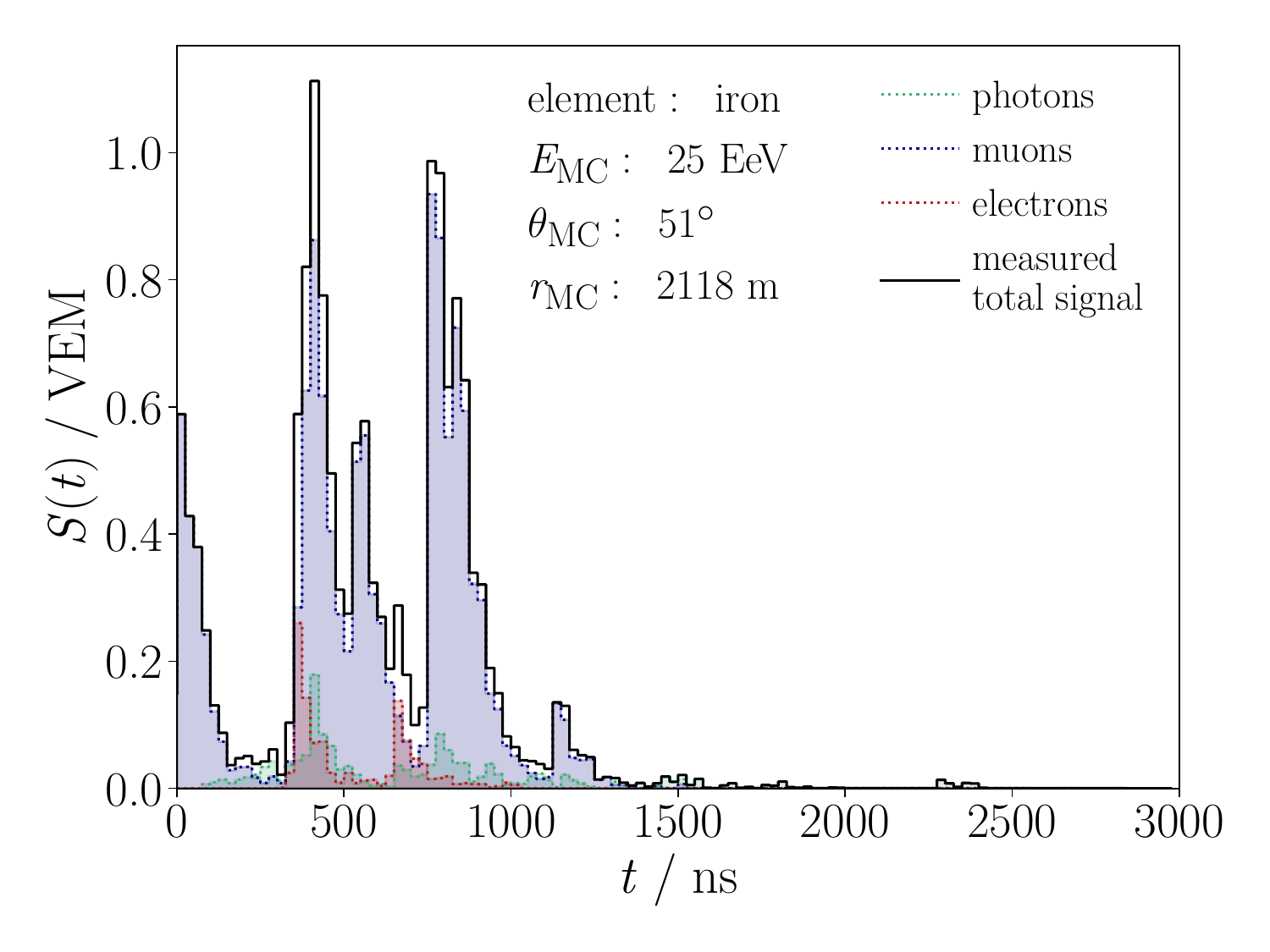}
                \subcaption{}
                \label{fig:trace_raw_3}
            \end{centering}
        \end{subfigure}
        \begin{subfigure}{0.489\textwidth}
            \begin{centering}
                \includegraphics[width=0.99\textwidth]{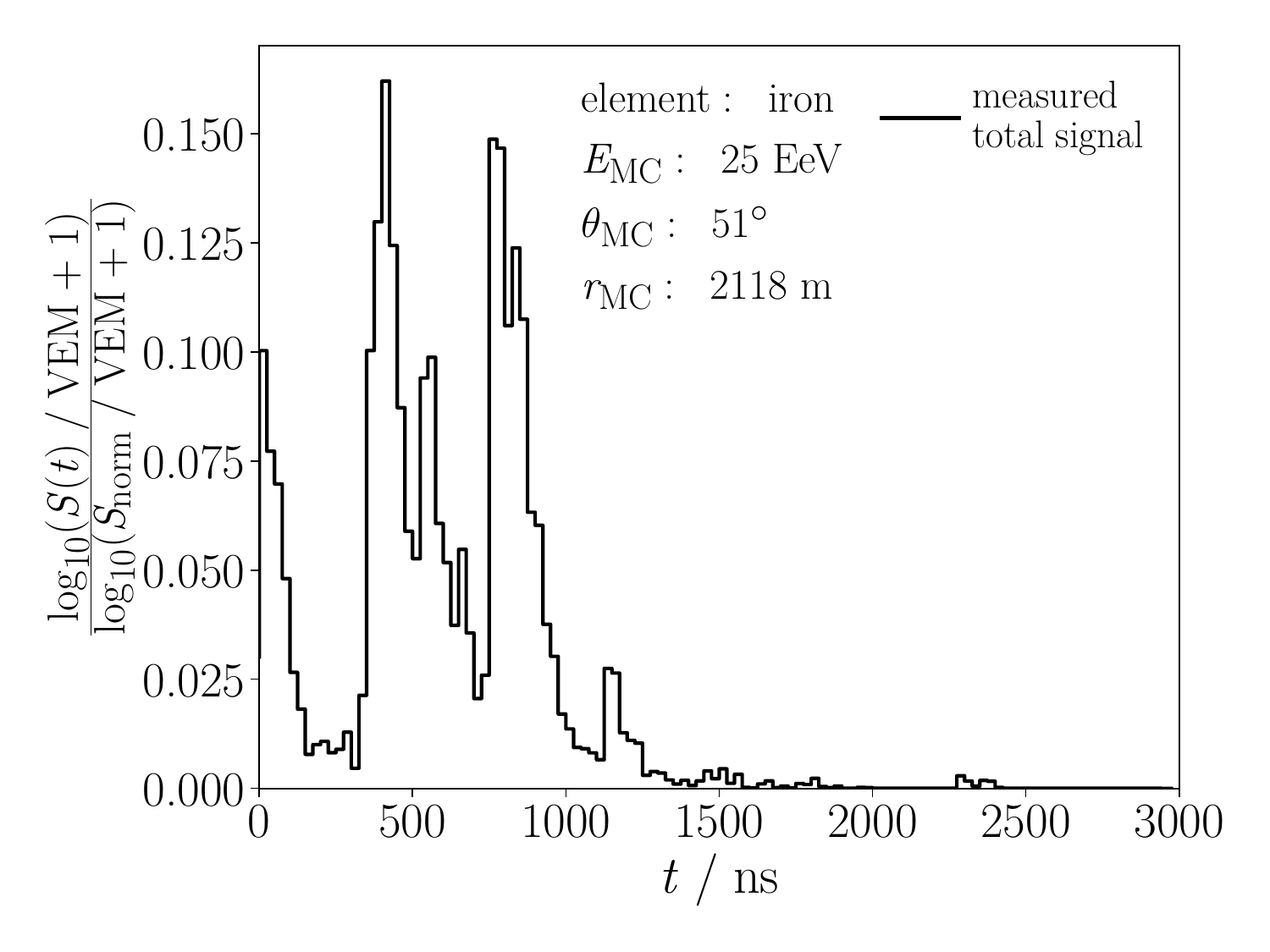}
                \subcaption{}
                \label{fig:trace_processed_3}
            \end{centering}
        \end{subfigure}
    \end{centering}
    \caption{Simulated signal traces of a cosmic-ray event in the SD stations before (left) and after the pre-processing (right). Simulated signal trace of a station close to the shower core (top), at a distance of around 1000\,m (middle), and at a distance of around 3500\,m to the shower core (bottom). Different colors indicate signals from different shower components. The black line denotes the total measured signal, including the saturation effects of the electronics that are only simulated for the sum of all shower components.}
    \label{fig:trace_data}
\end{figure*}

A typical air shower with a zenith angle below $60\grad$ and $E>10$~EeV induces a footprint with the size of tens of square kilometers at the Earth's surface, on average triggering around ten stations.
See \cref{fig:footprint}, for a simulated example event.
For each triggered station, three signal traces are recorded, one measured by each PMT.
The trace has a length of 768 time steps of 25~ns, resulting in a total length of  ${\sim} 20~\mathrm{\upmu s}$.
These traces are further processed with a peak finder to search for the signal window.
In this work, the signal window has a width of 120 time steps ($3~\mathrm{\upmu s}$), which includes more than $99\%$ of the signals.
Simulated example traces of the event shown in \cref{fig:footprint_data} are depicted in \cref{fig:trace_data} for stations located at three different distances to the shower core.
Note that, in contrast to the standard reconstruction, which integrates over the signal window to estimate the shower energy, in this work, we make use of the full signal trace for the \xmax reconstruction.

\clearpage

In addition to the three traces, the \emph{arrival time} of the shower front at each station is estimated based on the starting time of the signal window and the trigger time of each WCD station.
These arrival times, combined with the station positions, encode information on the arrival direction of the primary particle.
They are used in the standard reconstruction to determine the shower axis by fitting a model of an inflating sphere~\cite{sd_reco}.
The algorithm for the reconstruction of \xmax relies on the position of the triggered stations and on both the arrival times and the signal traces measured at each station.

\begin{table}
    \begin{center}
        \begin{tabular}{ l r r }
        \hline\hline
            Cut & Events & $\epsilon$~(\%) \\ 
            \hline
            reconstructed vertical event ($\theta<60^\circ$) & 5,994,712 & --- \\
            is 6T5 & 4,858,291 & 81.1 \\
            $\log_{10}{(E/\mathrm{eV})}> 18.5$ & 133,167 & 2.7 \\
            hardware status & 129,403 & 97.2 \\
            \hline
            station start slot & 128,308 & 99.2 \\
            $2.75 <A/P<3.45$ & 126,033 & 98.2 \\
            $\bar{S}_\mathrm{tot}>5$~VEM in surrounding hexagon & 125,828 & 99.8 \\
            $350~\mathrm{m} <$ core distance $< 1000~\mathrm{m}$ & 101,392 & 80.6  \\
            fiducial SD cut & 48,824 & 48.1 \\
        \hline\hline
        \end{tabular}
        \caption{Basic and analysis-specific selections (separated by a line) for the SD data set.}
        \label{tab:sd_cuts_sd1500}
    \end{center}
\end{table}

\subsection{Data selection}
The data set for the measurement of the cosmic-ray mass composition consists of air shower events recorded with the SD. Additionally, for the calibration and the validation of the reconstruction, hybrid measurements --- events detected by both the SD and the FD --- are utilized.

\subsubsection{SD data set}
The data selection for mass composition studies mostly follows the criteria used for determining the energy spectrum~\cite{Aab_2020_spectrum_prd} and is summarized in \cref{tab:sd_cuts_sd1500}.
As pre-selection criteria, we require a successful energy reconstruction, a zenith angle $<60\grad$ to consider vertical showers only, and exclude lightning-induced events.
We further require that the stations with the largest measured signals are surrounded by six working stations (a so-called 6T5 trigger) to ensure that the footprint is sufficiently sampled by the SD and that the events with shower cores outside the array are rejected.
In this analysis, we only consider events with $\log_{10}(E/\mathrm{eV})>18.5$, where the SD is fully efficient in the selected zenith angle range, and keep only events when the SD is properly operational.

\begin{table}
\begin{tabular}{ l r r }
    \hline\hline
    Cut & Events & $\epsilon$~(\%) \\
    \hline
    number of events & 25,076 & --- \\
    telescope cuts  & 19,733 & 78.7 \\
    hardware status & 16,916 & 85.7 \\
    aerosols/clouds & 9,822 & 58.1 \\
    hybrid geometry & 9,157 & 93.2 \\
    fiducial FoV cut & 3,497 & 38.2 \\
    profile cuts & 3,331 & 95.3 \\
    \hline
    passed SD selection & 3,086 & 92.6 \\
    analysis-specific cuts & 1,642 & 53.2 \\
    \hline\hline
\end{tabular}
\caption{Selections for hybrid data.}
\label{tab:hybrids}
\end{table}

In the analysis-specific post-selection, we remove a minor fraction of events where the starting bin of any single signal trace seems to be mis-reconstructed by the peak finder.
We only accept events with an average integrated
signal $\overline{S}_\mathrm{tot}>5$~VEM, for all triggered stations surrounding the station with the largest signal to ensure an adequate measurement of the signal trace.
Additionally, we reject events with very large and very low area-over-peak ($A/P$) ratios that cannot be linearly calibrated during the aging calibration, which is discussed in \cref{sec:sd_calib}.
We additionally remove events with small (350~m) and large (1000~m) distances of the reconstructed shower core to the station with the largest integrated signal to remove events with saturated stations and station multiplicities challenging to reconstruct, respectively.  
Since the multiplicity of triggered stations is small at low energies, and the sampling fluctuations of the WCDs are --- due to the smaller particle density --- high, the \xmax reconstruction bias depends on the zenith angle and the energy.
To obtain an unbiased dataset, we therefore only accept events where the composition bias is small.
This fiducial selection was derived using simulation and is discussed in \cref{sec:fiducial}.
It is strict at energies below 10~EeV, causing, due to the steeply falling energy spectrum, a low overall selection efficiency.
However, at high energies, it hardly causes any statistical disadvantage for the energy-dependent study of the UHECR mass composition.
In total, the SD data set contains, after selection, 48,824 events recorded between 1 January 2004 and 31 August 2018.

\subsubsection{Hybrid measurements}
For the calibration of the DNN to the \xmax scale of the FD, hybrid events with a high-quality reconstruction of the SD and FD data are used.
Thus, besides the selection of the SD events, FD cuts are applied to this data set that follow the selection used in previous composition analyses~\cite{aab_pierre_auger_collaboration_depth_2014}.
The selection is summarized in \cref{tab:hybrids}.
The pre-selection ensures good data-taking conditions by accepting only events with a stable gain calibration of the FD PMTs and adequate observation conditions, i.e., featuring a clear sky and a measurement of the vertical aerosol optical depth within the last hour that guarantees precise measurements.
To ensure an adequate air-shower reconstruction, we require a good ﬁt of the Gaisser--Hillas profile, a minimum observed track length of 200~\gcm, and the \xmax to be reconstructed in the field of view of the telescope with an \xmax uncertainty smaller than $40~\gcm$. 
Since the condition on \xmax to fall in the field of view constrains \xmax and, thus, results in an acceptance that depends on the mass of the primary particle, a fiducial field-of-view cut is applied.
The cut is derived in a data-driven fashion and ensures that only shower geometries that provide unbiased views of the bulk of the \xmax distribution are selected.
This strict criterion removing a significant fraction of events guarantees an unbiased, i.e., composition-independent data sample and has an efficiency of slightly less than $40\%$.
See Ref.~\cite{aab_pierre_auger_collaboration_depth_2014} for more details.
In addition, we remove events with holes in the profile that exceed $20\%$ of the observed track length, and those events with an uncertainty on the energy reconstruction above $12\%$.
In total, 3,331 events remain after the FD selection.
After applying the same SD selection as described above, the hybrid data set comprises 1,642 events measured between 1 January 2004 and 31 December 2017.

\begin{figure}[t!]
    \begin{center}
        \includegraphics[width=0.475\textwidth]{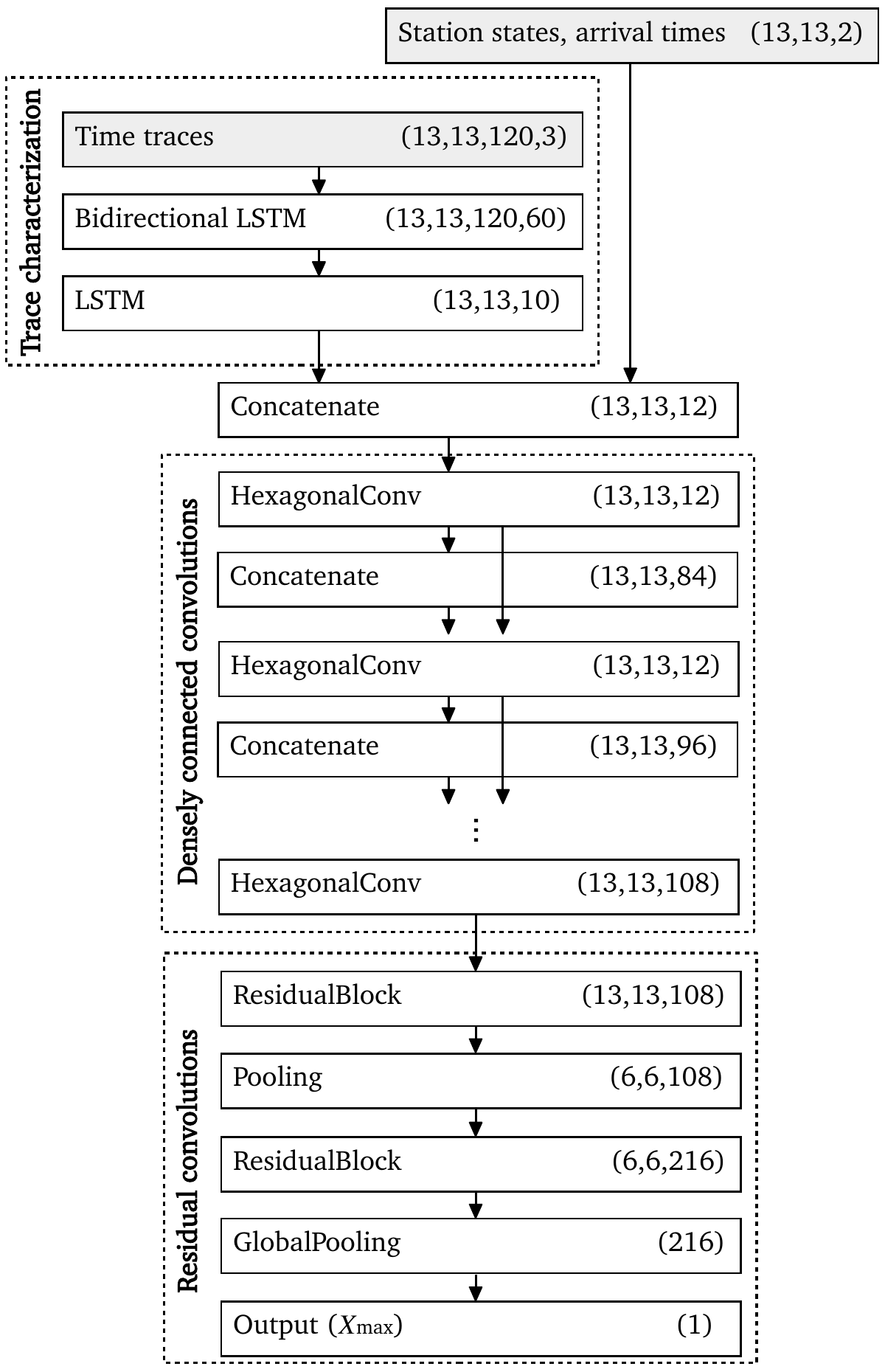}
        \caption{DNN Architecture used to reconstruct \xmax. The numbers in brackets denote the output shapes.}
        \label{fig:architecture}
    \end{center}
\end{figure}

\section{\label{sec:method}Reconstruction of the shower maximum using the Surface Detector and deep Learning}
The previous reconstruction of \xmax on an event-by-event basis was confined to fluorescence telescope data.
To obtain high-statistic measurements of the UHECR mass composition at the highest energies, the reconstruction of \xmax using the SD is a promising solution.
The reconstruction is challenging as, in contrast to the FD, the SD does not directly measure the longitudinal shower development --- enabling straightforward observations of \xmax --- but subsamples the particle density of the particle cascade at the ground.
To infer information on the shower development, the temporal structure of the particle footprint has to be exploited.
On the one hand, different particles induce different signals in the WCDs~\cite{AVE_univ, pao_muon_dnn}, e.g., a single muon typically induces a clear spike as it crosses the station in a straight line (cf.~\cref{fig:trace_raw_3}).
Thus, the SD signals contain information on the absolute density of the respective content.
Additionally, the temporal structure of the signals encodes information on the shower development.
For example, $\upgamma$, $\mathrm{e}^+$, and $\mathrm{e}^-$ that form the electromagnetic component undergo multiple scattering when penetrating the atmosphere, leading to delay and broadening of the signals (cf.~\cref{fig:trace_raw_1} and cf.~\cref{fig:trace_raw_2}), which scale with the distance to \xmax.
The temporal structure of the measured signal in a single station, however, is more complex as it further depends on additional kinematics like the energy and mass of the primary cosmic ray, the zenith angle, and the distance of the station to the shower core.
Previous approaches rely on measuring the signal risetimes~\cite{aab_pierre_auger_collaboration_inferences_2017} and thus provide insights into the muon content.
However, this data-driven approach does not consider all available information on the shower development. 
The complex temporal and spatial information in the SD signals are intractable to analyze using analytical models.
Therefore, complicated parameterizations are needed that rely on simulation libraries.
The phenomenological approach of air-shower universality~\cite{univ, Stadelmaier_univ} utilizes simplifications in order to parameterize and decompose the expected signals, limiting the performance of the algorithm, especially when exploiting the temporal structures of signals with strong fluctuations (signal spikes) beyond the average.
Thus, in this work, we use an alternative approach based on deep neural networks (DNNs).

\subsection{Deep-learning-based reconstruction}
The DNN trained for the reconstruction of \xmax is based on the signal traces measured using the WCDs of the triangular grid of the SD array and the arrival times.
To process the temporal and spatial structure of the particle footprint, the DNN uses the following architecture methodology, shown in \cref{fig:architecture}, separating the analysis in space and time.
Since the SD grid is triangular with a regular spacing of 1500~m, we use the axial representation for re-indexing into a Cartesian grid~\cite{erdmann_deep_2018}.
For a memory-efficient re-indexing, we use a cutout of $13\times13$ stations, where the station with the largest signal defines the center of this grid. 
The dimensions of $13\times13$ stations guarantee that, on average, more than $99.9\%$ of the triggered stations per event are contained within this sub-array.
See \cref{fig:footprint} that visualizes this process using an example SD event.

The time traces $S(t)$ at each time step $t$ are re-scaled using a logarithmic transformation
\begin{equation}
\tilde{S}_i(t) = \frac{\log_{10}\left(S_i(t) / \mathrm{VEM} + 1\right)}{\log_{10}\left(S_\mathrm{norm} / \mathrm{VEM}+1\right)}
\end{equation}
that maps stations with a large signal of $S_\mathrm{norm}=100~\mathrm{VEM}$ to 1 and maintains the physical property that non-triggered stations keep zero signals.
This normalization stabilizes the training process of the DNN.
In a similar way the shower arrival time $t_{0,i}$ at each WCD is normalized with respect to the arrival time $\tau_{\rm center}$ measured at the station with the largest signal, i.e., the center of the cutout, and the standard deviation $\sigma_{t, {\rm data}} = 48.97$~ns of the arrival times estimated over the whole training data set,
\begin{equation}
\label{eq:arr_time}
\tilde{t}_{0,i} = \frac{t_{0,i} - \tau_{\rm center}}{\sigma_{t, {\rm data}}}.
\end{equation}

\begin{figure*}[t!]
    \centering
    \includegraphics[width=0.95\linewidth]{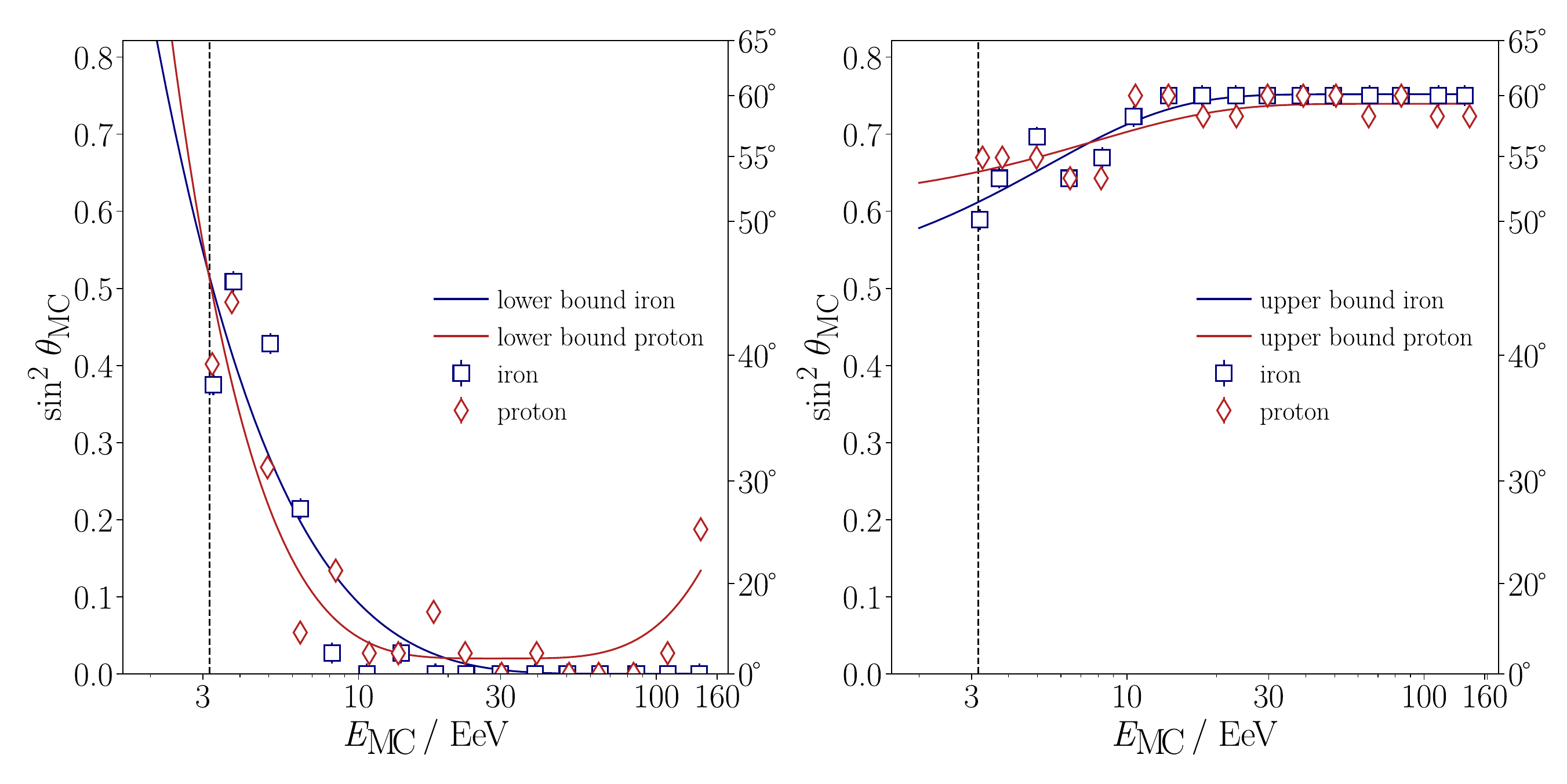} 
    \caption{Determination of the fiducial selection with events simulated using the \sibyllc hadronic interaction model. We show the lower (left) and upper (right) bounds of the selection as derived using a pure proton (red) and pure iron (blue) composition. The markers at a given energy indicate the minimum (lower bound) and maximum (upper bound) zenith angle bin where the reconstruction bias is less than $10$~\gcm. Derived parameterizations are shown as continuous curves. In the ongoing analysis, only events between the upper and lower bounds are accepted.}
    \label{fig:fiducial_cut}
\end{figure*}

To characterize the temporal structure of the signal traces, recurrent long short-term memory (LSTM) layers~\cite{hochreiter_long_1997} are utilized in the first part of the network.
The identical network subpart with the same adaptive parameters is applied to each signal trace, i.e., we apply \emph{weight sharing} along all stations as similar particles induce similar responses in the detector.
The output of this network can be interpreted as an image of $13\times13$ pixels (stations) with ten channels instead of three in a natural RGB image, as the recurrent network part characterizes the traces of each station into ten features.
These features are the input for the convolutional part and are concatenated with the channel of arrival times and an additional channel characterizing the detector states of the surrounding WCDs (working/not working).
The next stage is based on an advanced type of convolutional neural network (CNN)~\cite{lecun2015deep} to exploit the spatial structure of the event. 
We make use of the so-called HexaConv layers~\cite{hoogeboom_hexaconv_2018} and residual connections~\cite{he_deep_2015, huang_densely_2018}, which extends the principle of a filter sliding along an image by a rotation.
This is a meaningful extension as the induced signal patterns are to first order independent of the azimuth angle.
Finally, after a ResNet-like architecture, a single node for the prediction of \xmax forms the output of the DNN.
For a detailed description of the DNN architecture, we refer to Ref.~\cite{xmax_wcd}.

The network was trained using a library~\cite{mc_library, mc_library_2} of $400{,}000$ events with equal fractions of proton-, helium-, oxygen-, and iron-induced showers in an energy range of 1 to 160~EeV with a spectral index of $\gamma = -1$ simulated using CORSIKA~\cite{heck_corsika_1998} with the hadronic interaction model \epos~\cite{pierog_epos_2015} and the FLUKA model~\cite{fasso2003physics}.
We use only events with zenith angles $\theta < 60\grad$ and the full azimuth range ($0^\circ-360^\circ$).
During the training, we perform on-the-fly augmentation of the data using varying detector states\footnote{This includes malfunctioning stations, faulty PMTs, and varying saturation thresholds of the WCD electronics.} to increase the diversity of our data and mimic real operational conditions.
Technical details of the training and the model can be found in Ref.~\cite{xmax_wcd}.

\subsection{\label{sec:fiducial}Fiducial event selection}
The mass-composition analysis in this work relies on the first and second moments of the measured \xmax distributions and their energy evolution.
An unbiased selection of the reconstructed events has to be ensured for a precise determination of the moments.
In contrast to the FD, \xmax cannot directly be observed using the SD but needs to be inferred from the time-resolved particle density at the ground.
Due to the attenuation of the particle density for increased distances between the shower maximum and the detector plane, which further scales with the zenith angle, the amount of information encoded in the sampled signals depends on the shower geometry and the energy.
For example, at very low energies, there are fewer particles in the shower, and \xmax is farther away from the detector. 
This will lead to a smaller particle density at the ground, i.e., fewer triggered stations, and fewer particles arriving per station to be analyzed by the DNN, making the already challenging measurement of \xmax intractable.

\begin{figure*}[t!]
    \begin{centering}
    \includegraphics[width=0.99\textwidth]{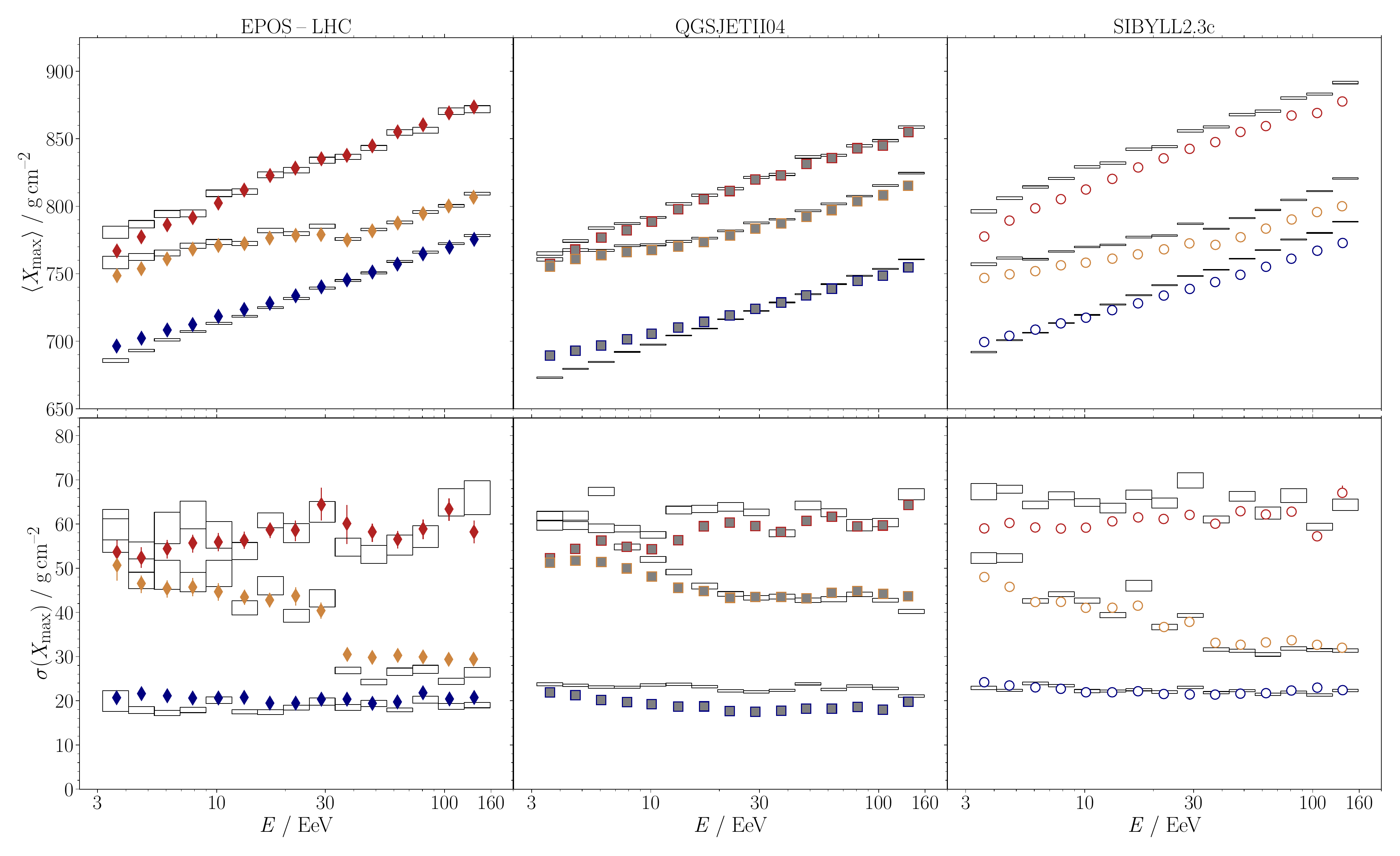}
    \end{centering}    
    \caption{Reconstructed (top) first moments and (bottom) second moments of the \xmax distributions as a function of energy using the SD for the scenarios of a pure proton (red), Auger mix (yellow), and pure iron (blue) composition for showers simulated using \epos (left, filled diamonds), \qgs (center, grey-filled squares), and \sibyllc (right, open circles). The injected (true) \xmax moments, prior to the energy and \xmax reconstruction, are shown as white boxes. The reconstruction of the DNN and systematic effects such as composition-dependent resolution and bias of the SD-based energy reconstruction are considered using forward folding. The fiducial SD selection is applied. Note that \epos was used as a hadronic interaction model for training the DNN.}
    \label{fig:comp_bias}
\end{figure*}

To avoid selections depending on \xmax, and thus the composition itself, we derive upper and lower zenith angle bounds for the selection of air-shower events as a function of energy.
We scan the reconstruction bias for proton and iron-induced showers\footnote{As the reconstruction of proton and iron showers is subject to the largest reconstruction biases. Also see \cref{fig:comp_bias} } as a function of energy and estimate the minimum (maximum) zenith angle at which the absolute reconstruction bias is below $|\Delta \xmax|<10~\gcm$ to derive a lower (upper) bound on the zenith angle.
This is visualized in \cref{fig:fiducial_cut}.

At low energies, for almost vertical showers, the number of triggered stations is small (around 6), and for events with large zenith angles, the signals decrease by up to 50\% due to the increased atmospheric attenuation\footnote{At 10~EeV, $S(1000)$ decreases from around 55\,VEM at $0^\circ$ to 25\,VEM at 60$^\circ$.} between shower maximum and ground level~\cite{auger_nim}.
Whereas the energy and arrival direction can be accurately reconstructed using the SD~\cite{sd_reco}, this leads to a reconstruction bias in \xmax.
At very high energies, events can be reconstructed for zenith angles up to $60\grad$, but for smaller angles, proton showers can develop the shower maximum below the ground, causing biased \xmax reconstructions. 
Therefore, we accept only events if they have zenith angles above the lower and below the upper iron \emph{and} proton bounds at a given energy.
We find the very same dependence across the investigated hadronic interaction models \epos, \qgs, and \sibyllc~\cite{sibyll} and use the selection derived using \sibyllc for further analyses, as it results in the most conservative cut.
The selection removes more than 50\% of events below 5~EeV and is very relaxed above 10~EeV.
Note that the cut is independent of the primary particle mass since only a cut on the zenith angle is performed.
However, due to the steep cosmic ray spectrum, strict cuts at lower energies still enable statistically powerful measurements.
Using this fiducial cut, high quality \xmax measurements are ensured, and a merit factor\footnote{The merit factor is defined as \begin{equation}
    f_\mathrm{MF} = \frac{|\langle X_\mathrm{max, P}\rangle - \langle X_\mathrm{max, Fe}\rangle|}{\sqrt{\sigma^2(X_\mathrm{max, P})+ \sigma^2(X_\mathrm{max, Fe})}},
\end{equation}}
of separating proton and iron close to 1.5 can be reached~\cite{glombitza_icrc_21}.

\subsection{\label{sec:foward_fold}Reconstruction of the $\mathbf{\textit{X}_\text{max}}$ moments}
The determination of the first two moments \xmaxmu and \xmaxsigma of the \xmax distribution and its evolution with energy relies on the \xmax reconstruction of the DNN and the energy estimator $S_{38}$\footnote{Defined as the signal a station measures at a distance of 1000~m to the shower core if the shower would have arrived at a zenith angle of $38^{\circ}$.} from the standard reconstruction of SD data~\cite{sd_reco}.
To examine the quality of the reconstructed \xmax moments, both resolution and bias must be considered.
Therefore, we study hereafter the reconstruction of the \xmax moments using a forward-folding approach.
The bias and resolution of the \xmax and energy reconstruction depend on the composition and energy.
The finite resolution of the energy estimator and its composition bias can cause a spillover of events into neighboring energy bins, depending on the underlying spectrum and the composition.
To handle this effect, we utilize the latest measurement of the UHECR spectrum~\cite{Abreu_2021_spectrum} and consider the trigger efficiency of the SD at low energies.
We investigate this forward-folding approach for the energy evolution of \xmaxmu and \xmaxsigma for three different composition scenarios following the Auger spectrum~\cite{Abreu_2021_spectrum}.

Since proton and iron showers feature the largest reconstruction biases, we study a pure proton and a pure iron composition. Note that this is a conservative approach since previous analyses strongly disfavor significant iron fractions at low and significant proton fractions at high energies~\cite{aab_pierre_auger_collaboration_depth_2014}.
As the most realistic scenario, we also use the \emph{Auger mix}, the composition fractions derived by fitting a template of simulations to the \xmax distributions measured using the FD~\cite{bellido_depth_2018}.
Since the measurement of the FD ends at about 50~EeV, we assume the composition remains unchanged from there onward.

To finally estimate the reconstruction performance, we compare the reconstructed \xmaxmu and \xmaxsigma after the forward-folding process with the injected moments from Monte-Carlo simulations.
To study the composition bias, we use bootstrapping in each bin to estimate \xmaxmu and its statistical uncertainty.
Since a composition bias in the \xmax reconstruction translates into an \xmax dependence of the reconstruction bias, the variance of the reconstructed distribution can be expressed as
\begin{align*}
\sigma^2(\xmaxdnn) &= \underbrace{\sigma^2(\xmax)}_\mathrm{phys.\;fluct.} + \underbrace{\sigma_\mathrm{res}^2(\xmaxdnn)}_\mathrm{resolution}   \\ &\quad + \underbrace{2\operatorname{Cov}(\xmax, \xmaxdnn - \xmax)}_{\xmax\;\mathrm{dependence\;of\;bias}}.
\end{align*}
To reconstruct \xmaxsigma, the resolution and the covariance term have to be considered since the reconstruction shows a composition, i.e., \xmax dependence.
In the absence of sufficient information, estimators like DNNs trained with the mean-squared-error objective function tend to predict samples close to the mean since the reconstruction is ambiguous.
Therefore, particularly at low energies, the DNN is likely to reconstruct, on average, iron with positive and proton with negative \xmax bias.
In turn, the covariance term is negative for the DNN, i.e., it will in part cancel the resolution term.
We studied the dependence of the sum of both terms, the covariance and the resolution for various hadronic models and compositions as a function of energy.
In the case of our trained DNN, we found that, to a good approximation, both terms cancel or are small in comparison to the physical fluctuations in \xmax. 
Therefore, the standard deviation of the distribution formed by the DNN predictions is used as the estimate for \xmaxsigma, and the statistical uncertainty on \xmaxsigma is obtained using bootstrapping. Deviations, i.e., scenarios with non-canceling contributions of the resolution and covariance, will translate into a composition bias of the second moment in this forward-folding approach and propagate into the systematic uncertainties of the \xmaxsigma measurement.

\begin{figure*}[ht!]
    \begin{center}
        \begin{subfigure}[b]{0.46\textwidth}
            \begin{centering}
                \includegraphics[width=0.93\textwidth]{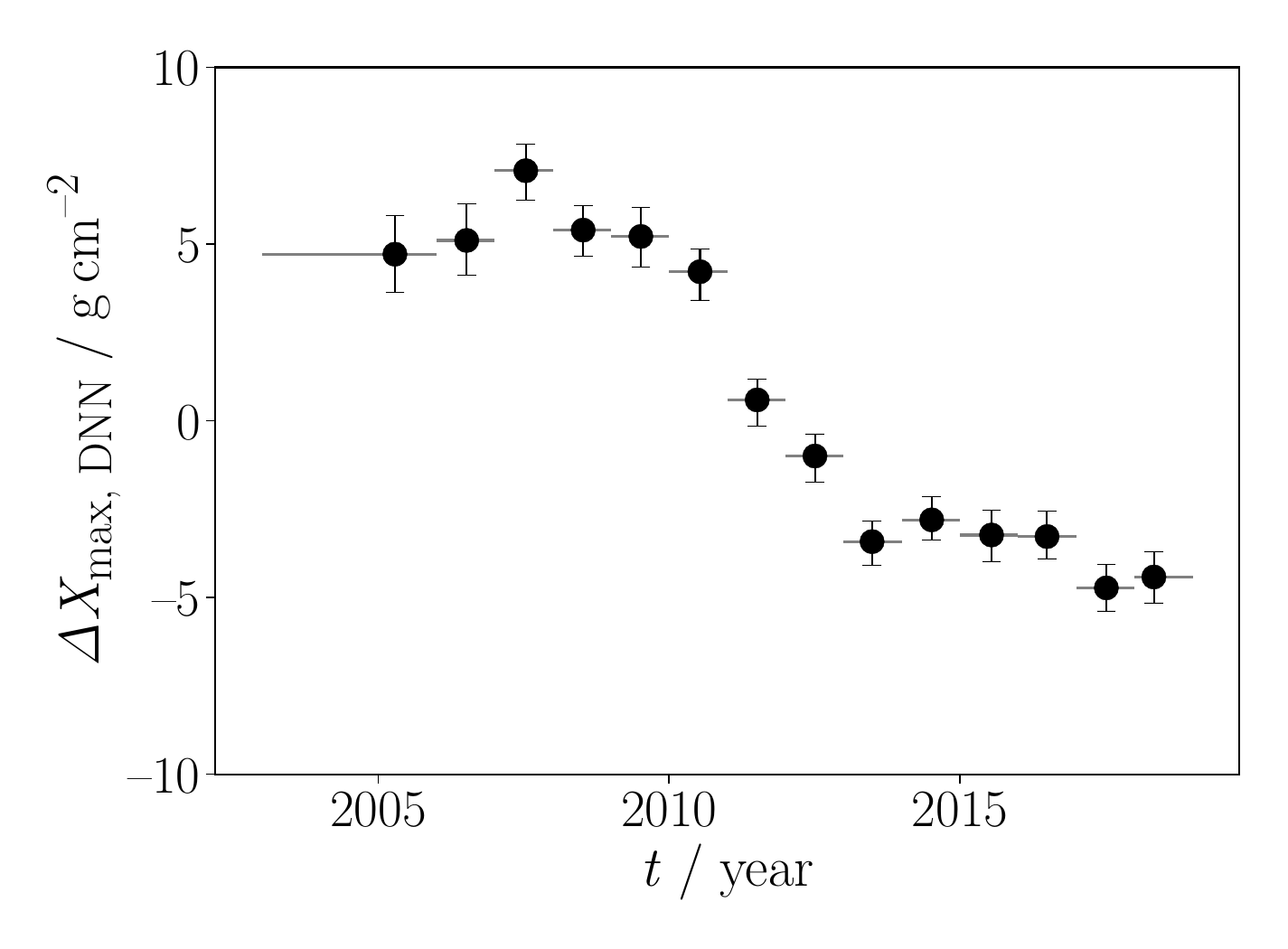}
                \subcaption{}
                \label{fig:aop_before_calib}
            \end{centering}
        \end{subfigure}
        \begin{subfigure}[b]{0.46\textwidth}
            \begin{centering}
                \includegraphics[width=0.93\textwidth]{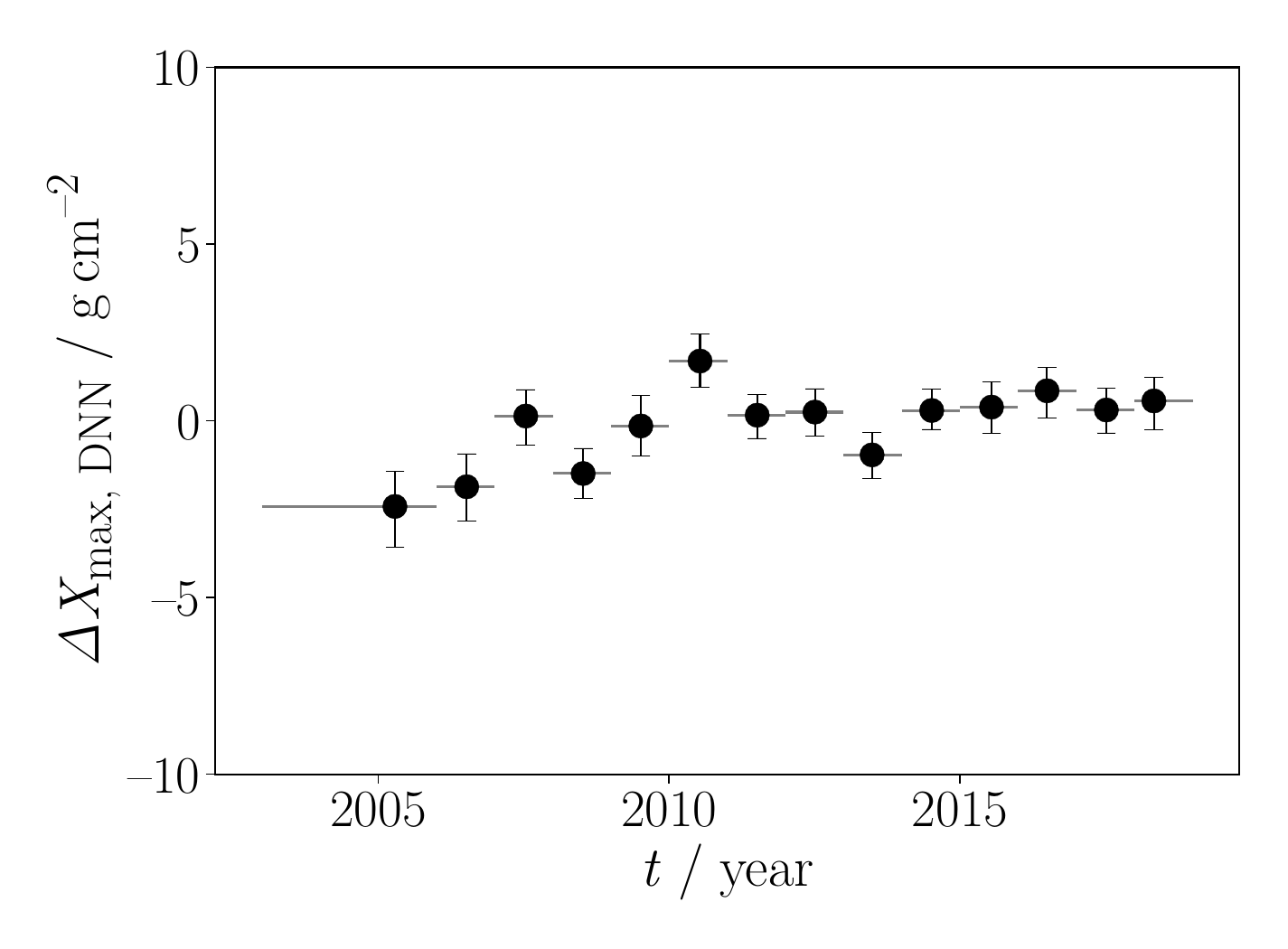}
                \subcaption{}
                \label{fig:aop_calib}
            \end{centering}
        \end{subfigure}     
        \begin{subfigure}[b]{0.46\textwidth} 
            \begin{centering}
                \includegraphics[width=0.93\textwidth]{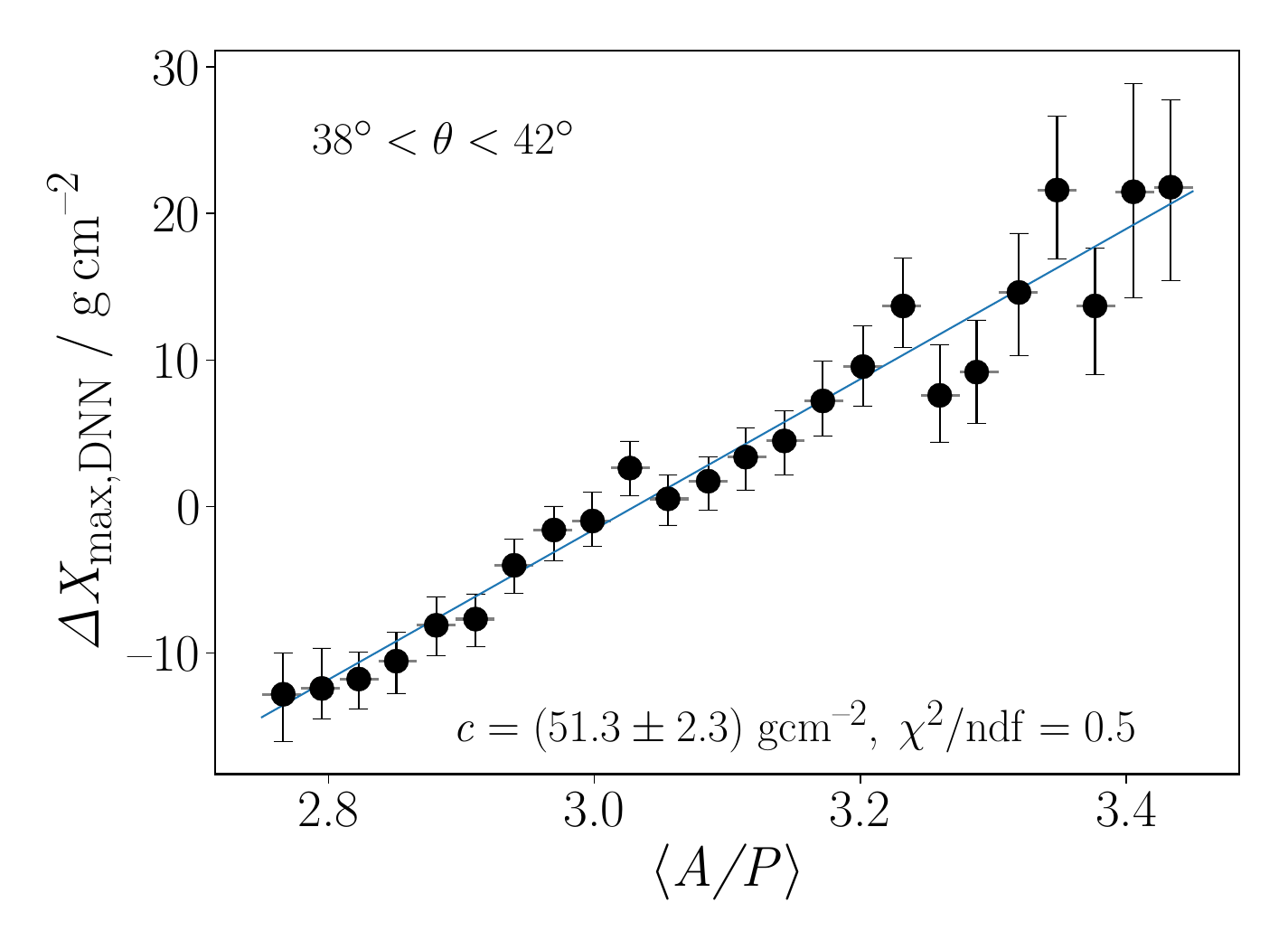}
                \subcaption{}
                \label{fig:aop_calib_zenith}
            \end{centering}
        \end{subfigure}
        \begin{subfigure}[b]{0.46\textwidth} 
            \begin{centering}
                \includegraphics[width=0.93\textwidth]{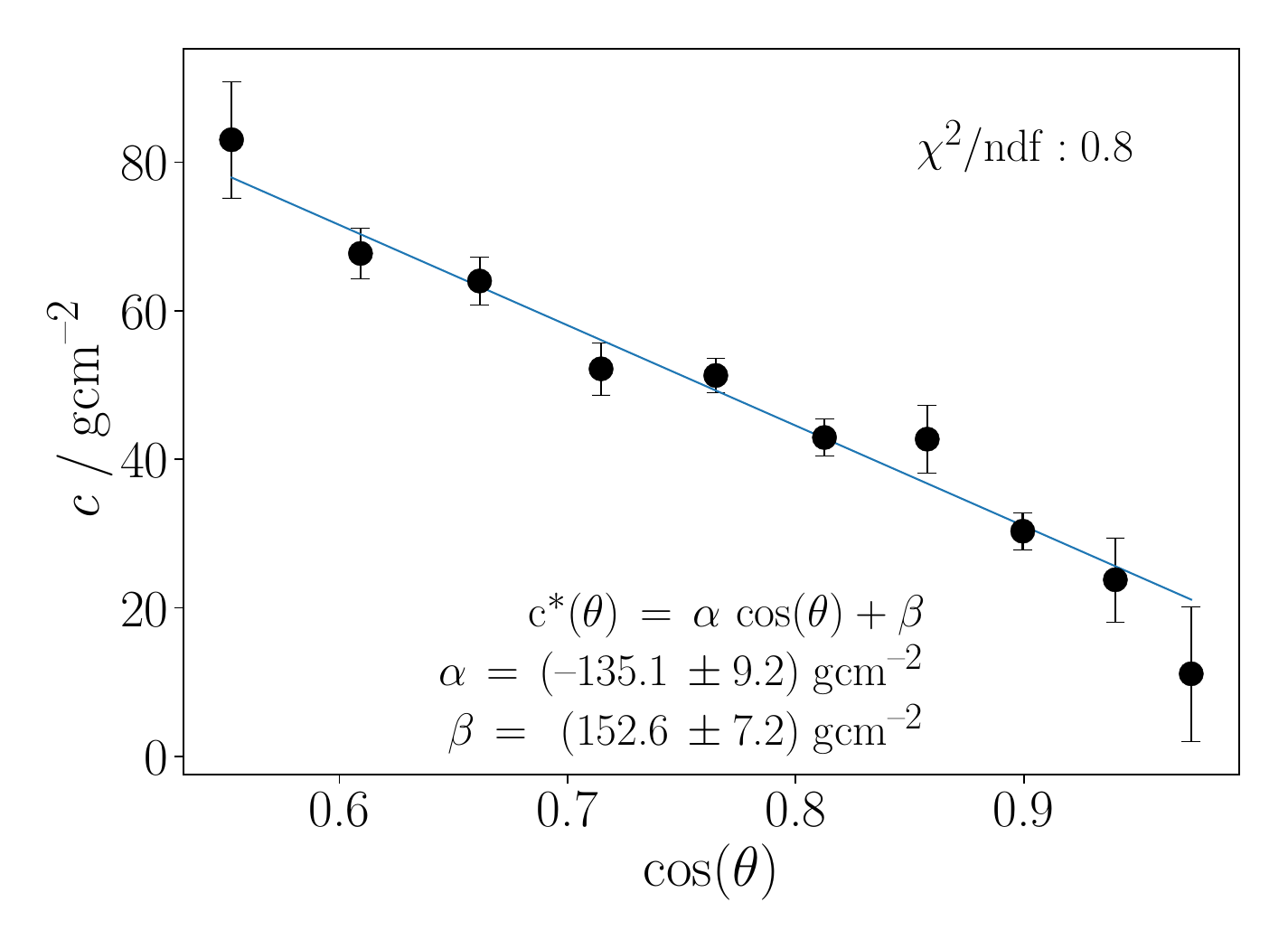}
                \subcaption{}
                \label{fig:aop_calib_zenith_param}
            \end{centering}
        \end{subfigure}
        \caption{\label{fig:aging_calib}Correction of the DNN \xmax predictions for detector aging effects of the SD.
        Decay of the predicted \xmax during the lifetime of the Observatory before (a) and after (b) the calibration. (c) Dependence of the \xmax predictions on the $\langle A/P \rangle$ for an example zenith angle bin. The fitted calibration function is shown in blue. (d) Obtained slope $c$ as a function of the zenith angle. The blue line denotes the fitted parameterization $c^{*}(\theta)$. The obtained parameter for plot (c) corresponds to $\cos(40\grad) \approx 0.77$.}
    \end{center}
\end{figure*}

In \cref{fig:comp_bias}, the performance in reconstructing the evolution of the moments \xmaxmu (top) and \xmaxsigma (bottom) using the SD is depicted for three different scenarios.
A pure proton composition is shown in red, pure iron in blue, and the Auger mix in yellow for the three hadronic interaction models \epos (filled diamonds), \qgs (grey-filled squares), and \sibyllc (open circles).
As a reference, the injected (true) moments are shown as white boxes where their vertical sizes indicate the statistical uncertainty prior to the \xmax reconstruction by the DNN and the energy reconstruction~\cite{sd_reco}.
Note that only \epos was used as a hadronic interaction model for training the DNN.
Since a large fraction of the \epos simulations was used for the DNN training, the statistical uncertainty is larger for \epos events than for \qgs and \sibyllc, where all simulations could be used for testing.
We find that the performance in the determination of \xmaxmu depends on energy and the hadronic interaction models.
An interaction-model bias, i.e., a systematic shift for all compositions, of $-5$~\gcm is visible for \qgs.
For \sibyllc, this bias amounts to $-12$~\gcm.
Because \epos was used for training, no such bias is visible for this model.
Above 10~EeV the performance differences across the models and mass composition scenarios are small~\footnote{A residual plot for the three investigated models can be found in \cref{fig:comp_bias_parameterizations}}.
The reconstruction bias shows a dependence on the composition at low energies.
At $3$~EeV, the pure iron composition is subject to a bias of up to $10\;\;\mathrm{to}\;15$~\gcm that reduces with energy.
For a pure proton composition, a similar dependence is visible, of up to $-10~\gcm$ decreasing with energy.

For the Auger mix, the composition bias is independent of energy for \qgs and \sibyllc. Only for \epos, a small composition bias of up to $-5$~\gcm can be seen below $6$~EeV.
Both the composition bias and the hadronic-interaction model bias would propagate into the systematic uncertainty of the \xmaxmu measurement.
However, since the Observatory features a hybrid detector design, both biases, and their energy dependence can be removed (as investigated in a simulation study) by re-calibrating the DNN measurement using fluorescence observations (as discussed in \cref{sec:hybrid_cal}).

\begin{figure*}[ht!]
    \begin{center}
        \begin{subfigure}[b]{0.46\textwidth}
            \begin{centering}
                \includegraphics[width=0.93\textwidth]{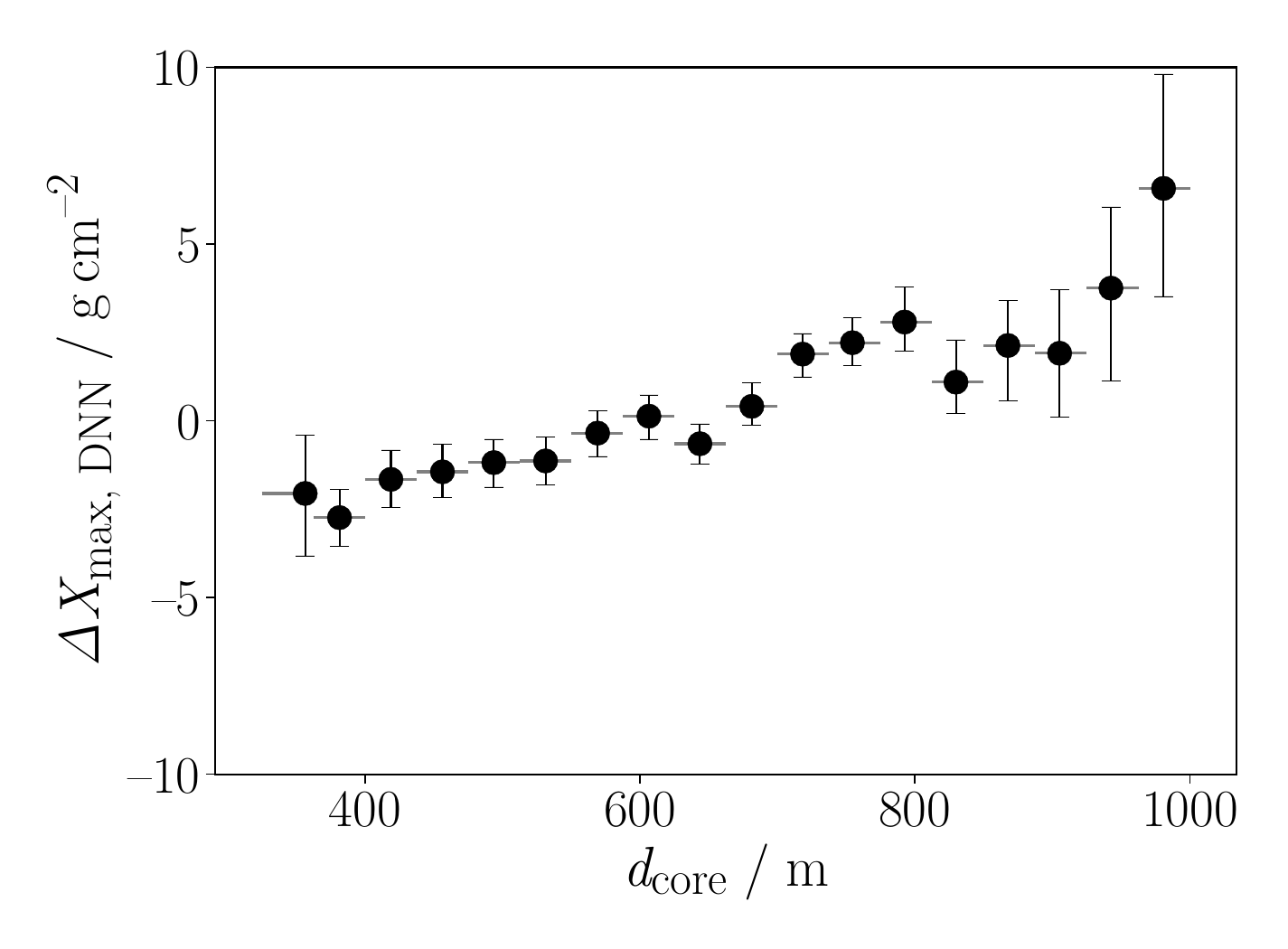}
                \subcaption{}
                \label{fig:core_before}
            \end{centering}
        \end{subfigure}
        \begin{subfigure}[b]{0.46\textwidth}
            \begin{centering}
                \includegraphics[width=0.93\textwidth]{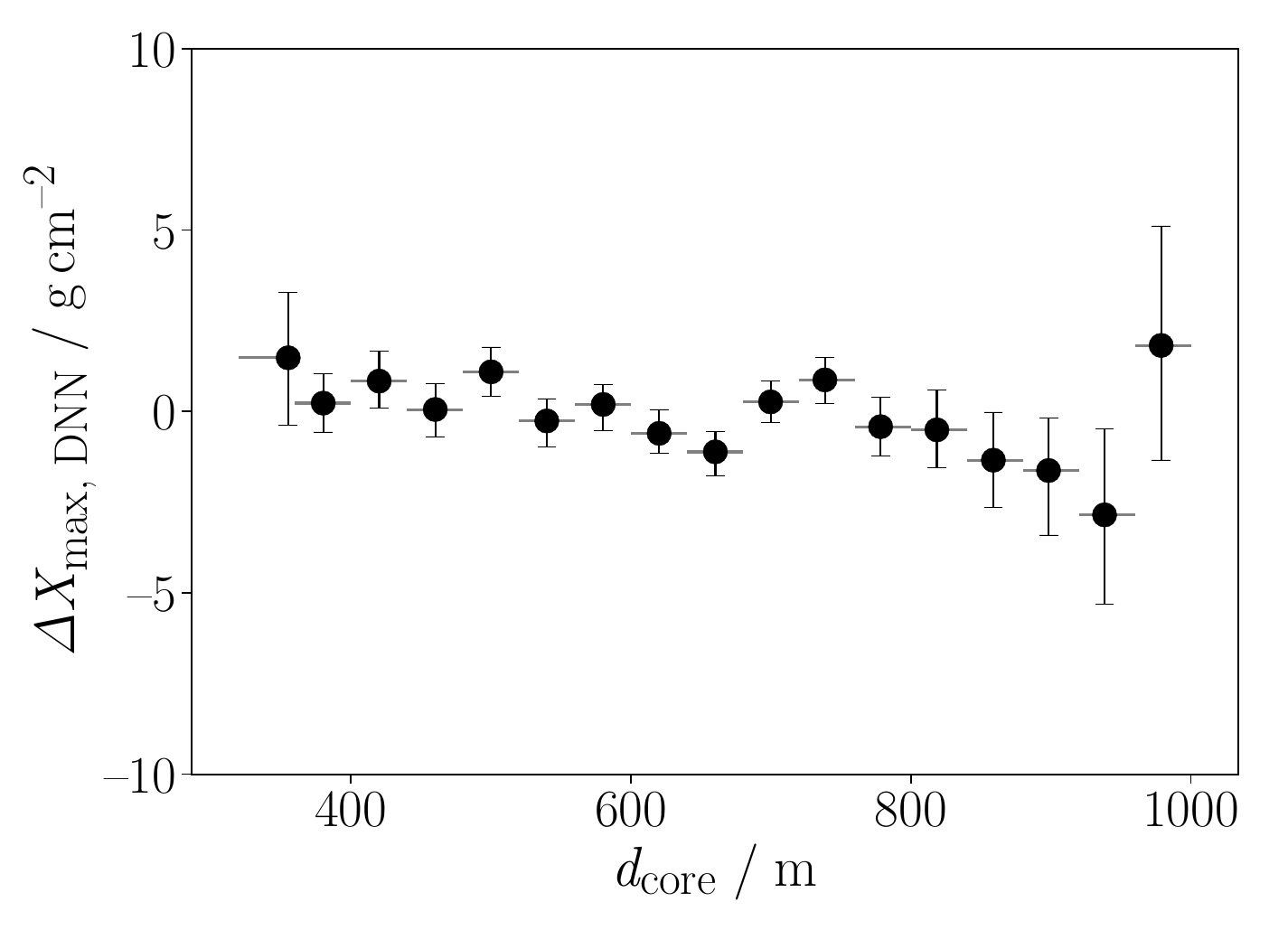}
                \subcaption{}
                \label{fig:core_after}
            \end{centering}
        \end{subfigure}
        
        \begin{subfigure}[b]{0.46\textwidth}
            \begin{centering}
                \includegraphics[width=0.93\textwidth]{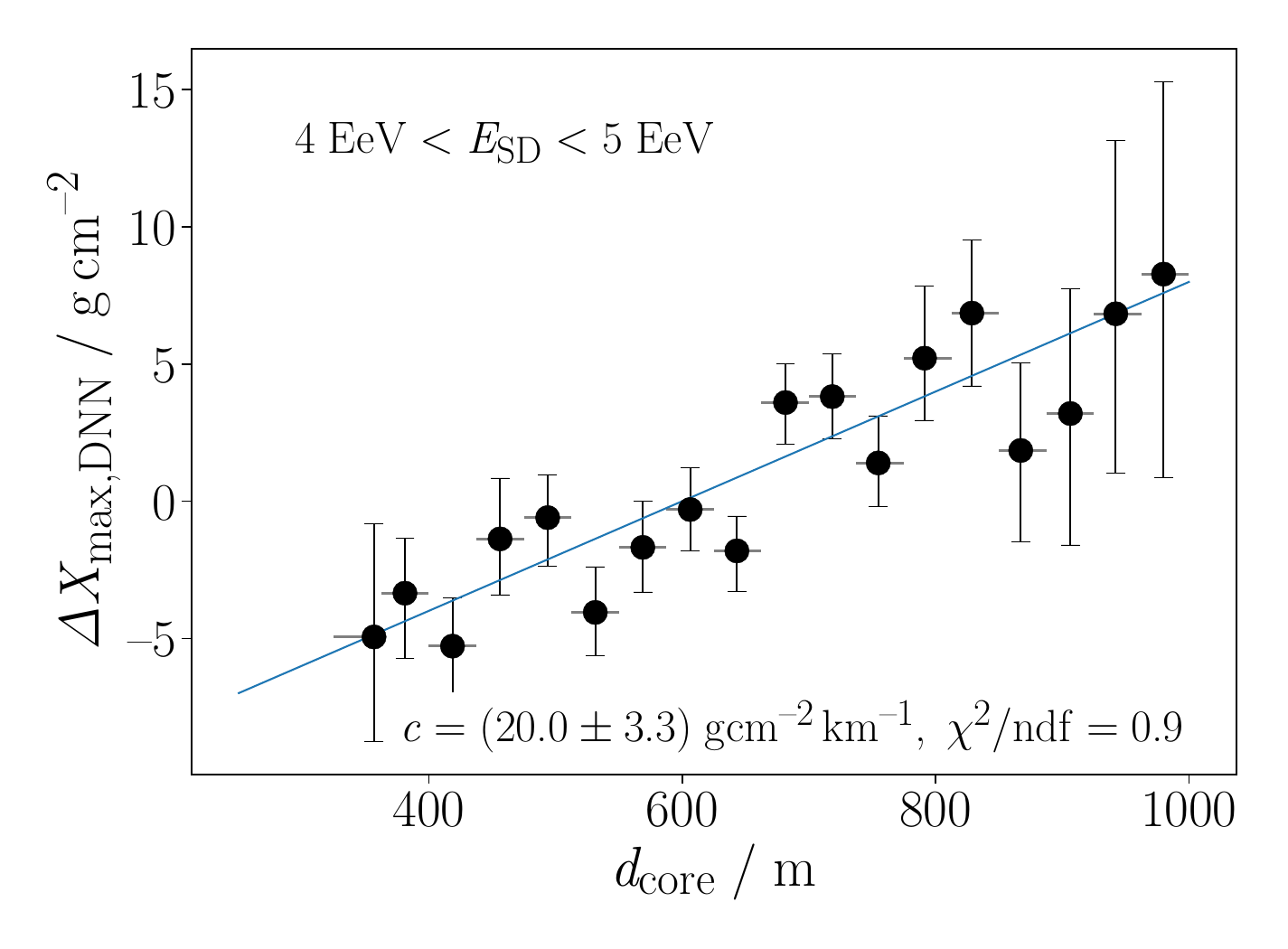}
                \subcaption{}
                \label{fig:core_cal_fit}
            \end{centering}
        \end{subfigure}
        \begin{subfigure}[b]{0.46\textwidth}
            \begin{centering}
                \includegraphics[width=0.93\textwidth]{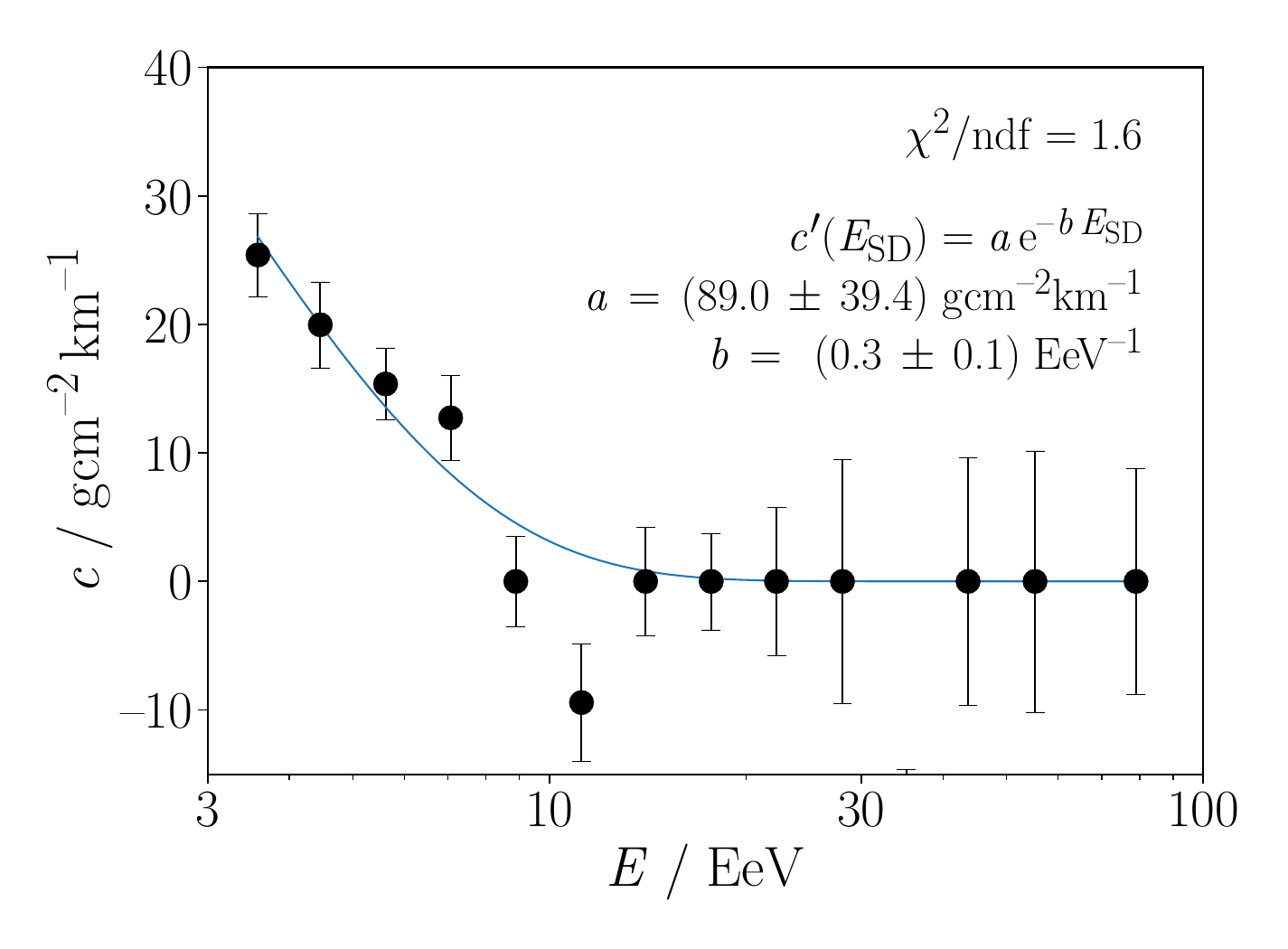}
                \subcaption{}
                \label{fig:core_cal_param}
            \end{centering}
        \end{subfigure}
    \caption{\label{fig:core_calibration}\small{Correction of the DNN \xmax predictions as a function of the core distance $d_\mathrm{core}$ to the station with the largest signal. (a) Reconstruction bias before the calibration. (b) Bias after performing the core calibration.
    (c) Linear calibration of the \xmax reconstruction bias as a function of the core distance for an example energy bin from 4 to 5~EeV. (d) Obtained slopes $c$ of the linear fit as a function of energy $E$. The determined parameterization of the slope is shown as a blue line.}}
    \end{center}
\end{figure*}

The composition bias of the \xmaxsigma reconstruction as seen in \cref{fig:comp_bias} (bottom) depends on energy and is below $5~\gcm$ above 10~EeV for proton and iron.
Only for iron in the \qgs model, a bias around 5~\gcm can be seen above 10~EeV.
For the Auger mix it is even lower at these high energies and not significant.
Overall, the biases observed for reconstructing \xmaxsigma are small over a large range of energies.
For this reason, no calibration using the FD will be performed.
Furthermore, the biases found here will be transferred to the systematic uncertainty of our composition measurements.
Different from the measurement of \xmaxmu, the estimation of \xmaxsigma is not subject to a strong dependence on the hadronic interaction model.
This can be explained by the fact that a large part of the shower fluctuations depends on the fluctuations of $X_1$, i.e., the traversed depth prior to the first interaction, owing to the given nuclear cross sections with air molecules.
Depending on the primary, the expected fluctuations in $X_1$ are in the order of 10~\gcm (50~\gcm) for iron (proton) nuclei.

\begin{figure*}[ht!]
    \begin{centering}
        \begin{subfigure}[b]{0.33\textwidth}
            \begin{centering}
                \includegraphics[width=\textwidth, trim={0 0 0.8cm 0},clip]{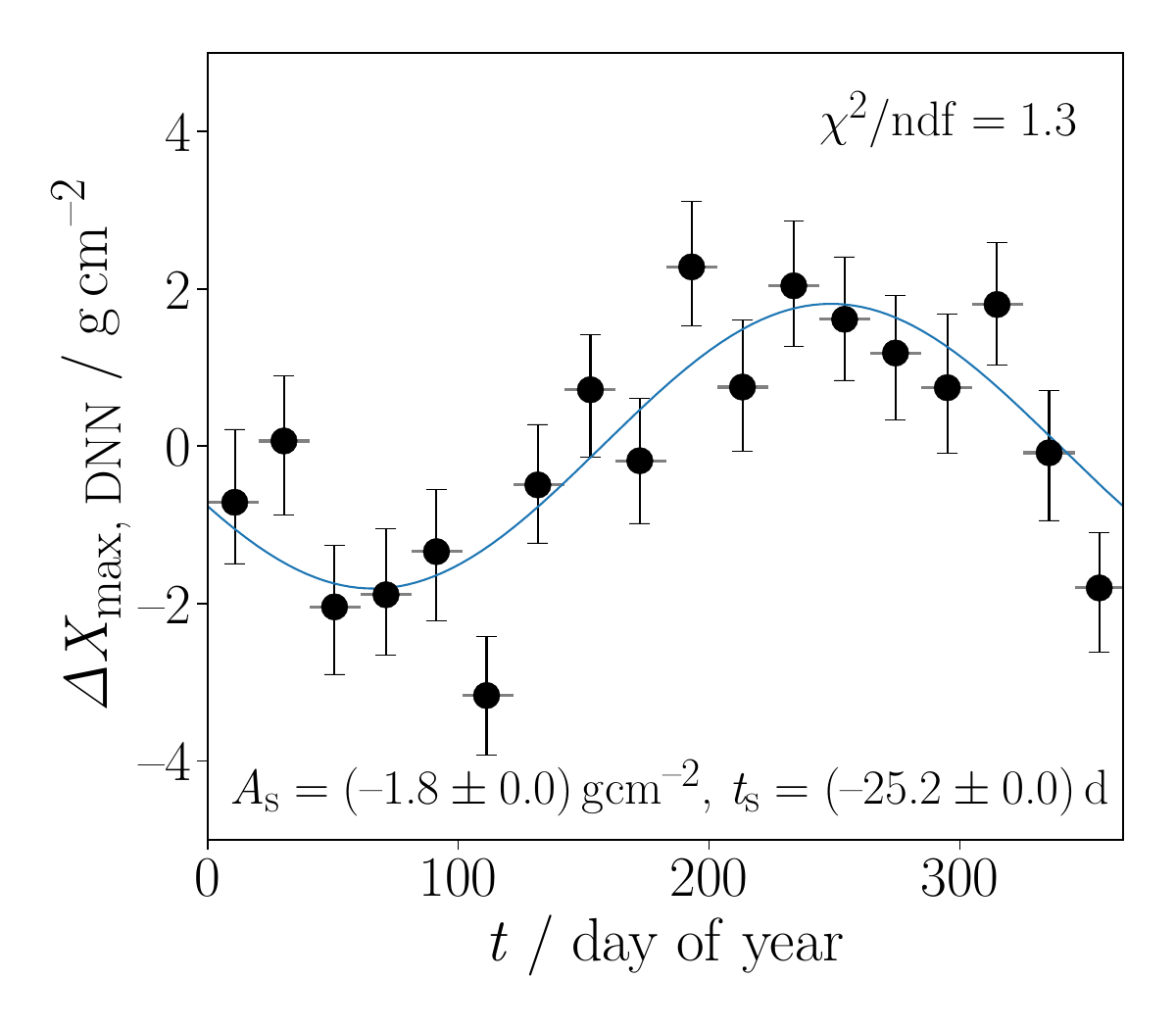}
                \subcaption{}
                \label{fig:seasonal_calib}
            \end{centering}
        \end{subfigure}
        \begin{subfigure}[b]{0.33\textwidth}
            \begin{centering}
                \includegraphics[width=\textwidth, trim={0 0 0.8cm 0},clip]{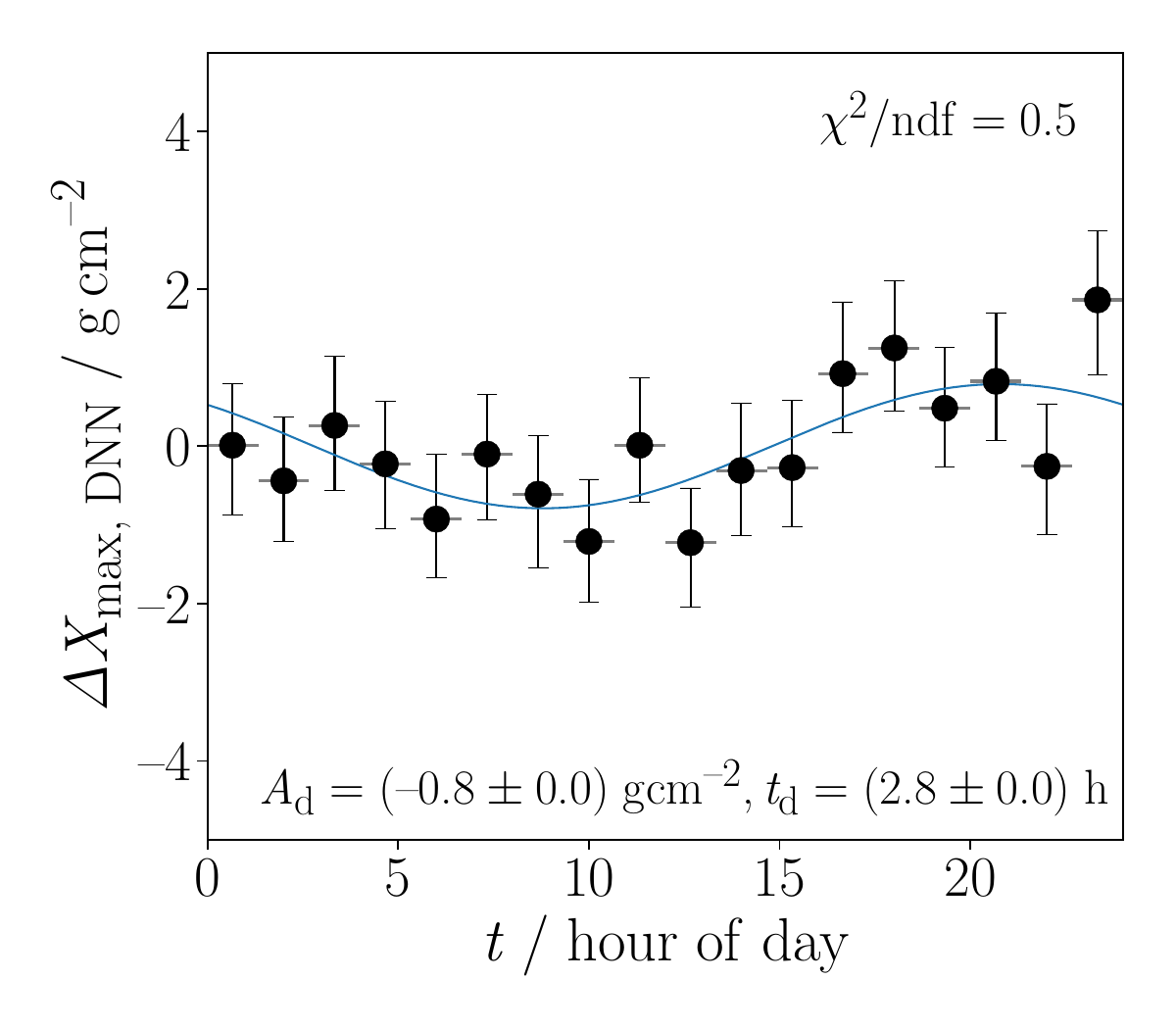}
                \subcaption{}
                \label{fig:diurnal_calib}
            \end{centering}
        \end{subfigure}
        \begin{subfigure}[b]{0.33\textwidth}
            \begin{centering}
                \includegraphics[width=\textwidth, trim={0 0 0.8cm 0},clip]{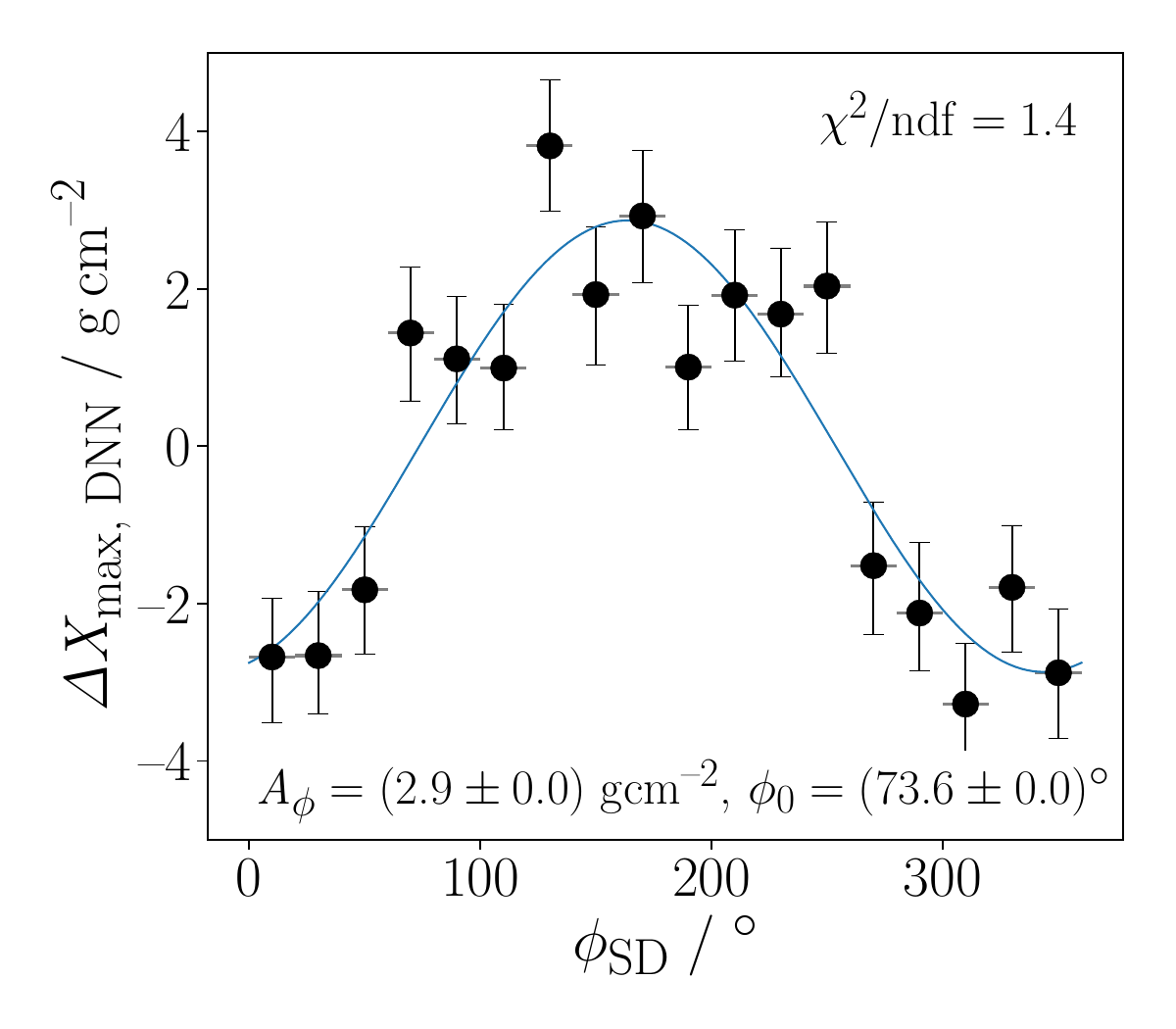}
                \subcaption{}
                \label{fig:az_calib}
            \end{centering}
        \end{subfigure}
        
        \begin{subfigure}[b]{0.33\textwidth}
            \begin{centering}
                \includegraphics[width=\textwidth, trim={0 0 0.8cm 0},clip]{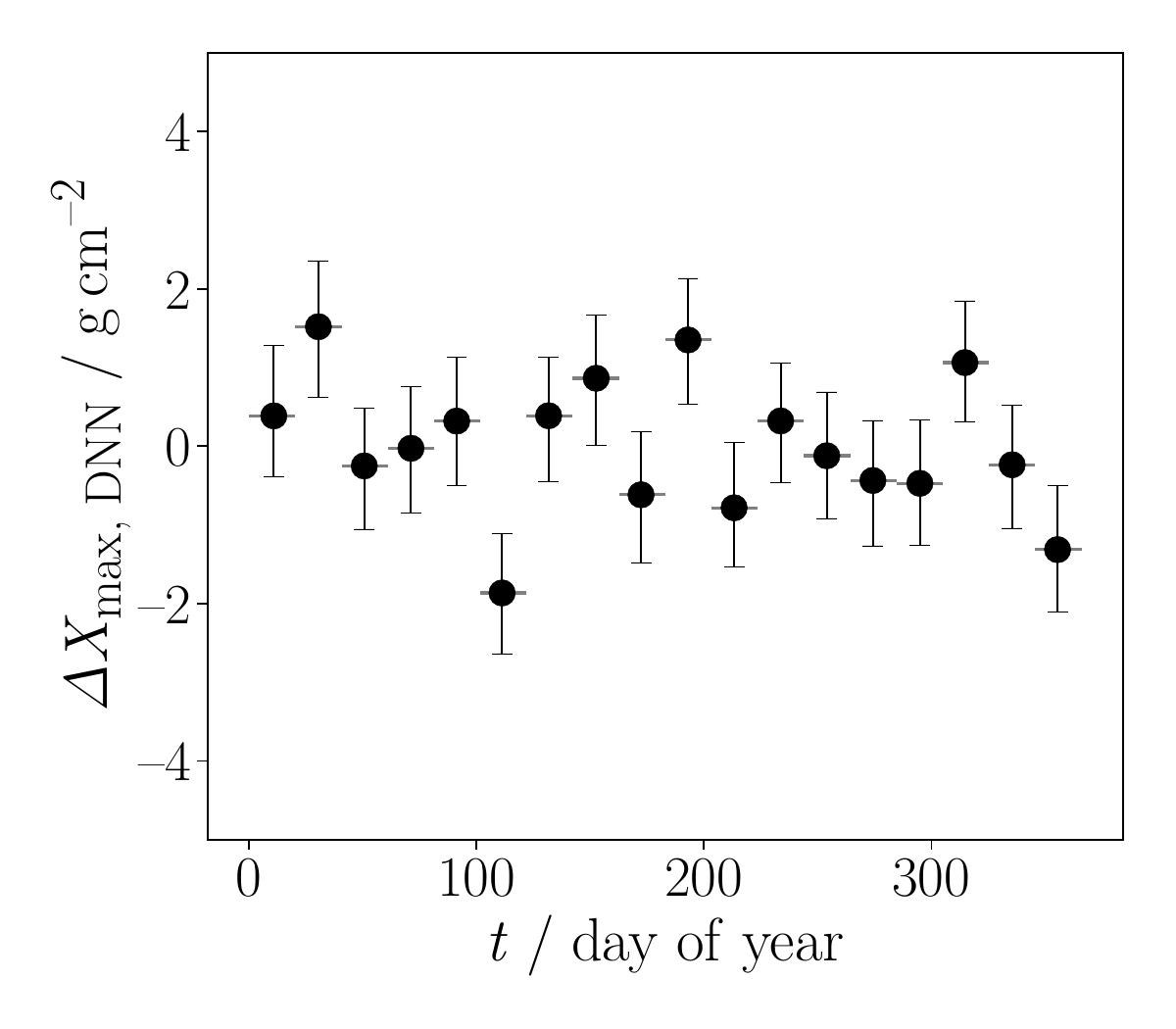}
                \subcaption{}
                \label{fig:seasonal_calib_after}
            \end{centering}
        \end{subfigure}
        \begin{subfigure}[b]{0.33\textwidth}
            \begin{centering}
                \includegraphics[width=\textwidth, trim={0 0 0.8cm 0},clip]{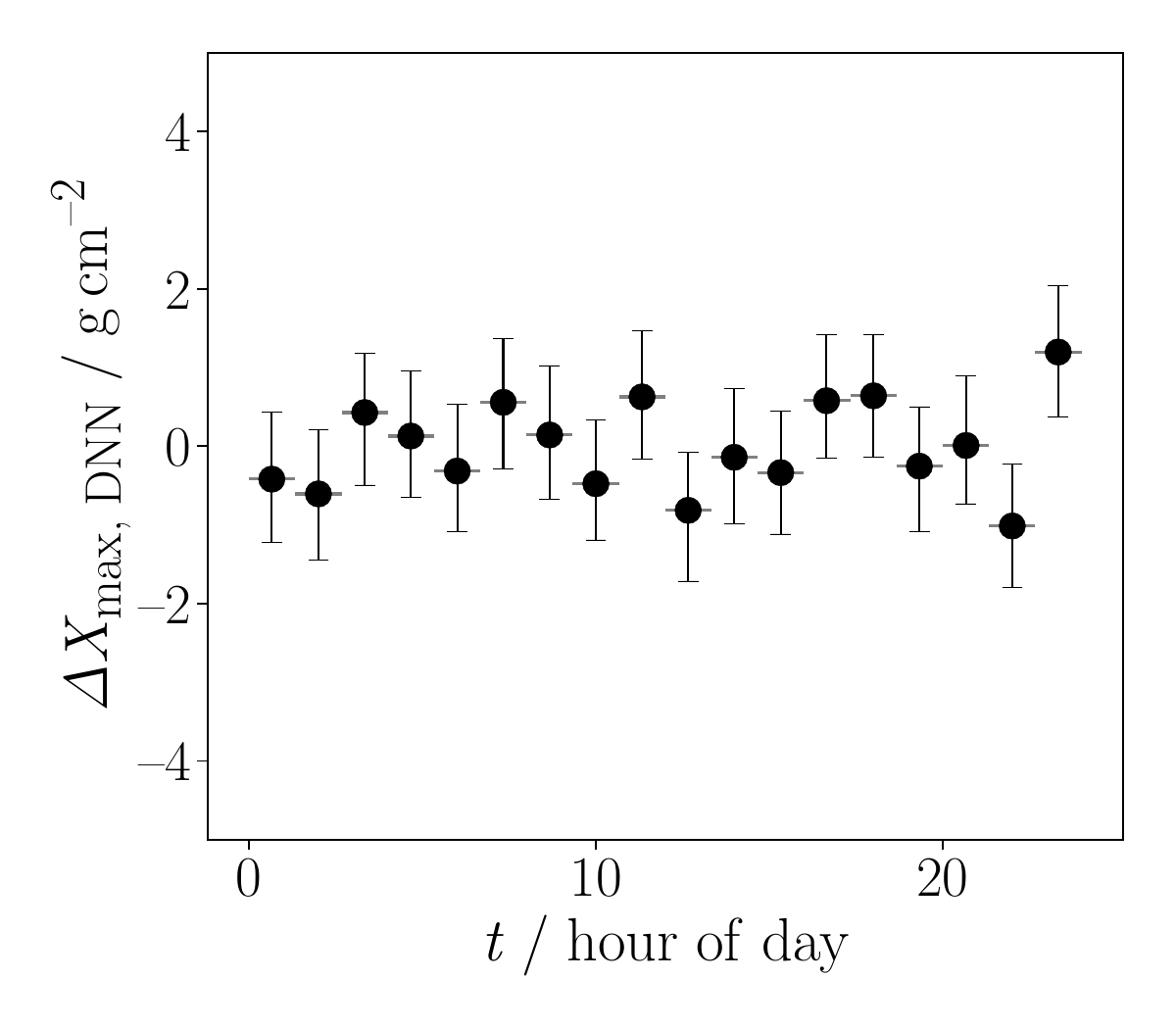}
                \subcaption{}
                \label{fig:diurnal_calib_after}
            \end{centering}
        \end{subfigure}
        \begin{subfigure}[b]{0.33\textwidth}
            \begin{centering}
                \includegraphics[width=\textwidth, trim={0 0 0.8cm 0},clip]{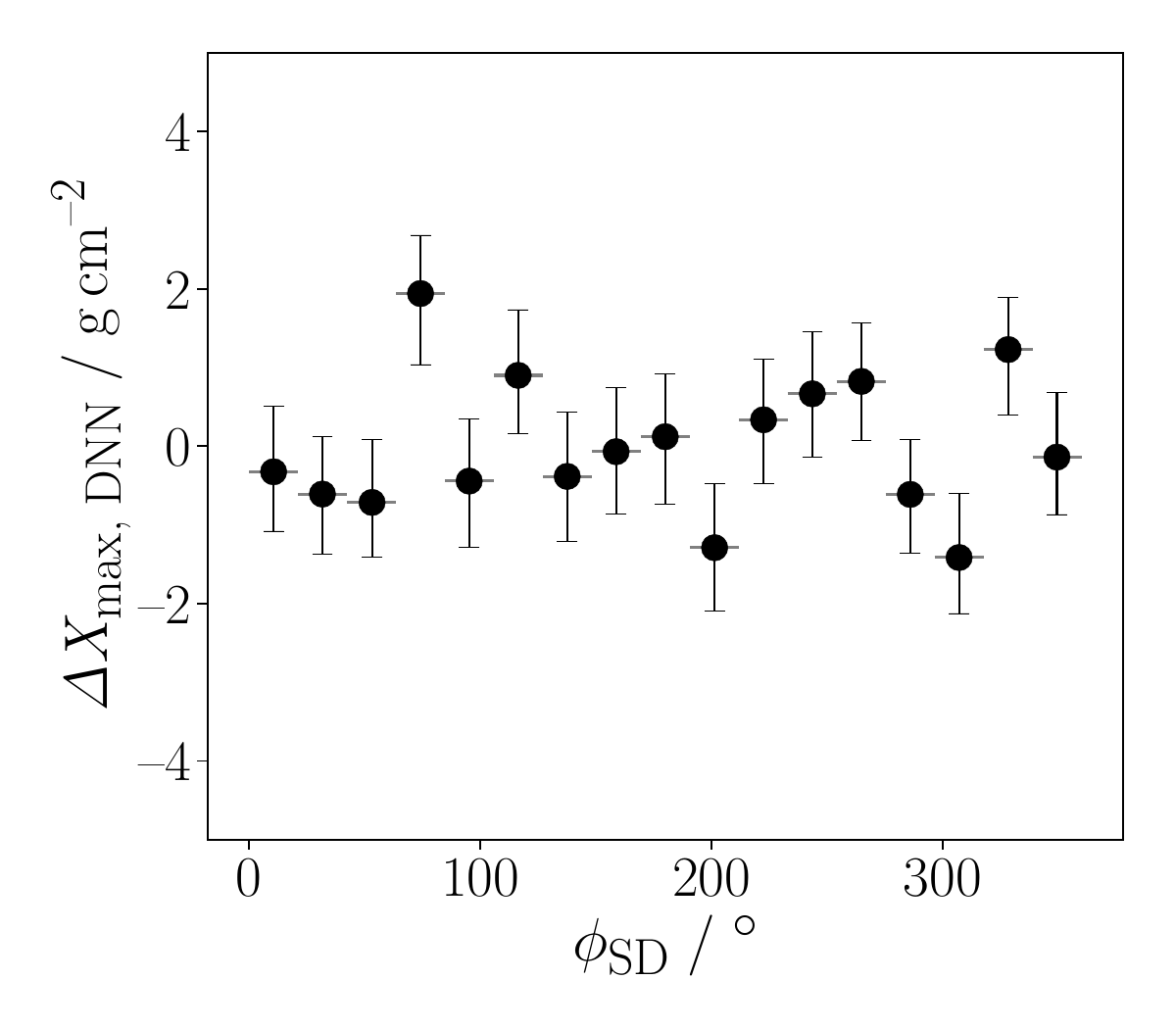}
                \subcaption{}
                \label{fig:az_calib_after}
            \end{centering}
        \end{subfigure}
    \end{centering}
    \caption{\label{fig:calibrations}\small{Data-based correction of the DNN \xmax prediction. Top: dependence of the \xmax reconstruction as a function of (a) season (time of year), (b) time of day (UTC), and (c) azimuth angle of the shower. The fitted calibration functions are shown in blue.
    Bottom: dependence of the reconstruction as a function of (d) season, (e) time of day (UTC), and (f) azimuth angle after the calibration.}}
\end{figure*}

\subsection{Calibration}
After training the DNN using simulations, the algorithm is applied to the measured data.
Even though the simulation is continuously developed and improved, important differences exist between the measured data and the simulations.
To remove such differences arising from inaccurate modeling, we perform calibrations using the SD data set by studying the \xmax reconstruction as a function of physics and monitoring observables.
We examine the reconstruction bias $\Delta \xmaxdnn = \xmaxdnn-\langle\xmaxdnn\rangle$, estimated with respect to the average \xmax prediction.
For each variable $y$ we intend to correct with, we perform an event-by-event correction with
\begin{equation}
    \xmax' = \xmax - f_{\Delta\xmax}(y),
\label{eq:calib}
\end{equation}
where $f_{\Delta\xmax}(y)$ denotes the dependence of the \xmax prediction on the variable $y$.
This approach performed for each event separately ensures meaningful corrections of the predictions beyond the first moment of \xmax.
Finally, using hybrid events, we calibrate the \xmax predictions to the scale of the FD and remove the dependence on the hadronic interaction model used during the algorithm training, the composition, and any remaining differences between the measured data and the detector simulation.

\subsubsection{Corrections using Surface Detector data\label{sec:sd_calib}}
The WCD stations of the SD are exposed to a harsh environment, with changes in temperature covering several tens of $\grad \mathrm{C}$.
During the many years of operation, the PMTs, read-out electronics, water, and reflective liner are subject to aging effects.
By utilizing muons that constantly cross the detector stations, the average shape of the single-muon signal is monitored using the area over peak ratio $A/P$, which relates the deposited charge in the detector (integrated pulse) to its height and is a rough measure of the signal duration.
Since $A/P$ is a monitoring observable summarizing the characteristics of the individual PMT responses, water quality, and liner reflectivity, it is specific to every station and changes with time.\\

\paragraph{Aging calibration.}
When monitoring the distribution of $A/P$ values of all stations, over the years, a decrease in its average $\langle A/P \rangle$ from $3.20$ to $2.95$ can be observed~\cite{PierreAugerlongterm, AbdulHalim:2023iN}. 
We assume this to be mainly caused by the decrease in the liner reflectivity or water transparency, leading to a drift of the \xmax predictions as a function of time (see \cref{fig:aop_before_calib}).
Since our simulation library is currently limited to simulated stations with $A/P=3.2$, the predictions of the DNN have to be calibrated as a function of the $\langle A/P \rangle$ --- the average $A/P$ of all triggered SD stations in a given event --- to remove possible time dependencies of the predictions.
As depicted in \cref{fig:aop_calib_zenith}, the dependence of the \xmax predictions on $\langle A/P \rangle$ can be modeled linearly and was found to not depend on energy.
Additionally, we find an increase of this dependence as a function of the zenith angle, likely caused by the fact that the average distance a particle travels through the detector rises with the zenith angle.
By parameterizing (blue line in \cref{fig:aop_calib_zenith_param}) this dependence as $c^{*}(\theta) = \alpha \, \cos(\theta) + \beta$, with $\alpha = (-135.1 \pm 9.2)$~\gcm, and $\beta = (152.6 \pm 7.2)$~\gcm, we calibrate the predictions using:
\begin{equation}
    f_{\Delta \xmax}(\theta, A/P) =\underbrace{c^{*}(\theta) \, \left(\langle {A/P} \rangle - \overline{\langle A/P\rangle}\right)}_{\mathrm{zenith\;\&\;}A/P\mathrm{\; dependence}} + \underbrace{\Delta X}_{\mathrm{scale}},
\end{equation}
where $\overline{\langle A/P\rangle}  = 3.03$ denotes the average of the distribution calculated over the full SD data set. Additionally, we introduce an absolute shift $\Delta X = \langle \xmax \rangle - \langle X_{\text{max}, A/P=3.2}\rangle = -9.5\,\text{g/cm}^2$ to adjust the \xmax scale by considering the different averages in the $A/P$ distributions in simulations $\overline{\langle A/P\rangle}_\mathrm{MC} = 3.2$ and data $\overline{\langle A/P\rangle}_\mathrm{data} = 3.03$.

\begin{figure*}[t]
    \begin{centering}
        \begin{subfigure}[b]{0.49\textwidth}
            \includegraphics[width=\textwidth]{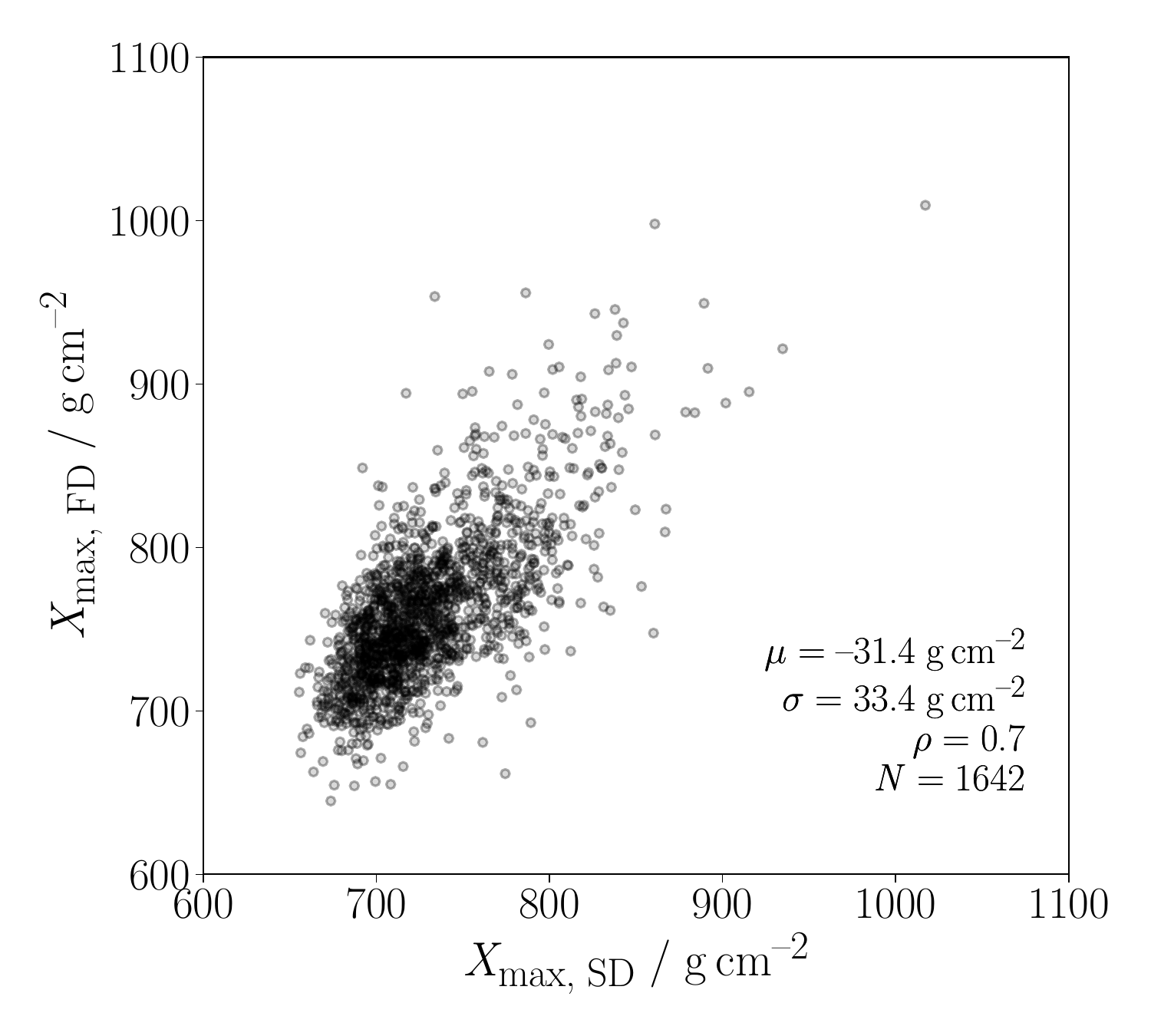}
            \subcaption{}
            \label{fig:dnn_fd_scatter}          
        \end{subfigure}
         \begin{subfigure}[b]{0.49\textwidth}
            \includegraphics[width=\textwidth]{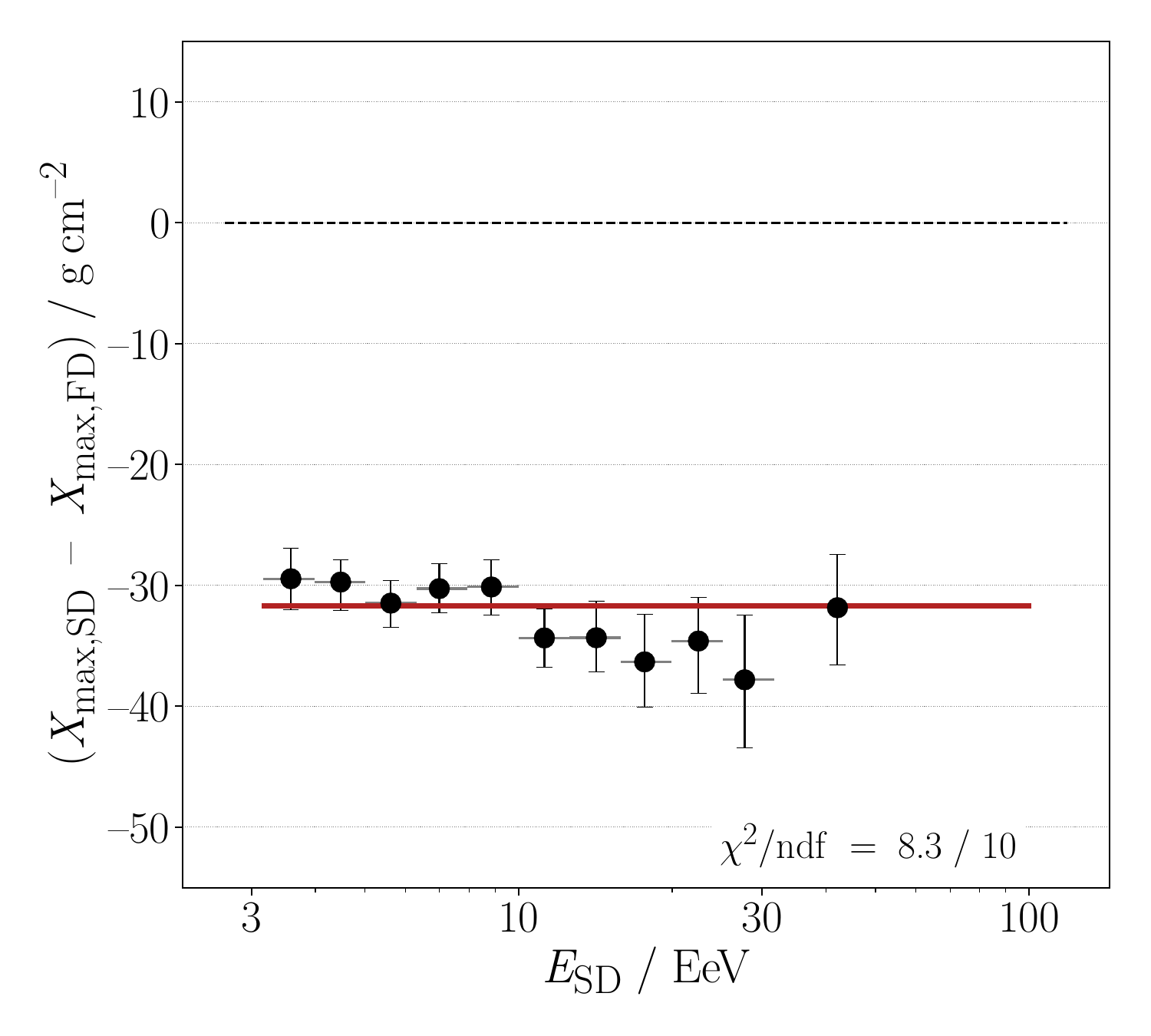}
            \subcaption{}
            \label{fig:dnn_fd_calib}            
        \end{subfigure}       
    \end{centering}
    \caption{\label{fig:hybrid_calibration}Application of the DNN \xmax estimation to hybrid data. (a) Correlation between FD observations and DNN predictions using SD data. Furthermore, bias $\mu$, resolution $\sigma$, Pearson correlation coefficient $\rho$, and the number of events $N$ are given. (b) The energy-dependent bias of the SD-based reconstruction of the DNN when compared to the reconstruction of the FD.}
\end{figure*}

The reconstruction bias of \xmax shows a dependence on the distance between the station with the largest signal and the shower core reconstructed using fitting of a lateral distribution function~\cite{sd_reco}.
At low and high energies, the biases are relatively large since at small distances, the stations with the largest signal are often saturated, and at low energies, the station multiplicity is small, making a detailed reconstruction more challenging.
Furthermore, the number of triggered stations is on average, lower for cores located close to one station since in this case the distances to the next stations of the grid all become close to $1500\,$m.
A similar effect applies to events with reconstructed shower cores far away from the station with the highest signal.
Simulation studies show that at a distance of roughly 600\,m, the bias is smallest. In general, events with core distances larger than $350$\,m and smaller than $1000$\,m feature a small bias and a dependence of the reconstruction bias on the core distance, as visible in \cref{fig:core_before}.
Events outside this regime that exhibit larger biases were rejected during the data selection.

The described effect mainly concerns events with a low multiplicity of triggered stations, i.e., showers produced by primaries having energies below ${\sim}10$~EeV.
Therefore, we apply an energy-dependent calibration to remove the \xmax bias at small energies.
In bins of energy, we perform a linear fit $f_{\Delta \xmax}(d) = c\,(d-0.6~\mathrm{km})$ of the reconstruction bias, as shown in \cref{fig:core_cal_fit} for the example energy bin from 4 to 5~EeV. Above 10~EeV, a constant fit usually shows a better $\chi^2$ and is preferred over a linear fit.
The dependence of the slope with energy is shown in \cref{fig:core_cal_param}, and parameterized by fitting the function $c'(E_{\mathrm{SD}}) = a\, e^{-b\, E_{\mathrm{SD}}}$. The obtained values are $a = (89.0\pm39.4)~\gcm/\mathrm{km}$ and $b = (0.3\pm 0.1)~\mathrm{EeV}^{-1}$.
The final calibration of the DNN is performed using
\begin{equation}
   f_{\Delta \xmax} = c'(E_\mathrm{SD})\,(d-0.6~\mathrm{km})
\end{equation}
and using \cref{eq:calib} on an event-by-event basis.\\

\paragraph{Temporal variations.}
The change in pressure and temperature due to diurnal and seasonal variations causes small influences on the detector response and the conversion of distance to \xmax, hence affecting its reconstruction.
To remove seasonal and diurnal variations, we investigated the reconstruction as a function of time on a yearly and daily basis, as shown in \cref{fig:seasonal_calib} and \cref{fig:diurnal_calib}.
We find small variations of the size of $2~\gcm$ or $1~\gcm$, respectively.
We first calibrate the \xmax predictions to remove the seasonal variation by fitting a sine function to the data and correcting it by using the time of year on an event-by-event basis.
These predictions are also used to remove the diurnal variations by again fitting a sine wave and correcting events using \cref{eq:calib}.
The dependencies after correction are depicted in \cref{fig:seasonal_calib_after} and \cref{fig:diurnal_calib_after}.\\

\paragraph{Angular dependence.}
In contrast to our simulation study, we find a dependence of the \xmax reconstruction on the azimuth angle (see \cref{fig:az_calib}).
The dependence is small, and its fluctuations are around $3~\gcm$ and possibly caused by a slight slope of the SD array tilted away from the Andes mountains.
We remove the dependence by fitting a cosine and calibrating the predictions using the azimuth angle on an event-by-event level.
The reconstruction after calibration is shown in \cref{fig:az_calib_after}.
We also tested the reconstruction for a possible dependence on the zenith angle.
Therefore, we studied \xmax as a function of the zenith angle for different energy intervals to account for the fiducial cut but could not find any indications for a dependence.

\begin{figure}[t!]
    \begin{centering}
            \includegraphics[width=0.490\textwidth]{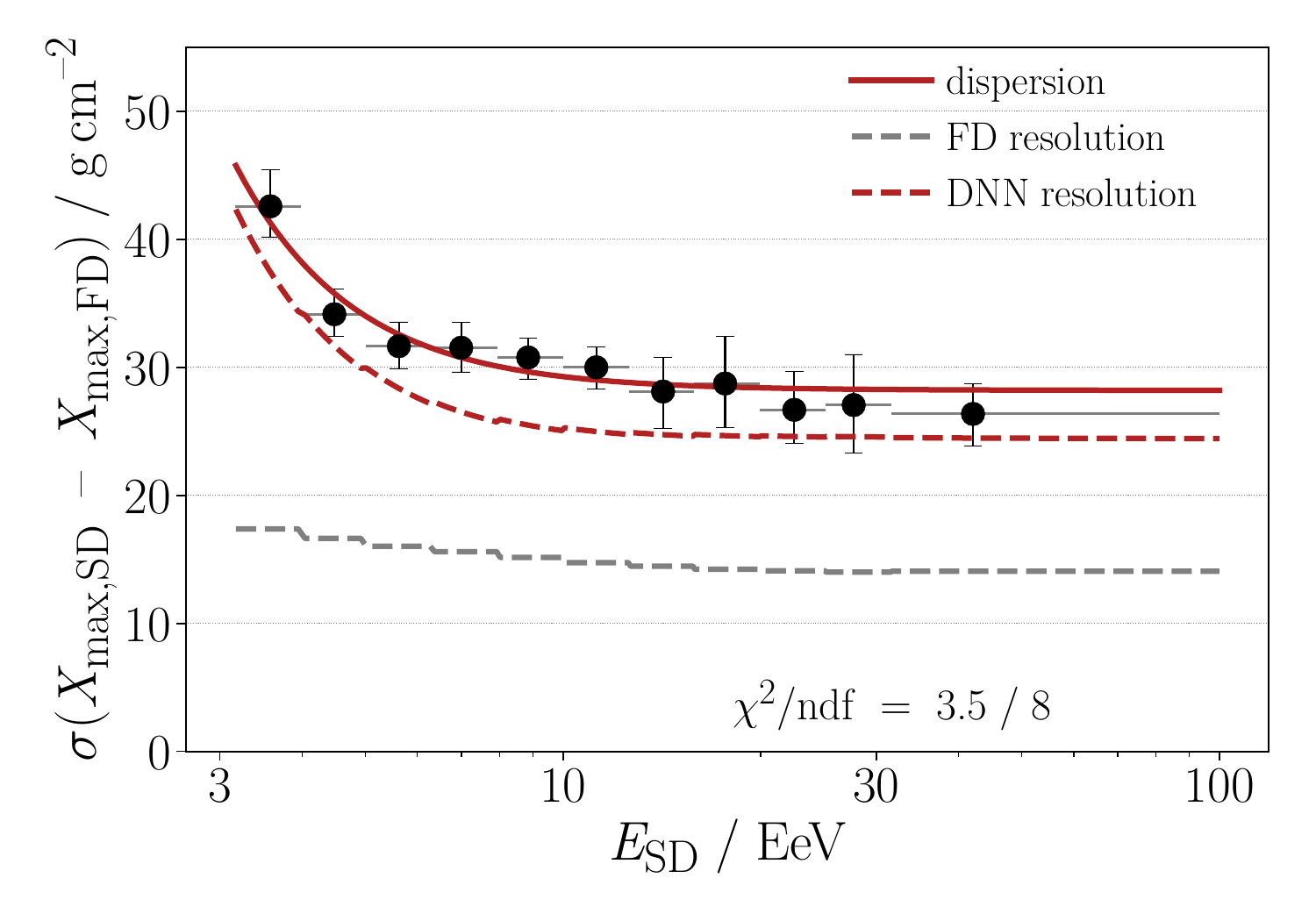}
    \end{centering}
    \caption{\label{fig:resolution}\small{Standard deviation of the distribution \xmaxdnn--\xmaxfd as a function of energy. The energy-dependent resolution of the DNN (dashed red line) is obtained by subtracting the FD resolution (dashed grey line) from the standard deviation of the FD and the SD \xmax reconstruction (red line) in the hybrid data set}.}
\end{figure}

\subsubsection{\label{sec:hybrid_cal}Calibration using hybrid events}
The hybrid design of the Pierre Auger Observatory enables a cross-calibration of the SD \xmax measurements with FD \xmax of hybrid events.
The dependence of the \xmax scale of the DNN on the hadronic interaction model can thus be eliminated by calibrating the DNN predictions with the FD \xmax scale that can be accurately determined~\cite{aab_pierre_auger_collaboration_depth_2014}.

The event-by-event correlation between the FD and the SD is shown in \cref{fig:dnn_fd_scatter}.
We find a Pearson correlation of $\rho=0.70\pm0.03$, which is in good agreement with the expectations from idealized simulations ($\rho_\text{MC}=0.73$).
The absolute bias, however, amounts to $-31.4 \pm 0.8$~\gcm.
This bias is larger than expected from simulation studies (up to $-15~\gcm$ assuming the Auger mix) with interaction models different from those used in the algorithm training (cf.~\cref{fig:comp_bias} and Ref.~\cite{xmax_wcd, glombitza_icrc_21}).
The observation of negative bias, i.e., a heavier composition in data (smaller \xmax values), is in line with findings in previous analyses, where the average signal footprint measured using surface detector arrays seems to favor a composition heavier than expected from simulations~\cite{aab_pierre_auger_collaboration_inferences_2017, Aab_2015_muons, the_pierre_auger_collaboration_measurement_muon_fluctuations}.
In particular, recent works indicate that the current generation of hadronic interaction models may not model the muonic component in full detail~\cite{Aab_2015_muons, Aab_2016_test_models, PhysRevD_90_012012}.
Additionally, adjustment of the longitudinal shower profile might be needed~\cite{jakub_2024testing}.
In contrast, the relative fluctuations in the muon component seem to be reasonably modeled~\cite{the_pierre_auger_collaboration_measurement_muon_fluctuations}.
Using the exotic hadronic interaction model \sibstar~\cite{sib_star} that features ad-hoc modifications of the shower content, a significant increase of the muon number can be accomplished. 
A test using \sibstar that predicts an increase of the muon number by $40\%$ for protons with respect to \sibylld shows that a bias of $-40$~\gcm could be reproduced, indicating that the observed scale of the bias could be explained by a mis-modeling of the muonic component of current interaction models.
However, it is unclear if such ad-hoc adjustments or data-based refinements~\cite{erdmann2018generating} offer a realistic solution.
In addition, note that a non-perfect detector simulation could cause deviations and that the systematic uncertainty on the FD \xmax scale amounts to roughly $10$~\gcm~\cite{pierre_auger_collaboration_depth_2014}.

\begin{figure}[t!]
    \begin{centering}
        \includegraphics[width=0.490\textwidth]{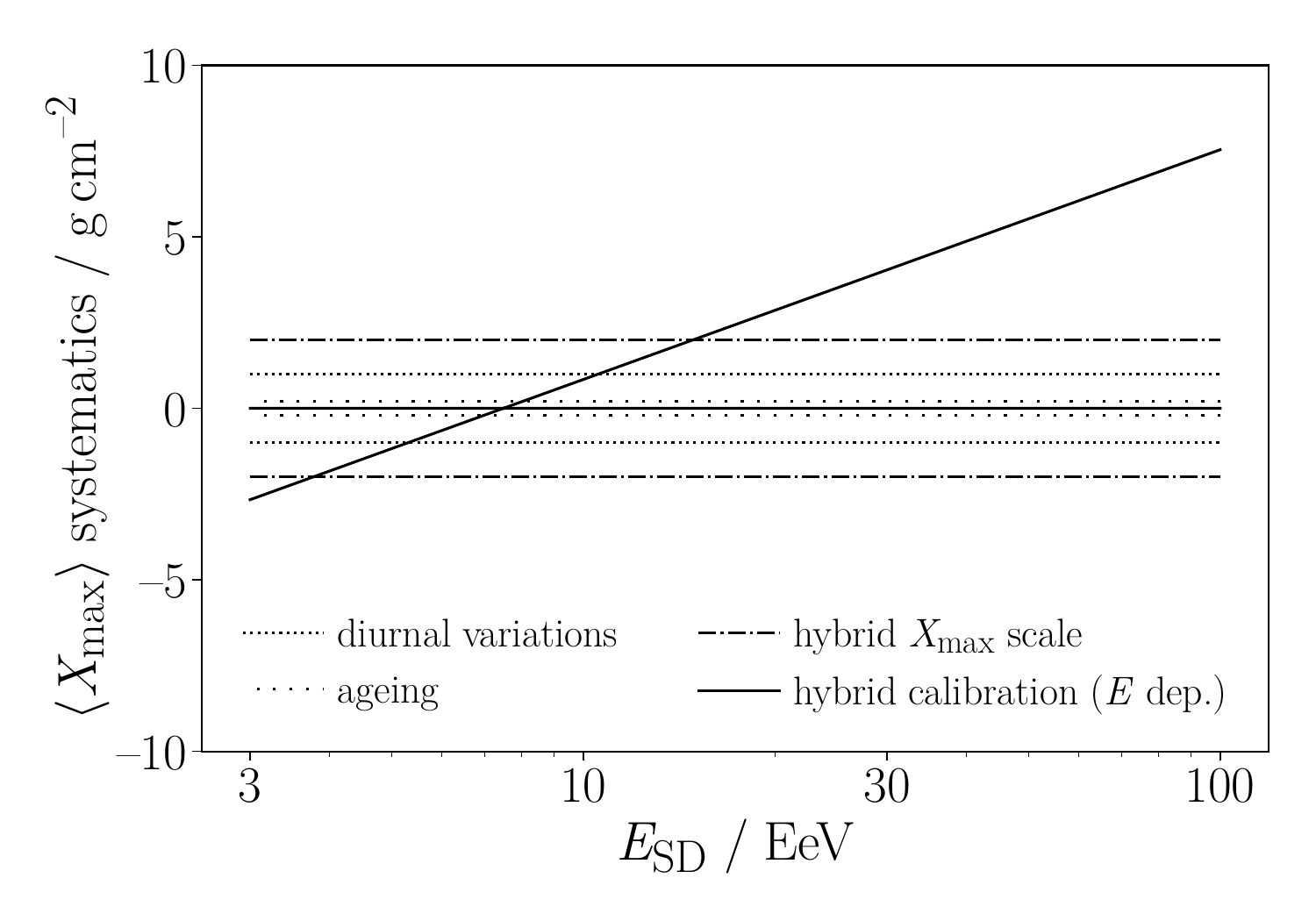}
    \end{centering}
    \caption{\label{fig:sd_only_unc}\small{Summary of the SD systematic uncertainties for the measurement of \xmaxmu after calibrating DNN to FD observations. Only the hybrid calibration shows a distinct energy dependence.}}
\end{figure}

\begin{figure*}[t!]
    \begin{centering}
        \begin{subfigure}[b]{0.495\textwidth}
            \begin{centering}
                \includegraphics[width=\textwidth]{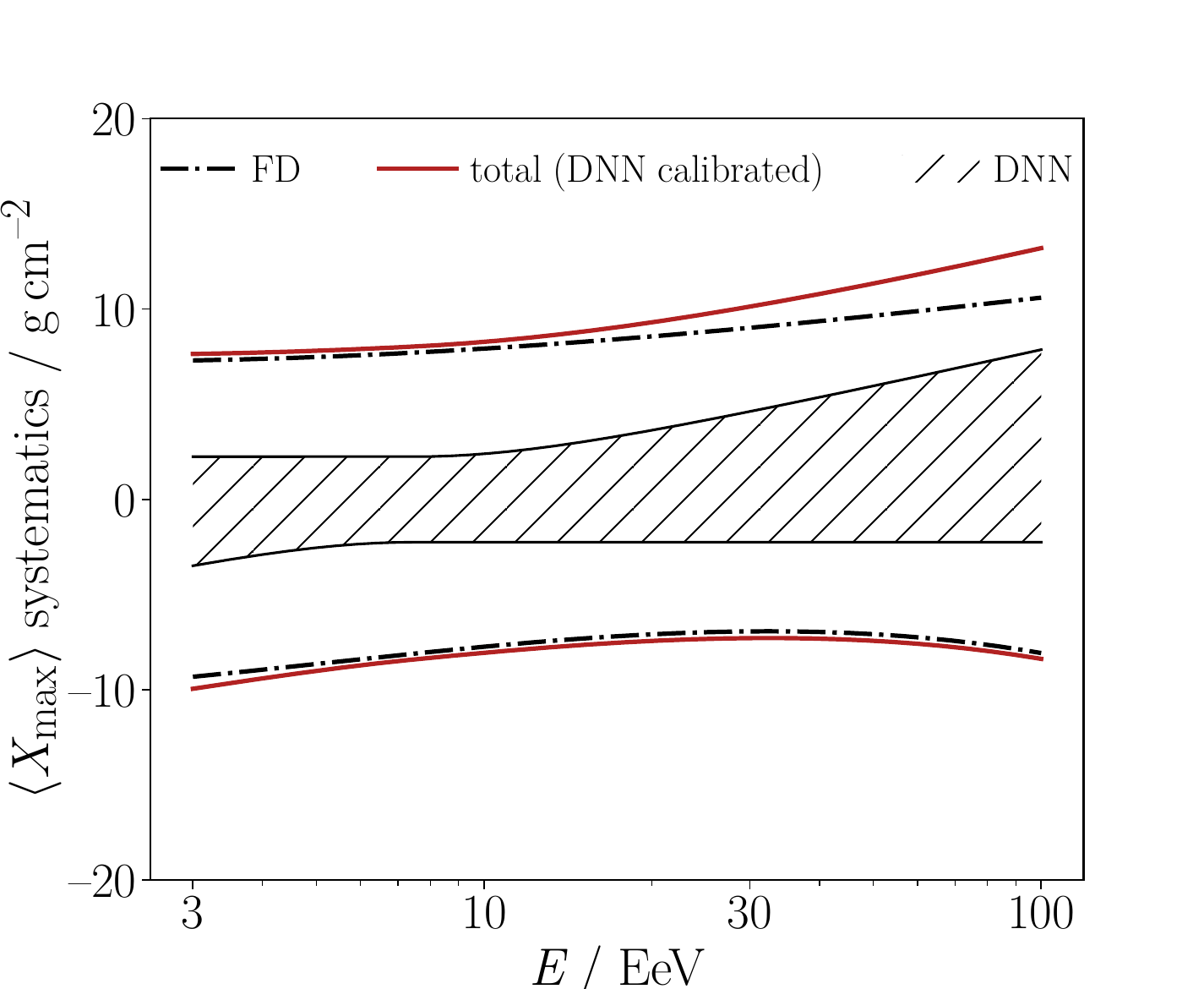}
                \subcaption{}
                \label{fig:mean_xmax_sys}
            \end{centering}
        \end{subfigure}
        \begin{subfigure}[b]{0.495\textwidth}
            \begin{centering}
                \includegraphics[width=\textwidth]{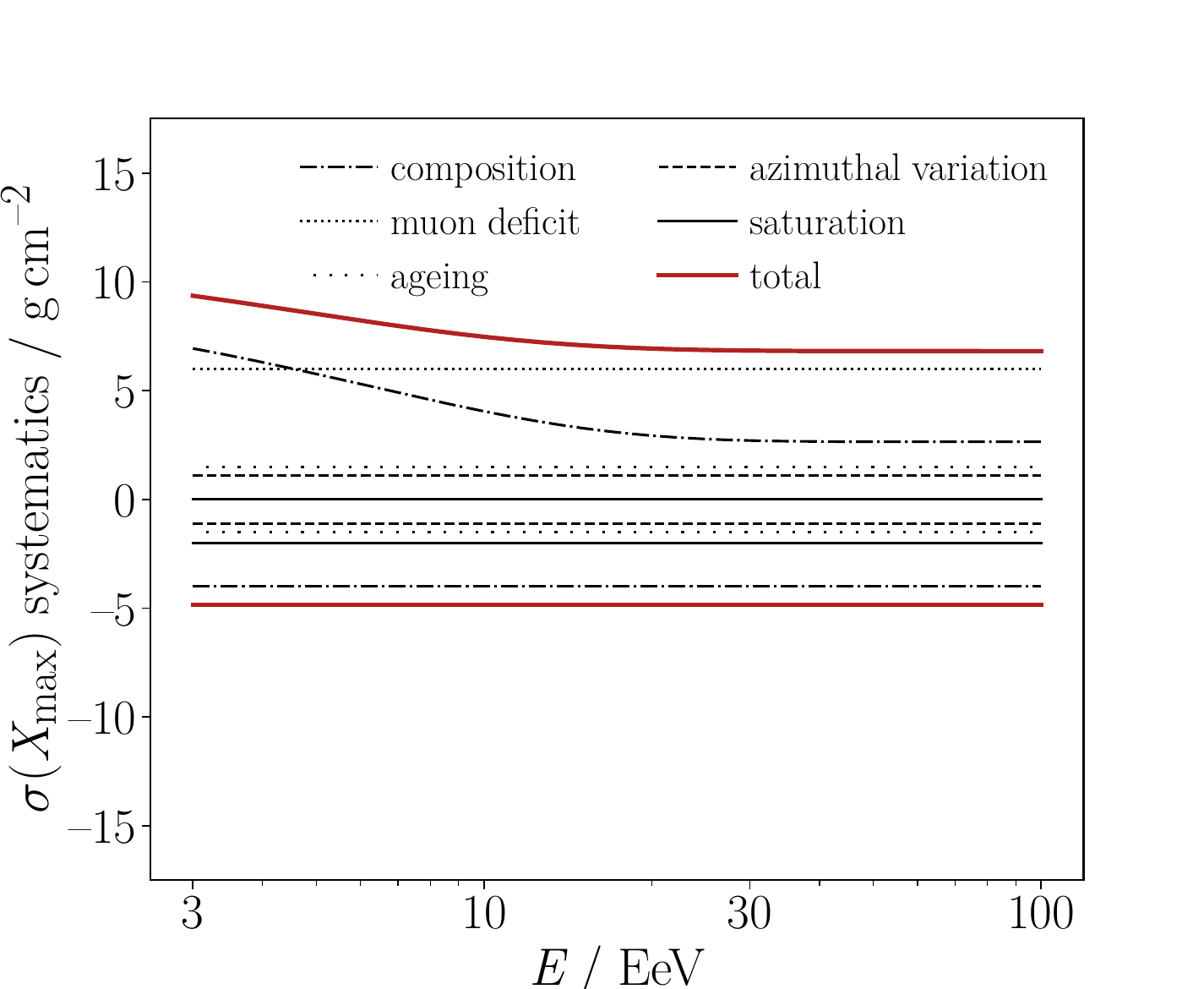}
                \subcaption{}
                \label{fig:sigma_xmax_sys}
            \end{centering}
        \end{subfigure}
    \end{centering}
    \caption{\label{fig:sys_unc}\small{Energy-dependent systematic uncertainties for the measurement of \xmaxmu and \xmaxsigma. (a) Total uncertainty of the calibrated DNN (continuous red lines) resulting from adding the uncertainty on the FD \xmax scale (dash-dotted line) and the DNN uncertainties (hatched region) after the hybrid calibration in quadrature. (b) Systematic uncertainties for the DNN reconstruction of \xmaxsigma as a function of energy. The total systematic uncertainty is denoted as the continuous red lines. Only contributions $>1~\gcm$ are presented.}}
\end{figure*}

We find no significant energy dependence when studying the bias as a function of energy (see \cref{fig:dnn_fd_calib}).
This is consistent with our simulation study since no strong energy dependence was found when studying the reconstruction bias of the most likely composition scenario, i.e., the Auger mix, for \sibyllc and \qgs (see \cref{fig:comp_bias}).
Since the \xmax scale can be precisely defined using the FD, we re-calibrate the predictions of the DNN for the SD events with a constant offset of $(-31.7 \pm 0.7)~\gcm$ obtained by the fit depicted as a red line in \cref{fig:dnn_fd_calib}.
Due to the calibration, we adopt the uncertainty of the FD \xmax scale to the systematic uncertainties on the \xmaxmu measurement.
This enables us to remove the composition and interaction-model-dependent contributions to the systematic uncertainty of the \xmaxmu measurement with the DNN and the SD.
Since with the low statistics in the hybrid data set at high energies, we cannot exclude a small energy dependence of the DNN reconstruction, deviations from a constant calibration offset are examined using energy-dependent calibrations.
We consider this in an energy-dependent systematic uncertainty on \xmaxmu measurements.
Note that we are using in this work the same data production as Ref.~\cite{yushkov_mass_2019}, which covers the same data-taking period.
Ongoing work on the FD reconstruction has led to refinements in the \xmax scale~\cite{Bellido_icrc23, Harvey_icrc23} that have not been considered, but remain below $5$~\gcm in \xmaxmu~\cite{fd_icrc_23}.

In \cref{fig:resolution}, we show the event-by-event resolution of reconstructing \xmax using the DNN (dashed red line) after subtracting the FD resolution~\cite{pierre_auger_collaboration_depth_2014} (dashed grey line) in quadrature from the standard deviation (continuous red line), found using the hybrid data.
The resolution improves from 40\,\gcm at low energies to $25$\,\gcm, which is in good agreement with simulations studies~\cite{xmax_wcd}.\\

\paragraph{Crosscheck of SD-based calibrations.}
We additionally checked the event-by-event correlation between the FD and the SD reconstruction before and after each calibration described in \cref{sec:sd_calib} to ensure its validity.
We found an increase in correlation with the FD \xmax measurement after performing each SD-based calibration.
Furthermore, the Pearson correlation coefficient increased from 0.62 to 0.7 by applying all the SD calibrations and the analysis-specific cuts, thus confirming the validity of the calibrations and the selection.

\subsection{\label{sec:sys_unc}Systematic uncertainties}
The systematic uncertainties of the \xmaxmu measurement using the SD are shown in \cref{fig:mean_xmax_sys}.
The \xmax-scale uncertainty of the FD, as inherited by the DNN during the calibration using hybrid measurements to remove the dependence on hadronic interaction models, is depicted as a dash-dotted line.
It contains uncertainties regarding the reconstruction, the atmosphere, and the calibration of the FD.
Whereas the latter is independent of energy, the energy dependence is caused by the former two contributions.
At low energies, reconstruction uncertainties of the FD dominate.
These are surpassed by atmospheric uncertainties with increasing energy since more distant showers can be detected with correspondingly larger corrections for the light transmission between the shower and the detector. 
For more details on the FD uncertainty, we refer to Ref.~\cite{pierre_auger_collaboration_depth_2014}. 
The uncertainties from the SD are denoted as a hatched region and are summarized in \cref{fig:sd_only_unc}.
They comprise the remaining uncertainties of the detector aging ($<0.5$~\gcm), diurnal variations (1~\gcm) --- since the FD calibration is performed at night --- and the uncertainty on the calibration using hybrid events.
The calibration uncertainty has two parts, the uncertainty of the definition of the absolute \xmax scale, which is estimated to be $\pm2~\gcm$, and the energy dependence of the calibration.
To estimate the energy-dependent uncertainty of the calibration, we compare the assumed constant calibration to a calibration function linear in $\log_{10}(E_\mathrm{SD} / \mathrm{eV})$ and use the observed differences as the upper and lower uncertainty on our calibration (compare \cref{fig:dnn_fd_calib_1} in the appendix).

\begin{figure*}[t!]
    \begin{centering}
        \begin{subfigure}[b]{0.495\textwidth}
            \begin{centering}
                \includegraphics[width=\textwidth,trim=0.45cm 0.4cm 2.05cm 0.15cm, clip]{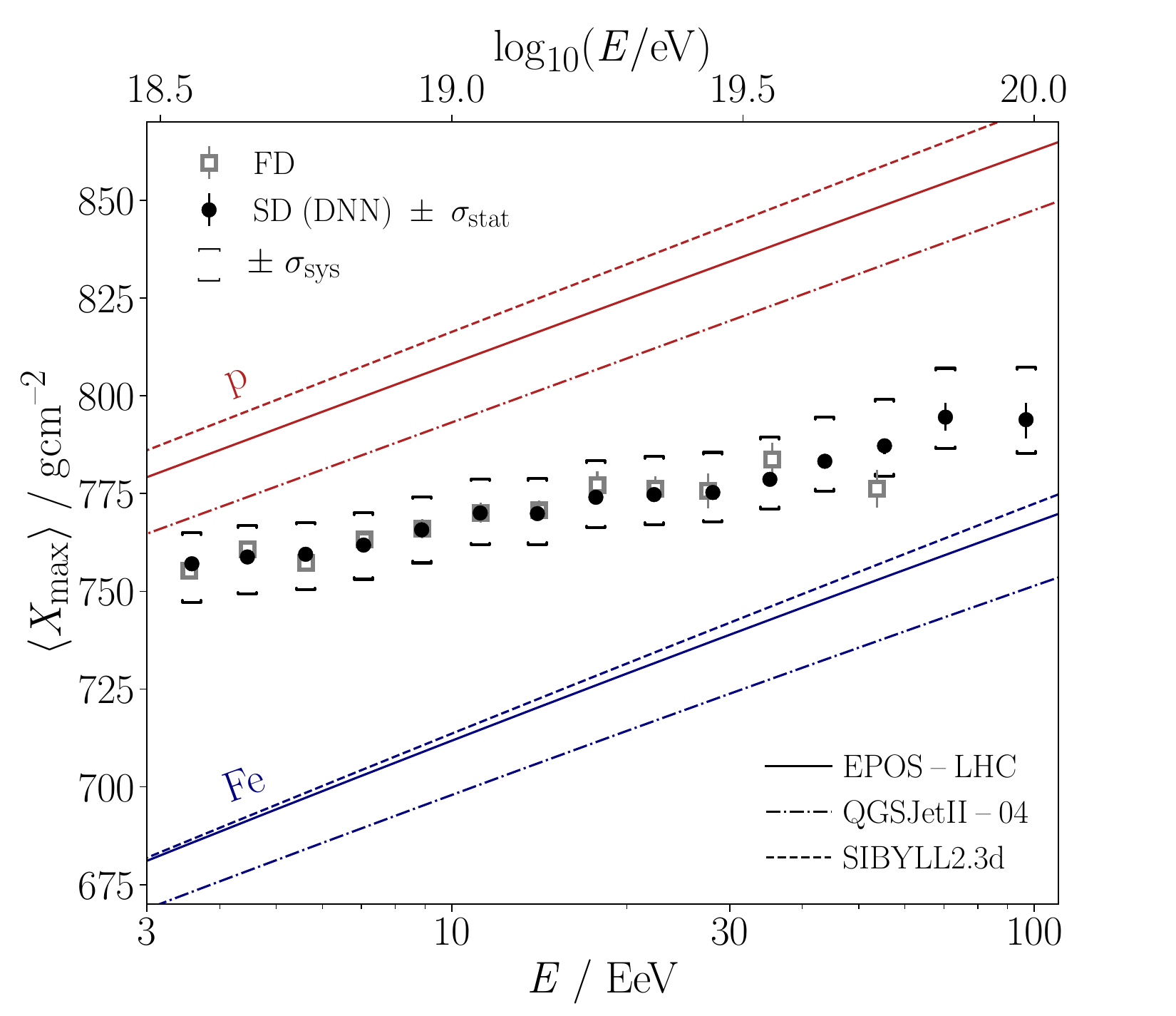}
                \subcaption{}
                \label{fig:1st_moment_dnn_fd}
            \end{centering}
        \end{subfigure}
        \begin{subfigure}[b]{0.495\textwidth}
            \begin{centering}
                \includegraphics[width=\textwidth,trim=0.45cm 0.4cm 2.05cm 0.15cm, clip]{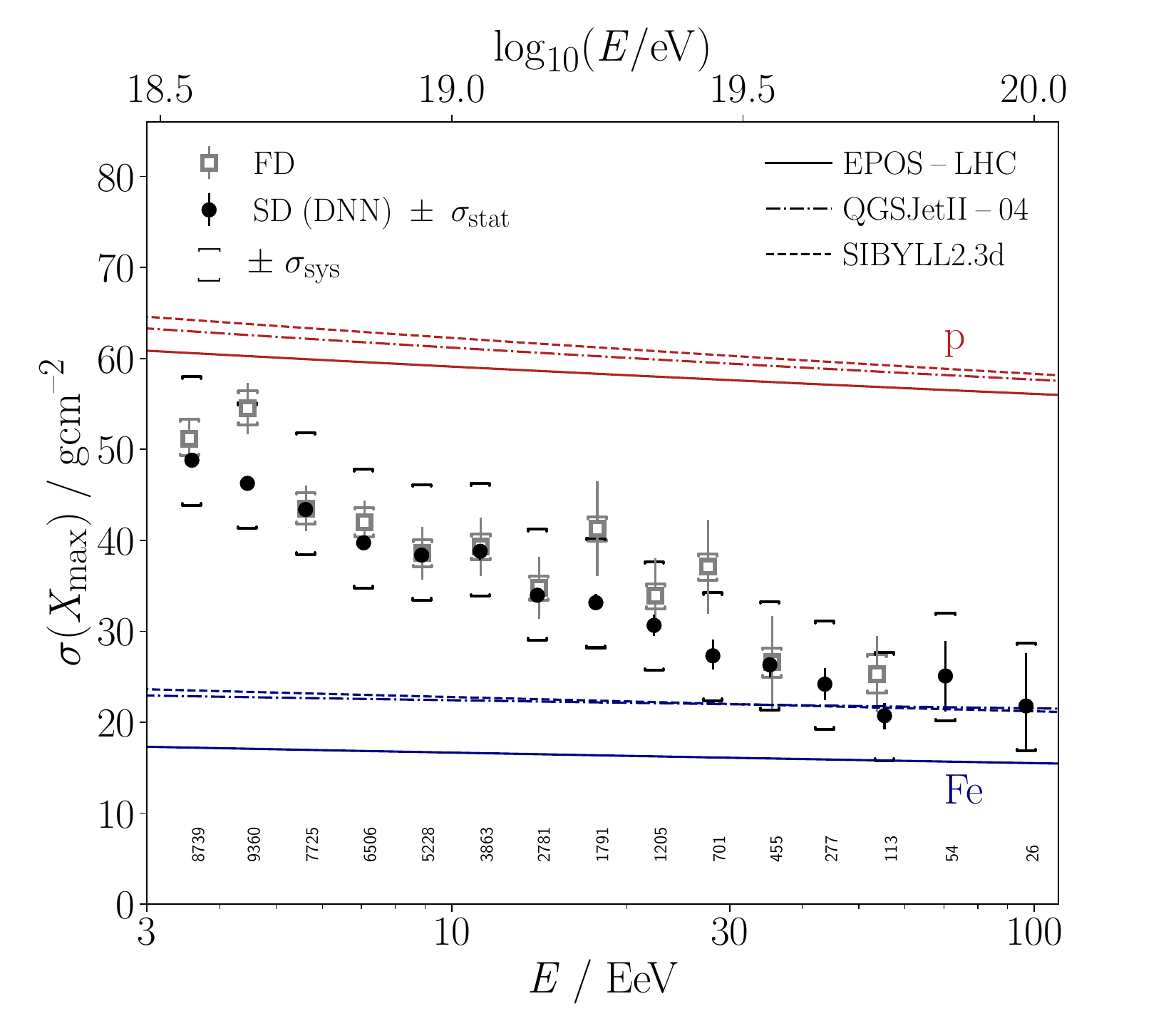}
                \subcaption{}
                \label{fig:2nd_moment_dnn_fd}
            \end{centering}
        \end{subfigure}
    \end{centering}
    \caption{\label{fig:moments_sd_fd}\small{Energy evolution of (a) the average depth of shower maximum \xmaxmu and (b) the fluctuations of the shower maximum \xmaxsigma as determined using the FD reconstruction~\cite{yushkov_mass_2019} (grey open squares) and the DNN \xmax predictions (black circles). Red (blue) lines indicate expectations for a pure proton (iron) composition for various hadronic models.}}
\end{figure*}

The resulting total uncertainties of the SD-based \xmaxmu measurement are of the order of $\pm 10~\gcm$ and shown as a continuous red line.
In general, the obtained uncertainty is very similar to the FD uncertainty. 
Only at high energies, due to the limited statistics of hybrid events, is the uncertainty on the calibration rising slightly.
Nevertheless, at high energies, substantial deviations from the applied calibration are not to be expected since the simulation study (see \cref{sec:foward_fold}) indicated only a very small reconstruction bias above 30~EeV.

In contrast with the measurement of the first moment, no strong dependence of \xmaxsigma on hadronic interaction models was found.
Therefore, no calibration is performed using the FD.
Hence, the measurement is independent of the FD and the systematic uncertainties contain only SD contributions.
\cref{fig:sigma_xmax_sys} displays the different contributions as a function of energy, where effects are only shown that contribute more than $1~\gcm$.
The largest source of uncertainty at low energies is the composition bias that was found to be independent of the interaction models.
It was derived from the simulation studies reported in \cref{sec:foward_fold} by assuming for each energy bin the largest reconstruction bias found in studies of a pure proton, a pure iron, and the Auger mix composition.
The assumed parameterizations are a conservative estimate as, in nature, a pure proton, a pure iron, and an Auger mix composition cannot exist at the same time.
Nonetheless, this provides an estimate for all potential scenarios, even though a substantial proton or iron fraction at high or low energies, respectively, are extremely unlikely.
In this work, we use the bias parameterization\footnote{No significant differences can be observed between different interaction models.
The obtained parameterization can be found in \cref{fig:comp_bias_parameterizations}.} obtained for the \epos interaction model.
To examine potential shortcomings of the modeling of the muon component as intensively discussed in the literature~\cite{Aab_2015_muons, Aab_2016_test_models, muons_amiga, the_pierre_auger_collaboration_measurement_muon_fluctuations}, we studied \sibstar with a significantly increased muon number and found a slight underestimation of the fluctuations of the order of $5$~\gcm, constant for all compositions and independent of energy.
In particular, at high energies, we account for a potential muon deficit, contributing more than all other factors to the uncertainty of \xmaxsigma.
Other contributions come from saturation effects ($-2$~\gcm) and detector aging ($\pm 1.5~\gcm$).

\section{\label{sec:measurement}Investigation of the UHECR mass composition using the Surface Detector}
In the following section, we present inferences on the UHECR mass composition based on the first two moments --- \xmaxmu and $\xmaxsigma~$--- of the reconstructed \xmax distributions.
The composition analysis is based on 48,824 events recorded using the SD and its evolution studied in bins of $\Delta \log_{10}{(E/\mathrm{EeV})} = 0.1$ with an integral bin above $10^{19.9}$~eV.
For comparison, we use FD data of Ref.~\cite{yushkov_mass_2019} covering the same data-taking period.
Since the full FD data set features --- in comparison to the hybrid data set used for calibrating the DNN, which requires a full SD and FD reconstruction as well as a specific geometry due to the different efficiencies of the two detectors as a function of zenith angle --- an increase in statistics of a factor of almost five, we allow for a constant shift of the SD \xmax scale when comparing the DNN and FD measurements.
The adjustment on top of the hybrid events study amounts to $-1.7$~\gcm and is within our statistical uncertainty (2~\gcm) of the calibration.

\begin{figure*}[ht!]
    \begin{centering}
        \begin{subfigure}[b]{0.48\textwidth}
            \begin{centering}
                \includegraphics[width=\textwidth,trim=0.45cm 0.4cm 2.05cm 0.15cm, clip]{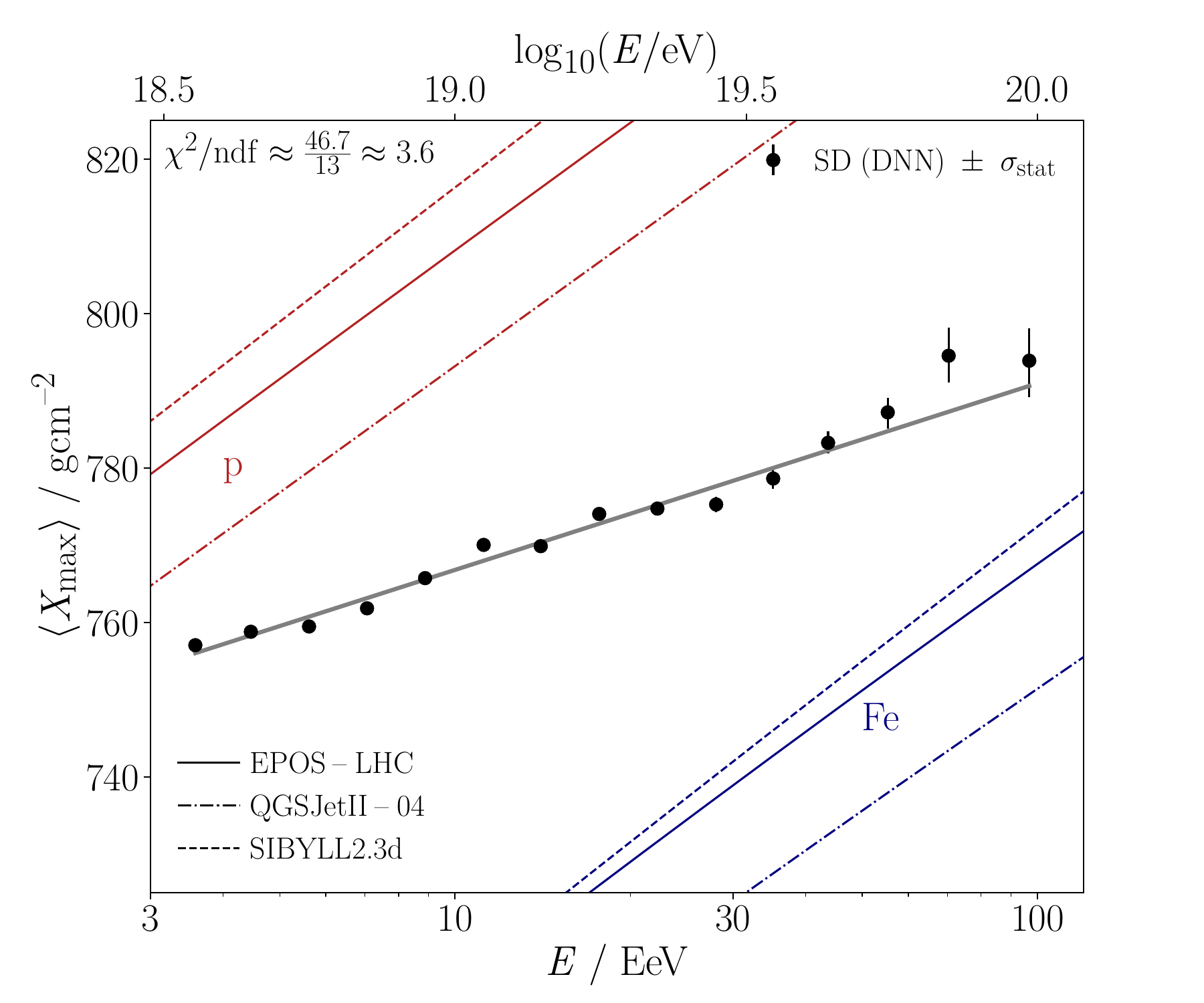}
                \subcaption{}
                \label{fig:const_elong}
            \end{centering}
        \end{subfigure}
        \begin{subfigure}[b]{0.48\textwidth}
            \begin{centering}
                \includegraphics[width=\textwidth,trim=0.45cm 0.4cm 2.05cm 0.15cm, clip]{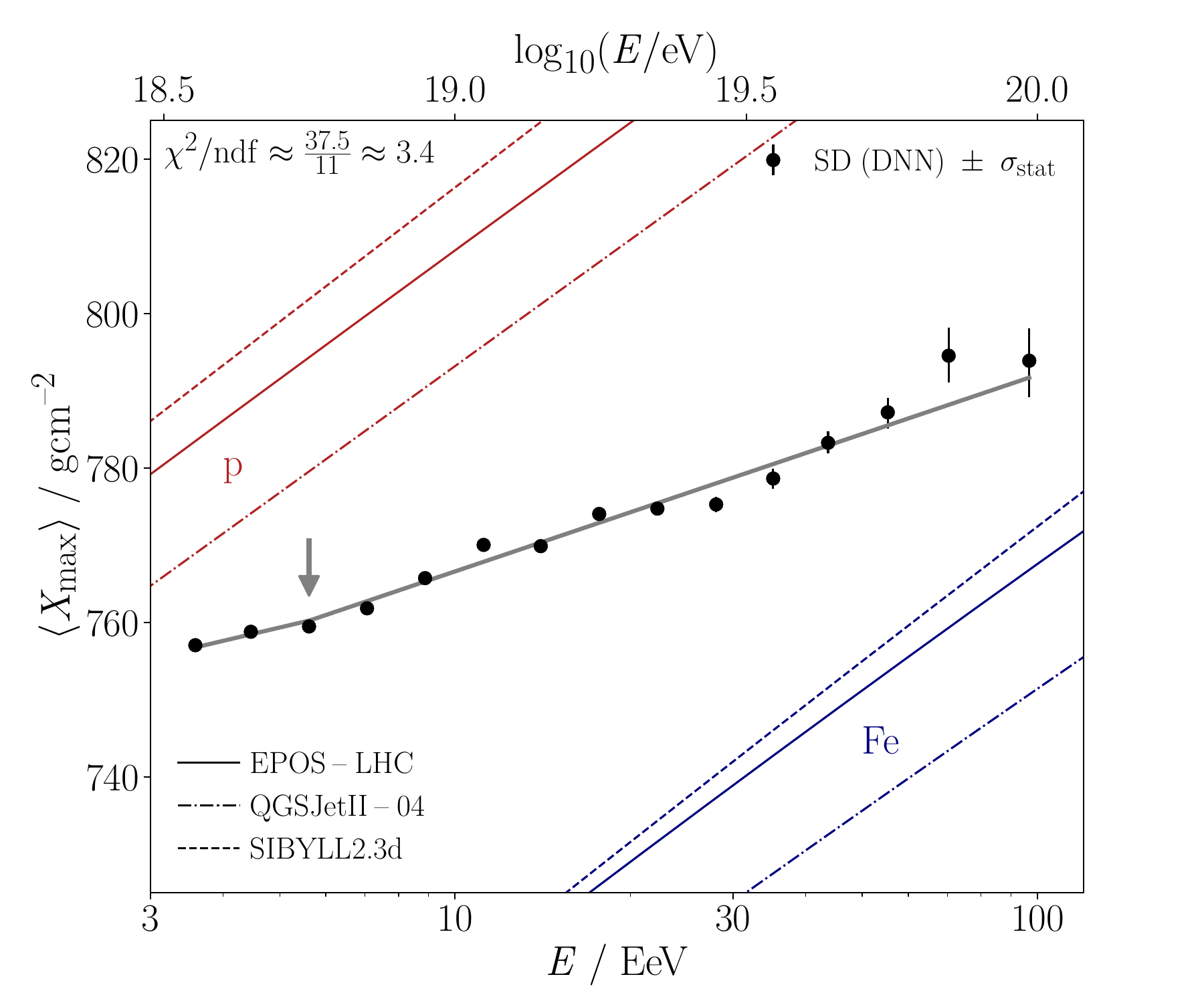}
                \subcaption{}
                \label{fig:1break}
            \end{centering}
        \end{subfigure}
    \end{centering}
    \begin{centering}
        \begin{subfigure}[b]{0.48\textwidth}
            \begin{centering}
                \includegraphics[width=\textwidth,trim=0.45cm 0.4cm 2.05cm 0.15cm, clip]{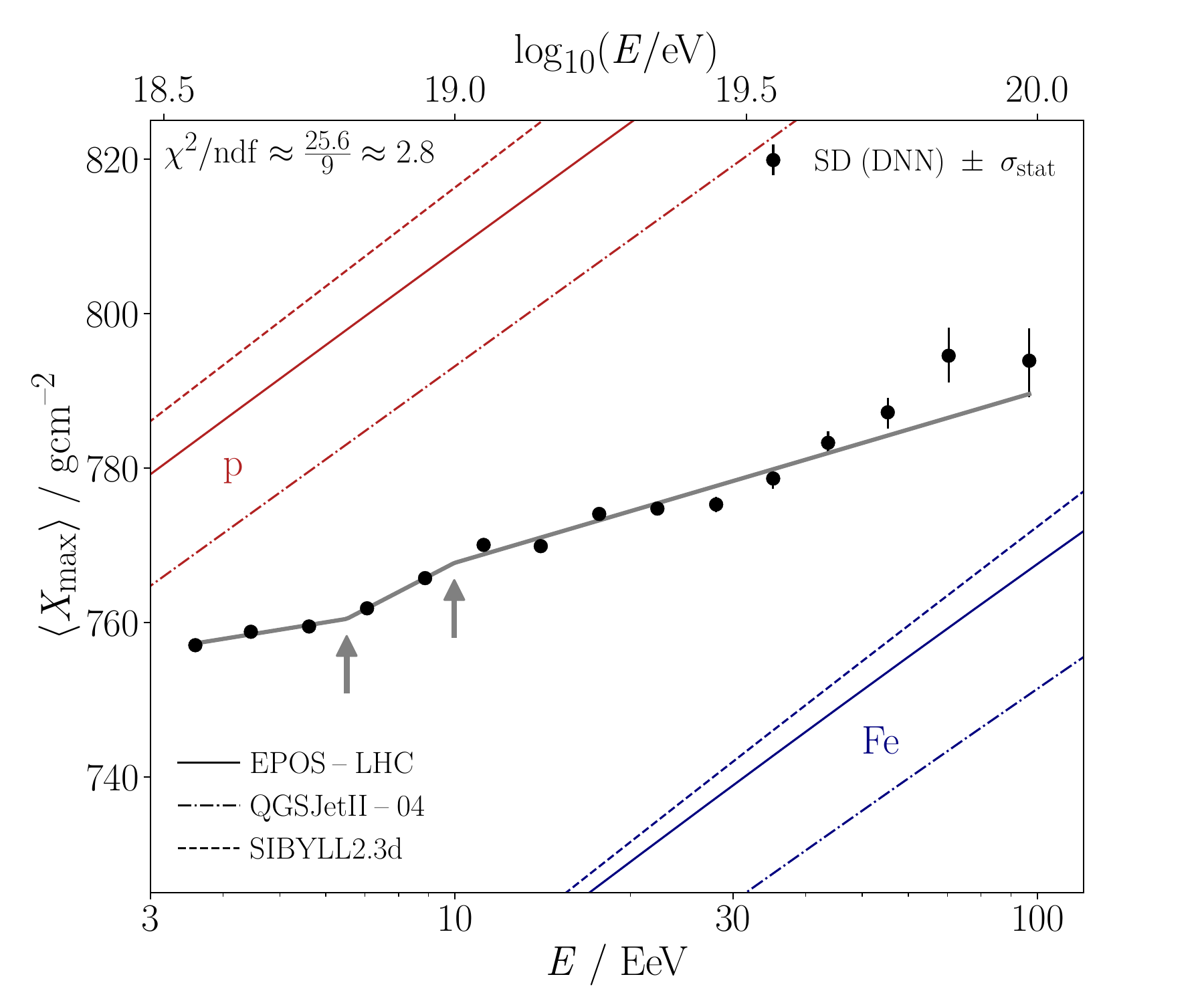}
                \subcaption{}
                \label{fig:2breaks}
            \end{centering}
        \end{subfigure}
        \begin{subfigure}[b]{0.48\textwidth}
            \begin{centering}
                \includegraphics[width=\textwidth,trim=0.45cm 0.4cm 2.05cm 0.15cm, clip]{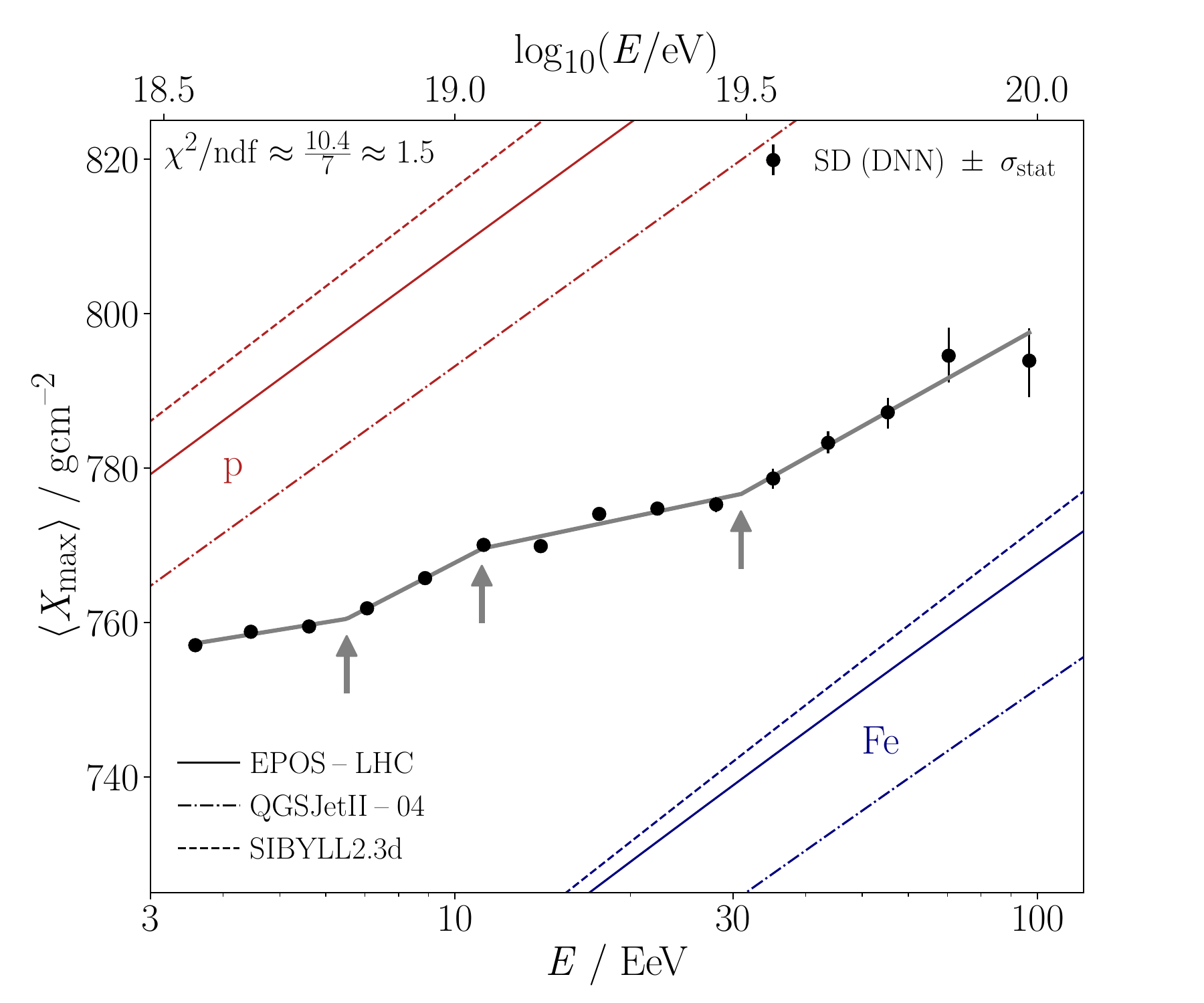}
                \subcaption{}
                \label{fig:3_breaks}
            \end{centering}
        \end{subfigure}
    \end{centering}
    \caption{\label{fig:fits_of_elong}{Investigated models (grey lines) describing the evolution of \xmaxmu as a function of energy $E$. The studied models are piecewise-linear in $\log_{10}(E/\mathrm{eV})$. (a) Fit of a constant elongation rate, as suggested by the FD data analyses above 3~EeV. More complex models describing a scenario beyond a constant evolution: piecewise-linear models with (b) one break, (c) two breaks, and (d) three breaks. The locations of the breaks are indicated by grey arrows.}}
\end{figure*}

In \cref{fig:moments_sd_fd}, we present the energy evolution of \xmaxmu and \xmaxsigma as reconstructed using the DNN based on SD data (black circles) and as obtained using the standard FD reconstruction (open grey squares).
The statistical uncertainties are estimated using bootstrapping and are shown as vertical lines, whereas the systematic uncertainties, discussed in \cref{sec:sys_unc}, are depicted as brackets\footnote{For the detailed table, see \cref{tab:xmax_dnn}}. 
We do not show the systematic uncertainty for the measurement of \xmaxmu using the FD as it is part of the SD uncertainty due to the cross-calibration we conducted. Since for \xmaxsigma no calibration is performed, we show the systematics for both measurements.
The measured data are compared to predictions~\cite{domenico_reinterpreting_2013} for protons (red) and iron (blue) of the three hadronic interaction models \epos, \sibylld, and \qgs, denoted by different line styles. In the right plot, we further show the number of events in each bin of the SD data, which is the same in both plots.
The evolution of \xmaxmu in \cref{fig:1st_moment_dnn_fd} as a function of energy shows an excellent agreement between the SD and the FD measurements with only very small deviations that can be explained purely by statistics.
This extends the \xmax measurements to 100~EeV and confirms the transition from a lighter to a heavier composition with increasing energy, also reported in previous SD-based studies using the risetime of signals in the WCDs~\cite{aab_pierre_auger_collaboration_inferences_2017}.

The \emph{elongation rate} $D_{10}$ is defined by the change of \xmaxmu per decade of energy
\begin{equation*}
D_{10} = \frac{\mathrm{d}\xmaxmu}{\mathrm{d}\log_{10}(E)} = \hat{D}_{10} \left( 1 - \frac{\mathrm{d}\langle \ln A\rangle}{\mathrm{d} \ln(E)} \right ),
\end{equation*}
where $A$ denotes the primary particle mass.
When measuring $D_{10}$, a deviation from the elongation rate $\hat{D}_{10}$, which is in a very good approximation, universal across all hadronic interaction models and primary nuclei, can be traced back to a change in the primary mass composition.
The elongation rate obtained using the SD over the whole energy range amounts to $D_{10}=(24.1\pm1.2)~\gcmd$ in good agreement with the FD result $\left((26\pm 2)~\gcm\right)$~\cite{yushkov_mass_2019}.
However, the reduced $\chi^2 / \mathrm{ndf}={46.7}/{13}$ obtained for the SD data indicates that another substructure exists, as will be comprehensively discussed in the next \cref{sec:features}.

The evolution in \xmaxsigma, sensitive to the composition mixing, is shown in \cref{fig:2nd_moment_dnn_fd}.
We find a decrease of \xmaxsigma as a function of energy and a very good agreement between the measurements of the SD and the FD.
This confirms for the first time the transition from a lighter and mixed composition into a heavier and purer composition with large statistics.
At the highest, previously inaccessible energies ($>50$~EeV), the fluctuations appear to stabilize and remain small.
However, more statistics are needed to examine the composition evolution at these energies in more detail.
Given the limited differences in the interaction model predictions of \xmaxsigma, the small fluctuations in \xmax beyond 30~EeV clearly exclude a scenario with a substantial fraction of protons and light nuclei in the UHECR composition.
Additionally, at around 10~EeV, the fluctuations appear to stay constant.

\subsection{\label{sec:features}Discussion of breaks in the elongation rate}
The observation of an elongation rate similar to the FD but obtained using the comprehensive SD data set that features $\chi^2/\mathrm{ndf} \approx 3.6$, indicates that a simple linear model is not describing the data well (see \cref{fig:const_elong}), suggesting the existence of a substructure to be analyzed.
The measurement of \xmaxsigma also shows a non-continuous decrease of fluctuations with energy.

In \cref{fig:fits_of_elong}, we study the evolution in the UHECR mass composition using different models. We analyze the evolution using broken-line fits with a different number of breaks.
The simplest model beyond a constant elongation rate is a broken-line fit with one fitted break point shown in \cref{fig:1break} that also cannot describe our data reasonably ($\chi^2/\mathrm{ndf} \approx 3.4$).
Considering Wilks' theorem, we compared the $\chi^2$ values of two nested models, in which the model of a constant elongation rate is used as the null hypothesis and test if it can be rejected with more complex models.
A model with two breaks in the elongation rate can reject the constant elongation rate hypothesis at a significance of $3.4\sigma$ (see \cref{fig:2breaks}).
In \cref{fig:3_breaks}, we show a model with three breaks in the elongation rate, where the slopes and the break position were determined by a fit.
This model can reject the hypothesis of a constant elongation rate at a level of $4.6\sigma$ and a single-break model at a level of $4.4\sigma$, which, on a statistical basis, indicates a substructure in the evolution of the UHECR composition.
The significance of rejecting the hypothesis of a two-break model using the three-break model amounts to $3.3\sigma$.

\begin{figure}[h!]
    \centering
        \includegraphics[width=0.99\linewidth]{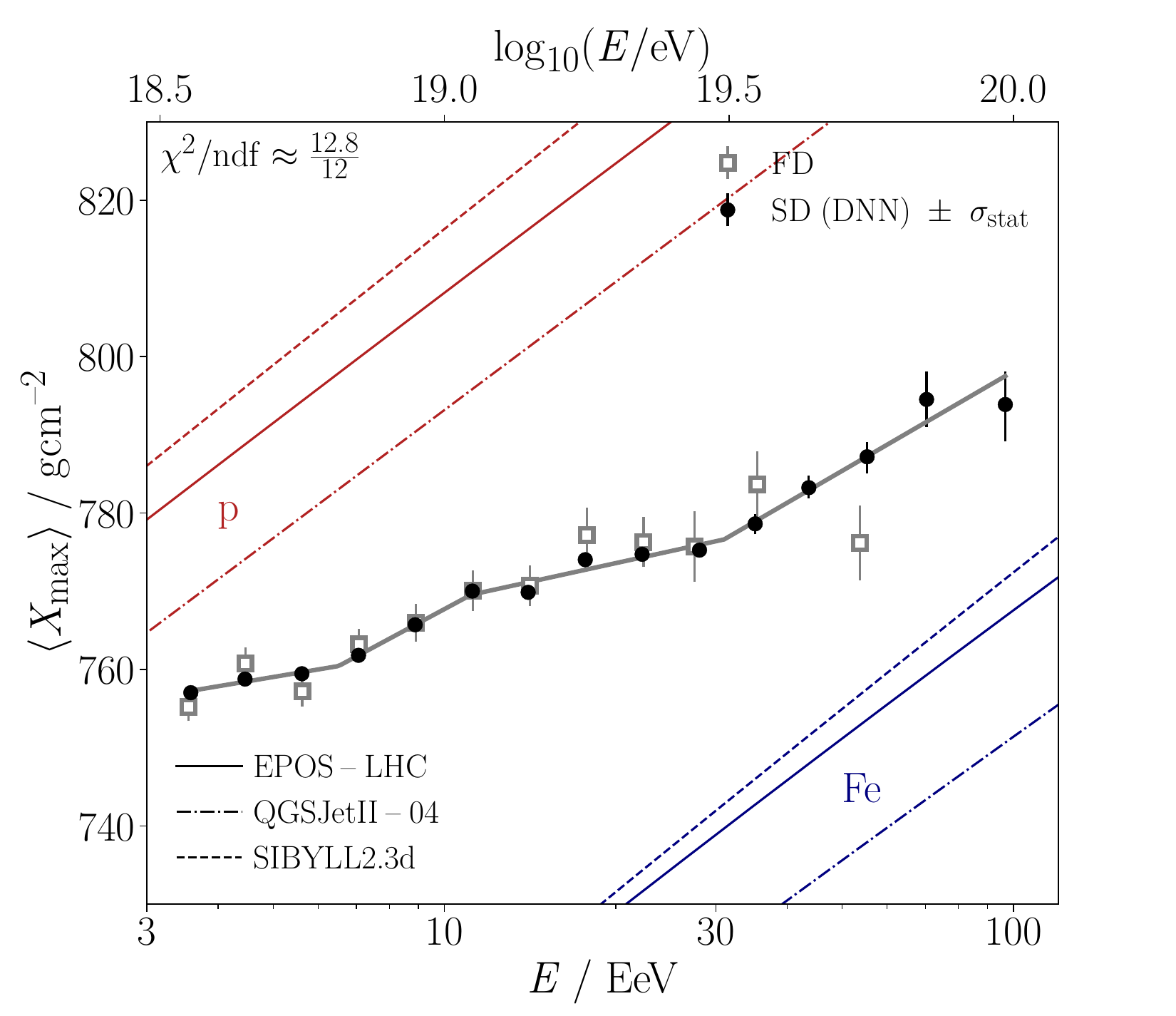}
    \caption{The found elongation rate model with three breaks obtained using SD data (continuous grey line) compared to the evolution of \xmaxmu as measured using the FD (open grey boxes) and the SD (black markers). The $\chi^2$ shown refers to the FD data.}
    \label{fig:elong_compare_FD_SD}
\end{figure}

\begin{table*}[ht!]
\begin{tabular}{ l c c c c c}
    \hline\hline
     Parameter & Const. elong. & 1-break model & 2-break model & 3-break model & Energy spectrum \\
               & $\mathrm{Val} \pm \sigma_\mathrm{stat} \pm \sigma_\mathrm{sys}$ & $\mathrm{Val} \pm \sigma_\mathrm{stat} \pm \sigma_\mathrm{sys}$  & $\mathrm{Val} \pm \sigma_\mathrm{stat} \pm \sigma_\mathrm{sys}$ & $\mathrm{Val} \pm \sigma_\mathrm{stat} \pm \sigma_\mathrm{sys}$ & $\mathrm{Val} \pm \sigma_\mathrm{stat} \pm \sigma_\mathrm{sys}$  \\
    \hline
    $b$~/~\gcm    & $743\pm5\pm13$& $743\pm5\pm13$ & $750.5\pm4\pm13$ & $750.5\pm3\pm13$ &  \\
    $D_0$~/~\gcmd & $24\pm1\pm4$  & $23\pm 2\pm 12$ & $12\pm 6\pm 5$ & $12\pm5\pm6$   &  \\
    $E_1$~/~EeV   &               & $35\pm12\pm16$& $6.5\pm0.9\pm1$& $6.5\pm0.6\pm1$& $4.09\pm0.1\pm0.8$  \\
    $D_1$~/~\gcmd &               & $39\pm 14\pm 12$ & $39\pm 12\pm 10$& $39\pm 5\pm14$ &  \\
    $E_2$~/~EeV   &               &                & $10\pm 2\pm 3$ & $11\pm 2\pm1$  & $14\pm1\pm2$  \\
    $D_2$~/~\gcmd &               &                & $22\pm 3\pm 8$& $16\pm3\pm6$   &  \\
    $E_3$~/~EeV   &               &                &                & $31\pm5\pm3$   & $47\pm 3\pm 6$  \\
    $D_3$~/~\gcmd &               &                &                & $42\pm9\pm12$  & \\

    \hline\hline
\end{tabular}
\caption{Best-fit parameters with statistical and systematic uncertainties for the studied elongation models that feature up to three changes at energies ($E_1, E_2, E_3$) in the elongation rate ($D_0, D_1, D_2, D_3$) and an offset $b$ (\xmaxmu at 1~EeV), without including a systematic uncertainty of 14\% on the energy scale. Also given are the positions of the energy spectrum features measured at the Pierre Auger Observatory~\cite{Abreu_2021_spectrum} in the same data-taking period.}
\label{tab:breaks}
\end{table*}

The investigated models and their parameters are summarized in \cref{tab:breaks}, including statistical and systematic uncertainties, and compared to the positions of the energy spectrum features identified at ultra-high energies.
To account for the two energy-dependent systematic uncertainties of the \xmaxdnn measurement, the inherited energy-dependent uncertainty on the FD \xmax scale (dash-dotted line in \cref{fig:mean_xmax_sys}) and potential energy dependence on the hybrid calibration were considered.
We assessed six different energy-dependent calibration functions (broken-line fits shown in \cref{fig:hybrid_calibration_calib_checks}) affecting the elongation rate and three different cases for the FD \xmax scale uncertainty: no shift, shift by the upper uncertainty, and shift by the lower uncertainty.
For each combination, fits and significance estimates were performed as described above.
The final significance was assigned to the minimum significance value found in this test.
The breaks in the evolution of \xmaxmu in all models are observed to be at similar energies as the features of the UHECR energy spectrum~\cite{Abreu_2021_spectrum}, i.e., the ankle at $\left(4.9\pm\,0.1(\mathrm{stat})\pm0.8\,(\mathrm{sys})\right)$~EeV, instep at $\left(14\pm1\,(\mathrm{stat})\pm2\,(\mathrm{sys})\right)$~EeV and suppression at $\left(47\pm3\,(\mathrm{stat})\pm6\,(\mathrm{sys})\right)$~EeV.
Note that, even for a joint astrophysical interpretation, features in the energy spectrum and the evolution of \xmaxmu do not have to coincide in energy, as, for example, the break in the elongation rate observed around 2~EeV~\cite{pierre_auger_collaboration_depth_2014} is physically interpreted in association with the ankle~\cite{combined_fit_auger, combined_fit_eleonora}, located at 5~EeV.

We analyzed the $\xmaxsigma$ measurement for characteristics similar to the ones found in the evolution of \xmaxmu.
Between $E_0=6.5$~EeV and $E_1=11$~EeV, where the observed elongation rate is within uncertainties compatible with a constant composition, also \xmaxsigma appears to stay constant.
Furthermore, beyond $E\approx30$~EeV (at $E_2$), the decrease in the fluctuations appears to stop, which would be consistent with the elongation rate that was found to be close to that of a constant composition at the highest energies.
Due to the increasing statistical uncertainties, more data are needed for a definite statement.
A quantitative test of a structure in \xmaxsigma with breaks at positions that agree with the ones found in the elongation rate study, however, is not significant.
The null hypothesis of a linear decrease of \xmaxsigma can be rejected at only a $2.2\sigma$ significance level, using a model with three break positions fixed to the ones found in the elongation rate study, which nonetheless seems to be compatible with the data ($\chi/\mathrm{ndf} = 10.3/10$).
Reduced uncertainties and more data are required to analyze the structure in the evolution of \xmaxsigma in detail.

Note that a one-to-one agreement of breaks and structures generally, in the measurements of \xmaxmu and \xmaxsigma, is not to be expected since a change in the mean logarithmic mass does not need to coincide with a similar change in the measurement of \xmaxsigma, i.e., the composition mixing~\cite{interpretation_auger_jcap}.
It would rather reveal a characteristic structure of the composition.
Interestingly, breaks at similar positions in the energy evolution of \xmax and \xmaxsigma can be obtained when fitting a simplified astrophysical model using the FD \xmax data and the Auger spectrum as measured by the SD (see Fig.~3 and Fig.~6 in Ref.~\cite{combined_fit_eleonora}).
It is worth mentioning that a simple transition between two primary species at a constant rate corresponds to a linear dependence of\xmaxmu on $\log E$ but a non-linear behavior of \xmaxsigma, for the interpretation of which the application of the broken-line fit model is generally inappropriate\footnote{For the evolution involving a larger number of primary species with unknown proportions, an analytical \xmaxsigma-model can not be defined.}.
A dedicated analysis focusing on the astrophysical interpretation and investigating the non-trivial interplay between the spectrum, \xmaxmu, and \xmaxsigma, is ongoing and will be discussed in a future publication.

\begin{figure*}
    \centering
    \includegraphics[width=0.99\textwidth]{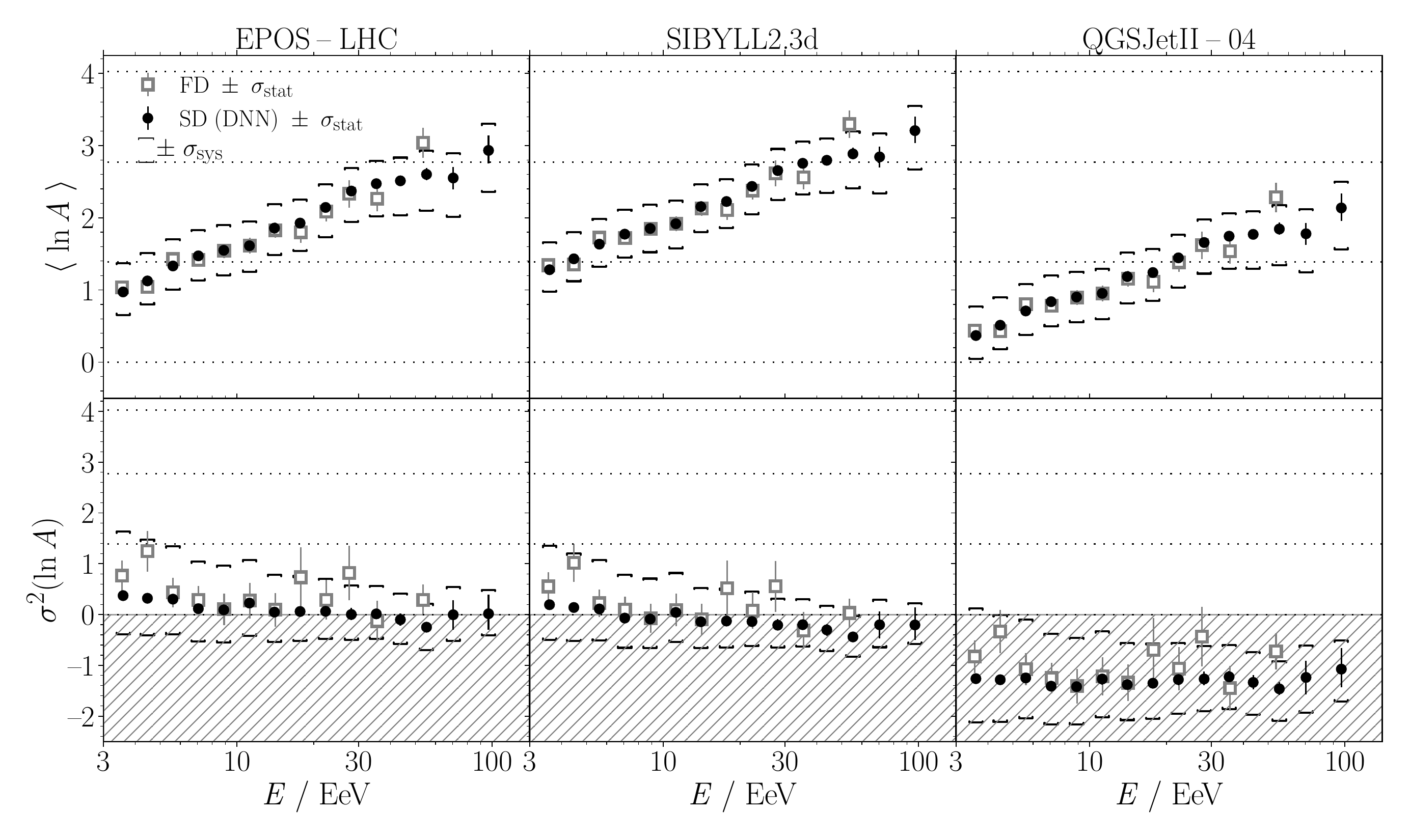}
    \caption{Evolution of the first two moments of the $\ln A$ distribution as a function of energy $E$ for the hadronic interaction models \epos, \sibylld, and \qgs, determined using the FD~\cite{yushkov_mass_2019} (grey open squares) and the DNN (black circles).}
    \label{fig:lna_moments}
\end{figure*}

\subsubsection{Crosscheck and comparison with the FD}
The obtained model exhibiting a characteristic structure beyond a constant change in the mean logarithmic mass has to be consistent with the FD measurements. 
The comparison of the elongation model with the FD and SD is presented in \cref{fig:elong_compare_FD_SD}.
The model describes the FD data adequately with $\chi^2 / \mathrm{ndf} = 1.1$, demonstrating the consistency of the model with FD data.
Additionally, we tested the obtained model by performing the fits using a different binning in energy and by dividing the data into different subsets binned in core distance, zenith angle, azimuth angle, and detector age, as well as season (winter/summer) and daytime (day/night).
In each of the subsets, the determined breaks agreed with uncertainties with the identified model, and no significant deviations could be found.
Next, we investigated the existence of the features before applying calibrations and without applying quality cuts and found a statistical significance level larger than $5\sigma$ using the model featuring three breaks.
Since our selection removes approximately half of the events, we estimated that the expected median significance for identifying the breaks with a data set with the quality of the full data set is $3.4\sigma$.
In this study, we performed an energy-dependent re-sampling of the full data to account for the energy-dependent efficiency of the fiducial cut, i.e., to ensure in each energy bin similar statistics as the fiducial data set.
The finding of a $4.6\sigma$ significance using our data selection confirms the expectation that the significance should increase with improving data quality.

We additionally tested energy-dependent calibrations of the DNN with the hybrid data by employing various broken-line model fits. The tested calibration functions are summarized in \cref{fig:hybrid_calibration_calib_checks}.
None of the tested calibration functions reduced the significance of rejecting a constant elongation rate but showed, due to the energy dependence of the calibration, a stronger rejection of a constant elongation.
In addition, for each studied hybrid calibration, we examined the energy-dependent FD \xmax scale uncertainty.
The significance of rejecting the constant elongation rate with the three-break model remains of the same order, with a minimum of $4.4\sigma$ observed for the cases where the total lower and upper uncertainty is applied to the measurement.
The two-break model can be rejected at a significance level of around $3\sigma$ in most cases.
Only for more complex functions (compare \cref{fig:dnn_fd_calib_5} and \cref{fig:dnn_fd_calib_6}), which cannot be strongly constrained due to the low statistics in the hybrid sample, the significance level drops to around $2\sigma$.
The rejection of a single-break model consistently remains above the $3\sigma$ level and is at the $4\sigma$ level in most cases.

Rejecting a constant elongation rate using the two-break model is very stable and above a significance of $3.4\sigma$ for all scenarios.
Applying instead of the FD calibration a correction of the SD \xmax reconstruction based on the expected composition bias of the Auger mix using simulations (compare \cref{fig:comp_bias}), a constant elongation rate can be rejected by more than $5\sigma$ assuming \epos, as the first break is strongly pronounced. 
Therefore, we find a robust indication at a $4.4\sigma$ level for structures beyond a constant elongation rate.
However, more statistics and/or a reduction in energy-dependent uncertainties are needed to confidently reject the two-break model, i.e., to investigate the existence and nature of the third break.

\begin{figure*}
    \centering
    \includegraphics[width=0.95\textwidth]{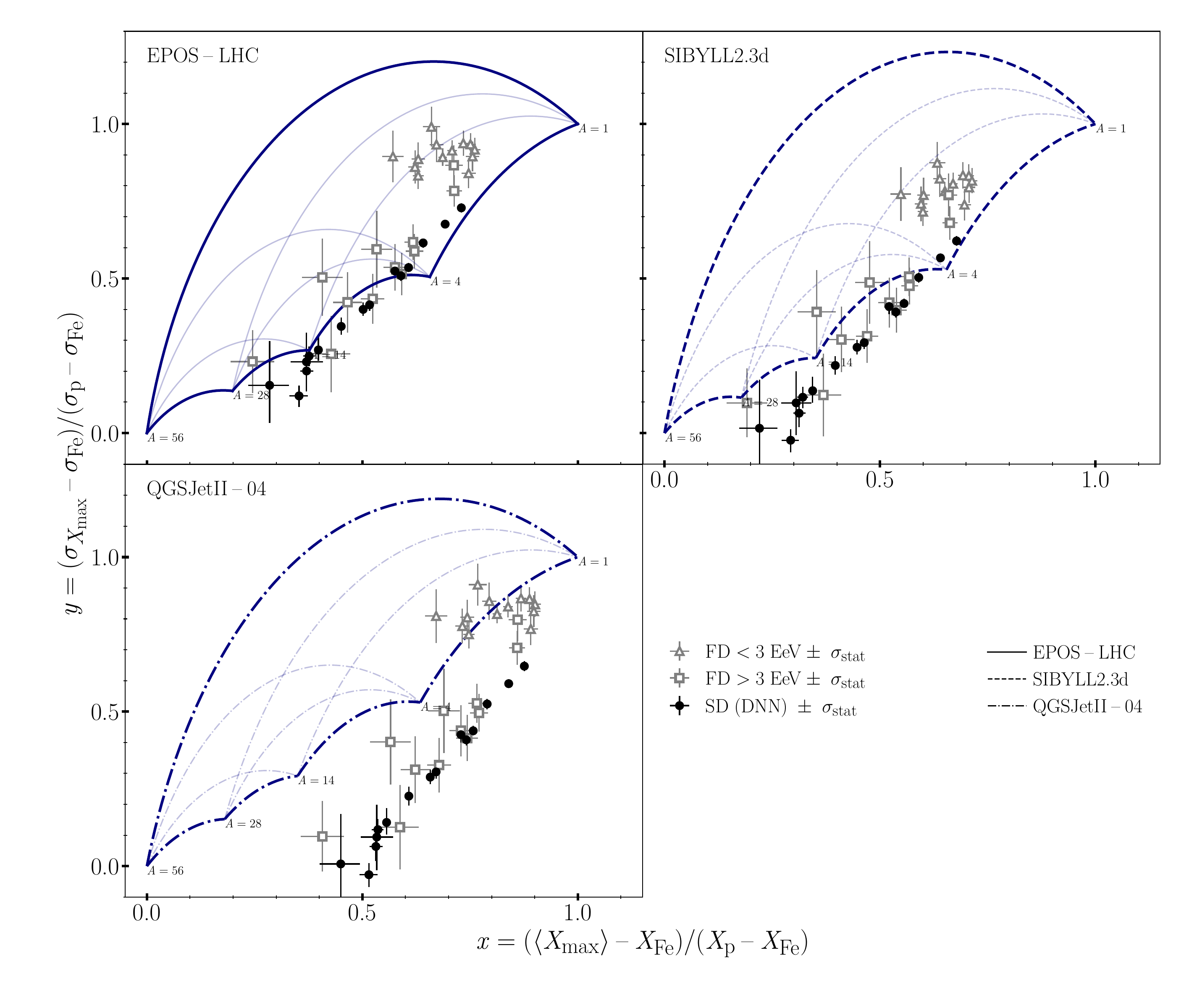}
    \caption{Evolution of the measurements determined using the FD~\cite{yushkov_mass_2019} (grey markers) and the DNN (black circles) in the re-scaled \xmaxmu vs. \xmaxsigma representation compared to specific composition predictions for the hadronic interaction models \epos, \sibylld, and \qgs at 10~EeV. The arc-like curves denoted by blue lines indicate transitions between pairs of pure compositions labeled with their mass numbers (resulting in a so-called ``umbrella'' plot).}
    \label{fig:umbrella}
\end{figure*}

\subsection{Interpretation using hadronic interaction models}
By interpreting the reconstructed moments \xmaxmu and \xmaxsigma using hadronic interaction models, the measurement can be converted into the first two moments of the distributions of the logarithmic mass~\cite{lna_paper, icrc_13_proc_auger}, its mean \lnamu and variance $\sigma^2(\ln\,A)$.
In \cref{fig:lna_moments}, the derived moments are shown using air-shower simulations based on the interaction models \epos, \sibylld, and \qgs.
The evolution of the mean logarithmic mass with energy shows a trend from a light composition towards a heavier composition, including the same characteristic breaks at three energies.
Likewise, at around 10~EeV and 30~EeV, the \lnamu shows indications of an almost constant composition.
For all interaction models, the fluctuations $\sigma^2(\ln A)$ in $\ln A$ are small, indicating a composition dominated by a single type of nucleus.
This observation exhibits a distinct characteristic that is quite compatible with the expectations for the Peters cycle. However, for quantitative results on the fluctuations of $\ln A$, the systematic uncertainties in the measurements, as well as the uncertainties in the interaction models, will need to be reduced.

Nonphysical negative fluctuations are found for \qgs across the whole energy range, strongly disfavoring the model, in line with previous studies~\cite{pierre_auger_collaboration_depth_2014, aab_pierre_auger_collaboration_inferences_2017, yushkov_mass_2019, fd_icrc_23}.
Negative fluctuations for \sibylld and \epos are also visible but are compatible with zero within uncertainties.
Note that this result does not state that the fluctuations are not correctly modeled in simulations but rather that the fluctuations expected from a composition derived from the \xmaxmu measurement are in tension with the model predictions.
In fact, the uncertainties from the interaction-model description of the fluctuations are rather small, and parts of the mismatch found could likely originate from differences in the \xmax scale in measured data and simulations.
Indications for such a tension in the \xmaxmu scale in simulation and data were previously reported in other studies~\cite{Aab_2016_test_models, jakub_2024testing}.

Another way of comparing the measured data to model predictions is the illustration of the data in a re-scaled \xmaxsigma vs. \xmax plane~\cite{Linsley_lna, Kampert_unger_2012, Lipari_2021}.
First, in this representation, the measurements of \xmaxmu are transformed into the scale of the respective model.
Thus, $x=0$ translates to a pure iron composition, and $x=1$ corresponds to a pure proton composition.
A similar transformation is applied to \xmaxsigma and denoted with $y$.
Note that extremely mixed compositions would feature values larger than $y=1$.
Since the elongation rate for pure beams is, to a good approximation, universal across all interaction models and the energy-dependence of \xmaxsigma is small, the representation allows for a concise interpretation in which transitions between two pure compositions follow arc-like curves in these ``umbrella'' plots.

In \cref{fig:umbrella}, we show the measurement of the SD (black dots) and the FD (open grey squares), including only statistical uncertainties, and compare them to the predictions of the hadronic models \epos, \sibylld, and \qgs.
The blue lines indicate transitions between pairs of pure compositions at an energy of 10~EeV.
FD measurements below 3~EeV are depicted as open grey triangles.
The statistical uncertainties are estimated using bootstrapping and shown as vertical lines.
In the matching energy range, the SD and FD data agree well and show the same evolution, demonstrating a consistent measurement of the two \xmax moments.

Our measurements are consistent with a relatively heavy and pure composition for \epos and \sibylld, within systematic uncertainties.
Again, \qgs shows a significant tension and is disfavored by our measurements.
Furthermore, it can be seen that the features we found in the energy evolution of \xmaxmu yield a consistent picture when including the \xmaxsigma measurement.
Energy regions with smaller changes in \lnamu would appear as clusters of points.
Two such regions are suggested by the elongation rate studies (cf.~\cref{fig:elong_compare_FD_SD}) and are also visible in \cref{fig:umbrella}.
For example, for \epos, the two regions at around 10~EeV and 50~EeV are close to the mass groups of $A\cong4$ and $A\cong14$.

\section{\label{sec:summary}Summary}
In this work, we have presented a study of the UHECR mass composition based on the first two moments of the distribution of depth of maximum, \xmax, of air shower profiles using surface detector data of the Pierre Auger Observatory recorded between 2004 and 2018.
With the use of deep learning, a novel reconstruction technique was developed, enabling for the first time a precise reconstruction of \xmax using the recorded time-dependent SD signals on an event-by-event level.
Our approach included cross-calibration with the complementary FD to remove mismatches between simulations and measured data and investigate systematic uncertainties, highlighting the importance of an independent data set for calibrating and validating machine learning algorithms.
After cross-calibrating the method using fluorescence observations, we have studied the energy evolution of \xmaxmu and \xmaxsigma from 3~EeV up to 100~EeV.
Due to the superior duty cycle of the SD in comparison to the FD, the statistics for composition studies using \xmax are increased by a factor of ten\footnote{A detailed comparison can be found in \cref{tab:sd_vs_fd_events}.} for energies above 5~EeV, enabling for the first time a measurement of \xmaxsigma, sensitive to the composition mixing, beyond 50~EeV.
We have found excellent agreement of the \xmaxmu measurement with previous studies using the FD and confirm the transition of \lnamu from a lighter to a heavier composition.
Furthermore, our \xmaxsigma measurement, which is independent of the FD calibration, agrees very well with previous studies using fluorescence telescopes.
The finding of a decrease in the fluctuations with energy is confirmed, indicating a transition to a heavier and purer composition.
The observation of very small fluctuations appears to exclude a large fraction of light nuclei at the highest energies, further excluding the flux suppression to be caused by a pure proton beam interacting with the cosmic microwave background.
However, this observation is insufficient to disentangle whether the suppression arises from the maximum injection energy at the sources, propagation effects, or a combination of both.

With the increase in statistics, we have found evidence at a level of $4.4\sigma$ for a characteristic structure in the evolution of the mass composition beyond a constant elongation rate, considering both statistical and systematic uncertainties.
The model describing our data best features three breaks in the energy evolution of the composition and is compatible with the FD measurements.
The locations of the identified breaks are found at energies similar to the ankle, instep, and suppression features identified in the UHECR energy spectrum.
While not statistically significant, in \xmaxsigma, two plateaus are visible, where the fluctuations seem to stay constant, interestingly, at energies at which the elongation rate is closer to that of a pure composition.
However, more statistics and reduced systematic uncertainties are needed to study the nature of the identified breaks and, in particular, investigate the existence of the third break.

The study presented here is one of the first that uses deep learning to analyze measured detector data in astroparticle physics, including a comprehensive study of systematic uncertainties.
The demonstrated performance, superior to previous approaches for mass composition studies using SD data, shows promising potential for machine-learning-based methods in astroparticle physics.
The ongoing AugerPrime upgrade, including the upgrade of the water-Cherenkov detectors~\cite{castellina_augerprime_2019}, as well as further improvements in machine-learning-based analysis strategies, opens up new and far-reaching prospects for understanding cosmic rays, their mass composition at ultra-high energies, and ultimately constraining astrophysical models of their origin.

\section*{Acknowledgments}

\begin{sloppypar}
The successful installation, commissioning, and operation of the Pierre
Auger Observatory would not have been possible without the strong
commitment and effort from the technical and administrative staff in
Malarg\"ue. We are very grateful to the following agencies and
organizations for financial support:
\end{sloppypar}

\begin{sloppypar}
Argentina -- Comisi\'on Nacional de Energ\'\i{}a At\'omica; Agencia Nacional de
Promoci\'on Cient\'\i{}fica y Tecnol\'ogica (ANPCyT); Consejo Nacional de
Investigaciones Cient\'\i{}ficas y T\'ecnicas (CONICET); Gobierno de la
Provincia de Mendoza; Municipalidad de Malarg\"ue; NDM Holdings and Valle
Las Le\~nas; in gratitude for their continuing cooperation over land
access; Australia -- the Australian Research Council; Belgium -- Fonds
de la Recherche Scientifique (FNRS); Research Foundation Flanders (FWO),
Marie Curie Action of the European Union Grant No.~101107047; Brazil --
Conselho Nacional de Desenvolvimento Cient\'\i{}fico e Tecnol\'ogico (CNPq);
Financiadora de Estudos e Projetos (FINEP); Funda\c{c}\~ao de Amparo \`a
Pesquisa do Estado de Rio de Janeiro (FAPERJ); S\~ao Paulo Research
Foundation (FAPESP) Grants No.~2019/10151-2, No.~2010/07359-6 and
No.~1999/05404-3; Minist\'erio da Ci\^encia, Tecnologia, Inova\c{c}\~oes e
Comunica\c{c}\~oes (MCTIC); Czech Republic -- GACR 24-13049S, CAS LQ100102401,
MEYS LM2023032, CZ.02.1.01/0.0/0.0/16{\textunderscore}013/0001402,
CZ.02.1.01/0.0/0.0/18{\textunderscore}046/0016010 and
CZ.02.1.01/0.0/0.0/17{\textunderscore}049/0008422 and CZ.02.01.01/00/22{\textunderscore}008/0004632;
France -- Centre de Calcul IN2P3/CNRS; Centre National de la Recherche
Scientifique (CNRS); Conseil R\'egional Ile-de-France; D\'epartement
Physique Nucl\'eaire et Corpusculaire (PNC-IN2P3/CNRS); D\'epartement
Sciences de l'Univers (SDU-INSU/CNRS); Institut Lagrange de Paris (ILP)
Grant No.~LABEX ANR-10-LABX-63 within the Investissements d'Avenir
Programme Grant No.~ANR-11-IDEX-0004-02; Germany -- Bundesministerium
f\"ur Bildung und Forschung (BMBF); Deutsche Forschungsgemeinschaft (DFG);
Finanzministerium Baden-W\"urttemberg; Helmholtz Alliance for
Astroparticle Physics (HAP); Helmholtz-Gemeinschaft Deutscher
Forschungszentren (HGF); Ministerium f\"ur Kultur und Wissenschaft des
Landes Nordrhein-Westfalen; Ministerium f\"ur Wissenschaft, Forschung und
Kunst des Landes Baden-W\"urttemberg; Italy -- Istituto Nazionale di
Fisica Nucleare (INFN); Istituto Nazionale di Astrofisica (INAF);
Ministero dell'Universit\`a e della Ricerca (MUR); CETEMPS Center of
Excellence; Ministero degli Affari Esteri (MAE), ICSC Centro Nazionale
di Ricerca in High Performance Computing, Big Data and Quantum
Computing, funded by European Union NextGenerationEU, reference code
CN{\textunderscore}00000013; M\'exico -- Consejo Nacional de Ciencia y Tecnolog\'\i{}a
(CONACYT) No.~167733; Universidad Nacional Aut\'onoma de M\'exico (UNAM);
PAPIIT DGAPA-UNAM; The Netherlands -- Ministry of Education, Culture and
Science; Netherlands Organisation for Scientific Research (NWO); Dutch
national e-infrastructure with the support of SURF Cooperative; Poland
-- Ministry of Education and Science, grants No.~DIR/WK/2018/11 and
2022/WK/12; National Science Centre, grants No.~2016/22/M/ST9/00198,
2016/23/B/ST9/01635, 2020/39/B/ST9/01398, and 2022/45/B/ST9/02163;
Portugal -- Portuguese national funds and FEDER funds within Programa
Operacional Factores de Competitividade through Funda\c{c}\~ao para a Ci\^encia
e a Tecnologia (COMPETE); Romania -- Ministry of Research, Innovation
and Digitization, CNCS-UEFISCDI, contract no.~30N/2023 under Romanian
National Core Program LAPLAS VII, grant no.~PN 23 21 01 02 and project
number PN-III-P1-1.1-TE-2021-0924/TE57/2022, within PNCDI III; Slovenia
-- Slovenian Research Agency, grants P1-0031, P1-0385, I0-0033, N1-0111;
Spain -- Ministerio de Ciencia e Innovaci\'on/Agencia Estatal de
Investigaci\'on (PID2019-105544GB-I00, PID2022-140510NB-I00 and
RYC2019-027017-I), Xunta de Galicia (CIGUS Network of Research Centers,
Consolidaci\'on 2021 GRC GI-2033, ED431C-2021/22 and ED431F-2022/15),
Junta de Andaluc\'\i{}a (SOMM17/6104/UGR and P18-FR-4314), and the European
Union (Marie Sklodowska-Curie 101065027 and ERDF); USA -- Department of
Energy, Contracts No.~DE-AC02-07CH11359, No.~DE-FR02-04ER41300,
No.~DE-FG02-99ER41107 and No.~DE-SC0011689; National Science Foundation,
Grant No.~0450696; The Grainger Foundation; Marie Curie-IRSES/EPLANET;
European Particle Physics Latin American Network; and UNESCO.
\end{sloppypar}

\bibliographystyle{apsrev4-2}
\bibliography{PRD}

\vspace{4em}
\clearpage
\section*{Appendix}
\FloatBarrier
\hspace{0.1em}

\renewcommand\arraystretch{1.5}
Derivation of the formula for the reconstruction of \xmaxsigma. Here, \xmax denotes the true depth of the shower maximum and \xmaxdnn the reconstruction of the DNN. 
\begin{align}
    \begin{split}
        \sigma^2(\xmaxdnn) = &\,\sigma^2(\xmax + \xmaxdnn - \xmax)\\
        = &\,\sigma^2(\xmax) + \sigma^2(\xmaxdnn - \xmax) \\ &+ 2\operatorname{Cov}(\xmax, \xmaxdnn - \xmax)\\
    \end{split}
\label{eq:covariance_2nd_moment}
\end{align}

\begin{table}[]
    \begin{center}
        \begin{tabular}{c c c c }
            $\log_{10}E/\mathrm{eV}$\,bin & $\langle \log_{10}E/\mathrm{eV} \rangle$& \xmaxmu\,/\;\gcm & \;\xmaxsigma\,/\;\gcm\\
            \hline
            18.5--18.6 & 18.55 & 757.0 $\pm\;0.5_{-9.6}^{+7.7}$   &\;\; 48.8 $\pm\;0.6$ $_{-4.8}^{+9.1}$ \\
            18.6--18.7 & 18.65 & 758.8 $\pm\;0.5_{-9.2}^{+7.8}$   &\;\; 46.3 $\pm\;0.6$ $_{-4.8}^{+8.7}$ \\
            18.7--18.8 & 18.75 & 759.5 $\pm\;0.5_{-8.8}^{+7.9}$   &\;\; 43.4 $\pm\;0.7$ $_{-4.8}^{+8.3}$ \\
            18.8--18.9 & 18.85 & 761.8 $\pm\;0.5_{-8.5}^{+8.0}$   &\;\; 39.7 $\pm\;0.6$ $_{-4.8}^{+8.0}$ \\
            18.9--19.0 & 18.95 & 765.7 $\pm\;0.5_{-8.2}^{+8.2}$   &\;\; 38.4 $\pm\;0.7$ $_{-4.8}^{+7.6}$ \\
            19.0--19.1 & 19.05 & 770.0 $\pm\;0.6_{-7.9}^{+8.4}$   &\;\; 38.8 $\pm\;1.0$ $_{-4.8}^{+7.3}$ \\
            19.1--19.2 & 19.15 & 769.9 $\pm\;0.6_{-7.7}^{+8.7}$   &\;\; 34.0 $\pm\;0.8$ $_{-4.8}^{+7.1}$ \\
            19.2--19.3 & 19.25 & 774.0 $\pm\;0.8_{-7.5}^{+9.1}$   &\;\; 33.2 $\pm\;0.9$ $_{-4.8}^{+7.0}$ \\
            19.3--19.4 & 19.35 & 774.7 $\pm\;0.9_{-7.4}^{+9.5}$   &\;\; 30.7 $\pm\;1.2$ $_{-4.8}^{+6.9}$ \\
            19.4--19.5 & 19.45 & 775.3 $\pm\;1.0_{-7.3}^{+10.0}$  &\;\; 27.3 $\pm\;1.6$ $_{-4.8}^{+6.8}$ \\
            19.5--19.6 & 19.55 & 778.6 $\pm\;1.3_{-7.3}^{+10.5}$  &\;\; 26.3 $\pm\;1.3$ $_{-4.8}^{+6.8}$ \\
            19.6--19.7 & 19.64 & 783.2 $\pm\;1.4_{-7.3}^{+11.0}$  &\;\; 24.2 $\pm\;1.8$ $_{-4.8}^{+6.8}$ \\
            19.7--19.8 & 19.74 & 787.2 $\pm\;2.0_{-7.5}^{+11.6}$  &\;\; 20.7 $\pm\;1.4$ $_{-4.8}^{+6.8}$ \\
            19.8--19.9 & 19.85 & 794.5 $\pm\;3.6_{-7.8}^{+12.2}$  &\;\; 25.1 $\pm\;3.9$ $_{-4.8}^{+6.8}$ \\
            $>19.9$   & 20.00  & 793.9 $\pm\;4.5_{-8.3}^{+13.1}$  &\;\; 21.8 $\pm\;5.4$ $_{-4.8}^{+6.8}$ \\
        \end{tabular}
        \caption{First two moments of the \xmax distributions. Energies are given in $\log_{10}E/\mathrm{eV}$ and \xmaxmu and \xmaxsigma are given in \gcm followed by their statistical and systematic uncertainties.}
        \label{tab:xmax_dnn}
    \end{center}
\end{table}

\begin{table}[]
    \begin{center}
        \begin{tabular}{c c c}
            $\log_{10}E/\mathrm{eV}$ bin\; & \;\;\xmaxfd & \;\;\xmaxdnn  \\
            \hline
            18.5--18.6 & 1,347 & 8,739\\ 
            18.6--18.7 & 1,007 & 9,360\\ 
            18.7--18.8 & 707 & 7,725\\ 
            18.8--18.9 & 560 & 6,506\\ 
            18.9--19.0 & 417 & 5,228\\ 
            19.0--19.1 & 312 & 3,863\\ 
            19.1--19.2 & 253 & 2,781\\ 
            19.2--19.3 & 159 & 1,791\\ 
            19.3--19.4 & 122 & 1,205\\ 
            19.4--19.5 & 80 & 701\\ 
            19.5--19.6 & 50 & 455\\ 
            19.6--19.7 & \multirow{4}{*}{35} & 277\\ 
            19.7--19.8 & & 113\\ 
            19.8--19.9 & & 54\\
            $>19.9$ & & 26\\ 
        \end{tabular}
        \caption{Available statistics for determining the UHECR composition using the FD and the DNN. FD data are taken from Ref.~\cite{yushkov_mass_2019}.}
        \label{tab:sd_vs_fd_events}
    \end{center}
\end{table}

\newpage

\begin{figure*}[t!]
    \begin{centering}
        \begin{subfigure}[b]{0.495\textwidth}
            \begin{centering}
                \includegraphics[height=5.7cm]{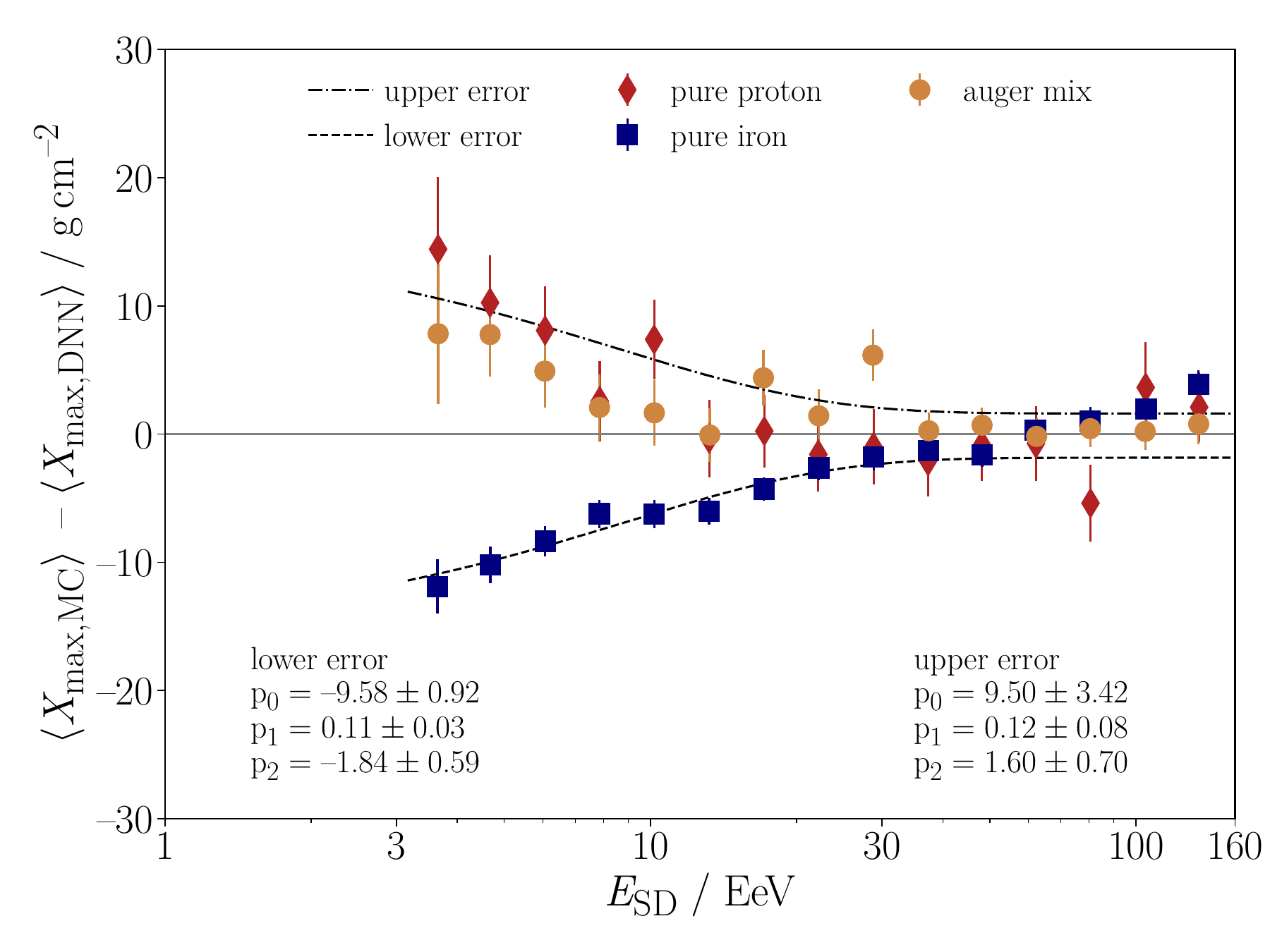}
                \subcaption{\epos}
                \label{fig:comp_bias_mean_epos}
            \end{centering}
        \end{subfigure}
        \begin{subfigure}[b]{0.495\textwidth}
            \begin{centering}
                \includegraphics[height=5.7cm]{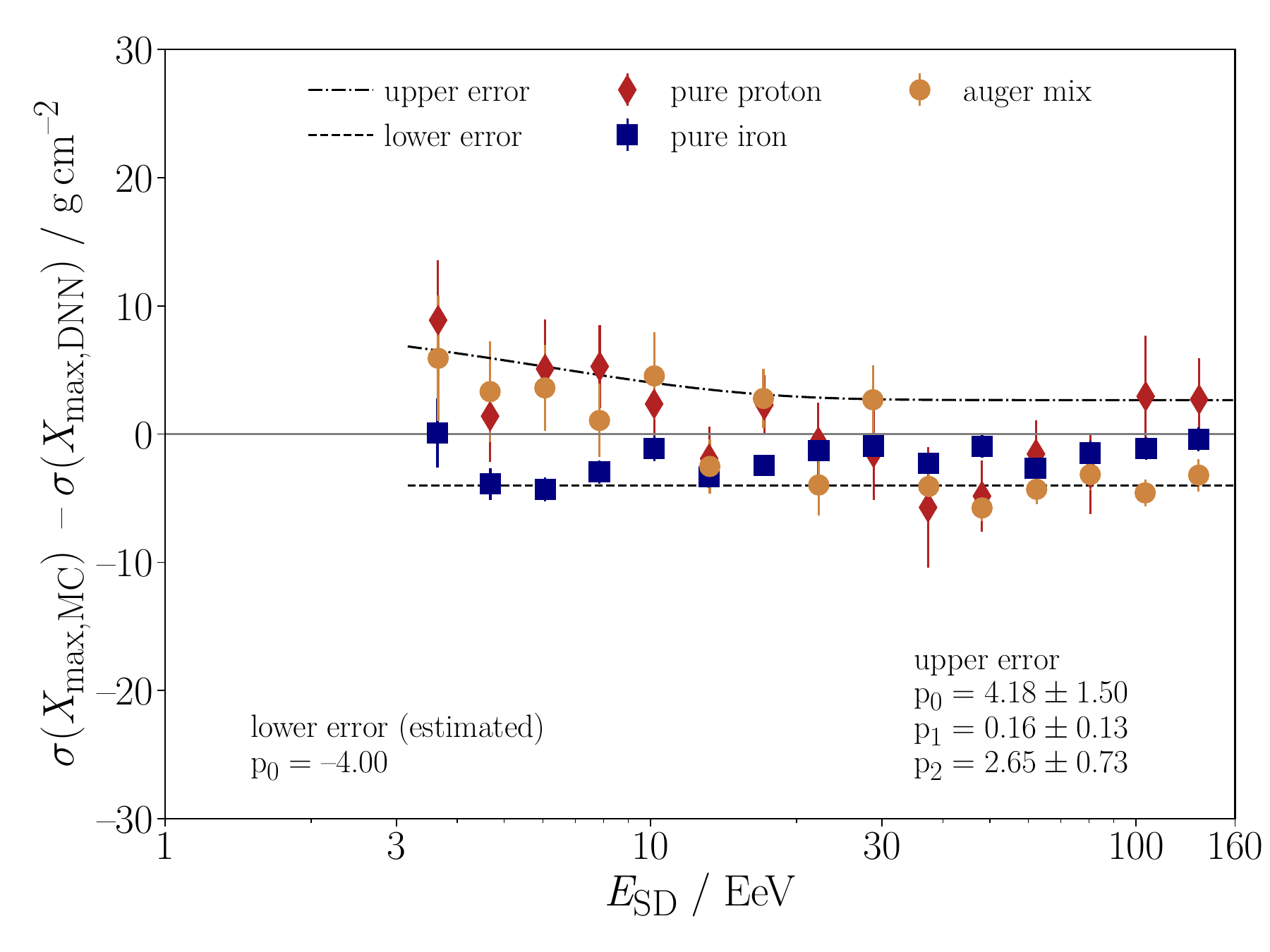}
                \subcaption{\epos}
                \label{fig:comp_bias_std_epos}
            \end{centering}
        \end{subfigure}
        
        \begin{subfigure}[b]{0.495\textwidth}
            \begin{centering}
                \includegraphics[height=5.7cm]{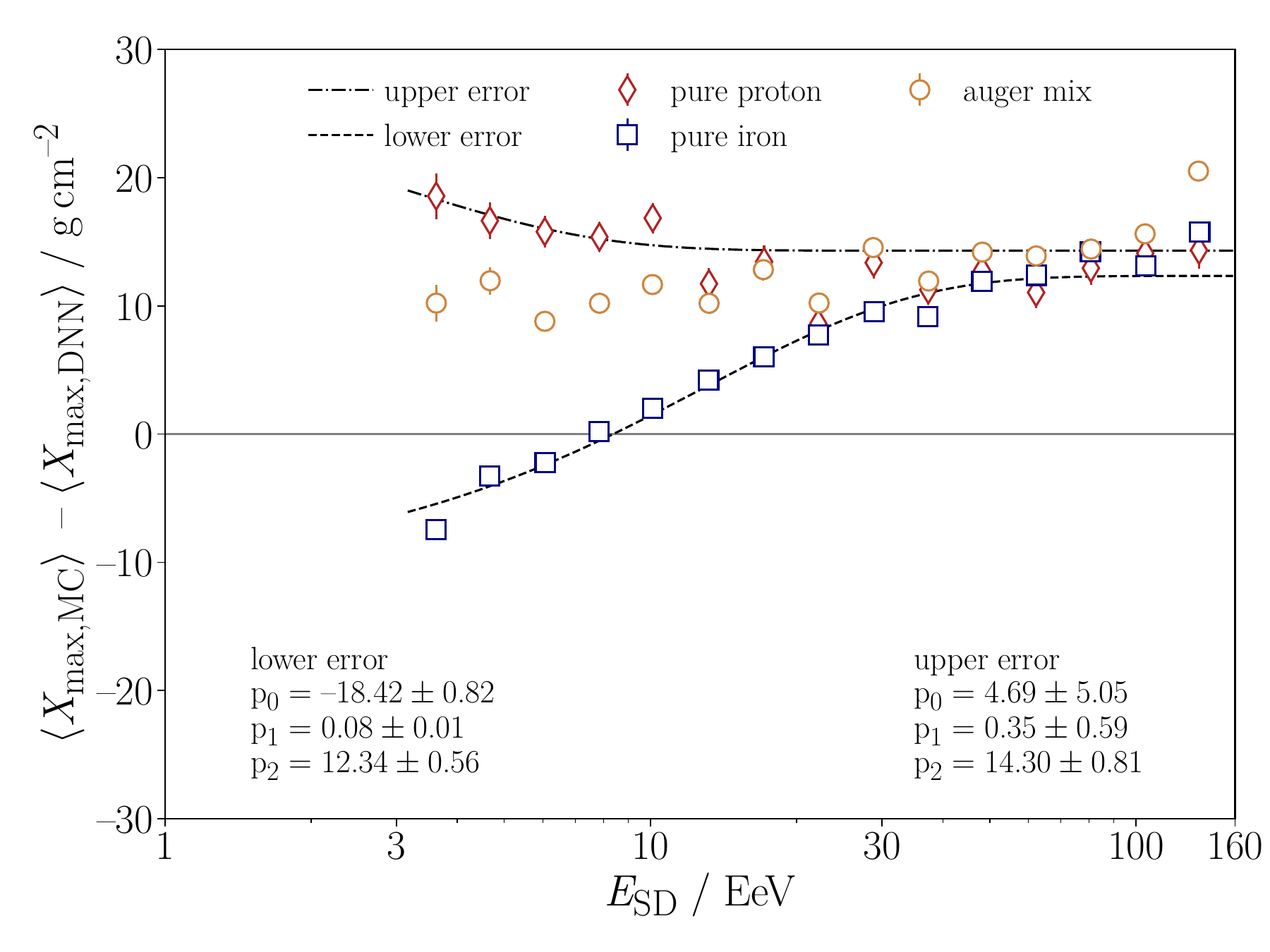}
                \subcaption{\sibyllc}
                \label{fig:comp_bias_mean_sib}
            \end{centering}
        \end{subfigure}
        \begin{subfigure}[b]{0.495\textwidth}
            \begin{centering}
                \includegraphics[height=5.7cm]{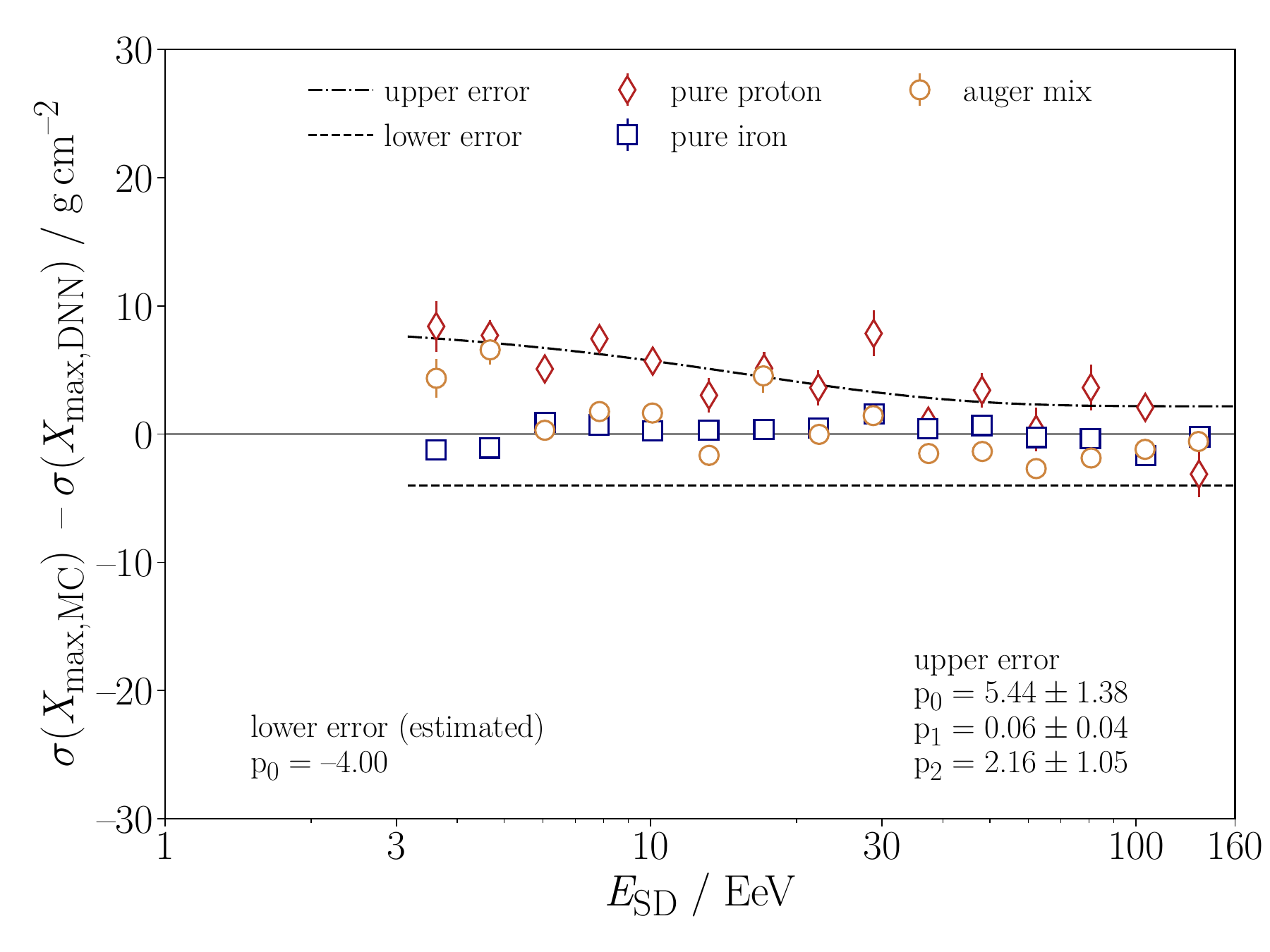}
                \subcaption{\sibyllc}
                \label{fig:comp_bias_std_sib}
            \end{centering}
        \end{subfigure}

        \begin{subfigure}[b]{0.495\textwidth}
            \begin{centering}
                \includegraphics[height=5.7cm]{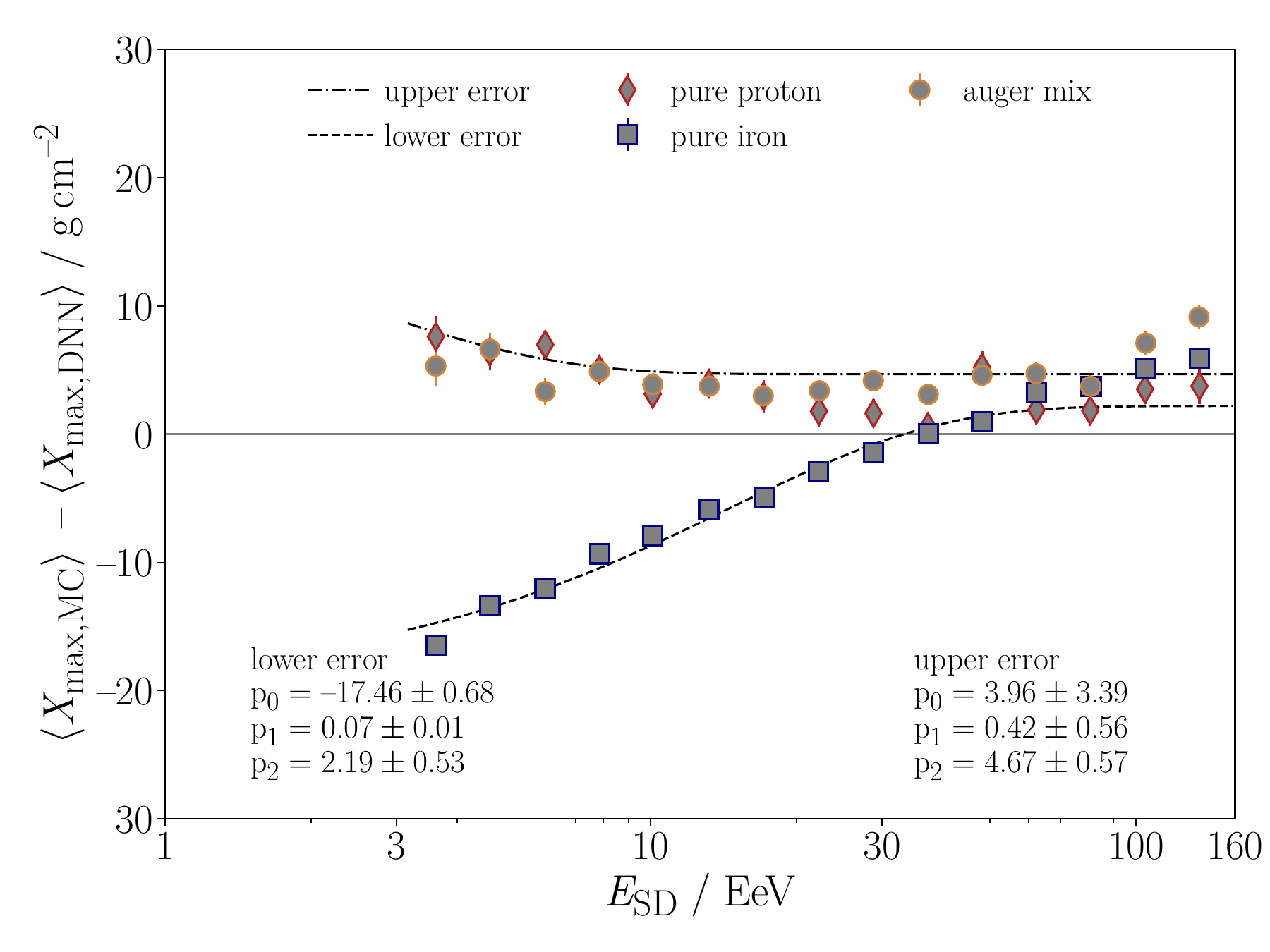}
                \subcaption{\qgs}
                \label{fig:comp_bias_mean_qgs}
            \end{centering}
        \end{subfigure}
        \begin{subfigure}[b]{0.495\textwidth}
            \begin{centering}
                \includegraphics[height=5.7cm]{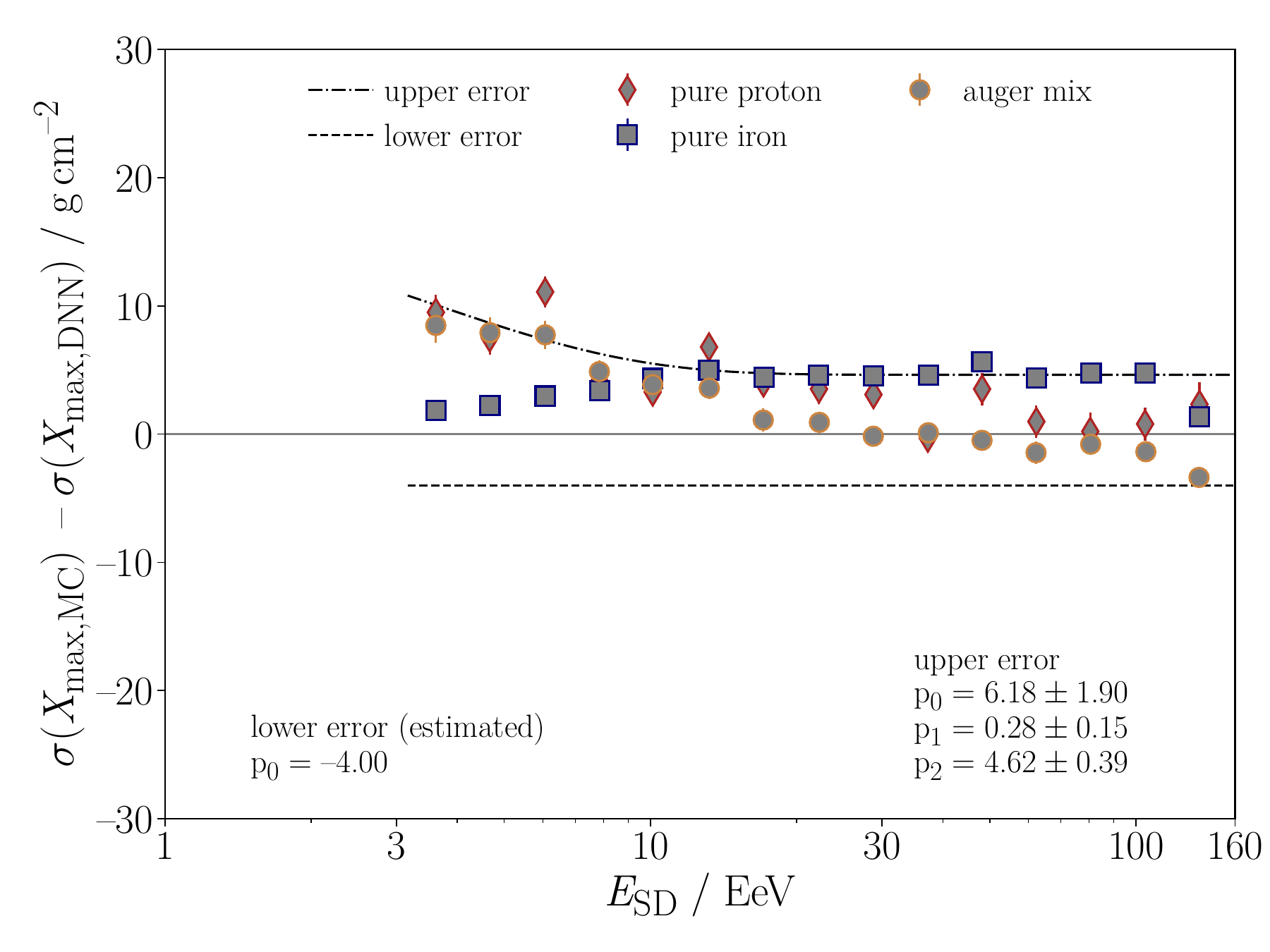}
                \subcaption{\qgs}
                \label{fig:comp_bias_std_qgs}
            \end{centering}
        \end{subfigure}        
    \end{centering}    
    \caption{Expected composition bias for measuring the first moment \xmaxmu (left) and the second moment \xmaxsigma (right) of \xmax distributions as a function of energy for \epos, \sibyllc, and \qgs (from top to bottom) after forward-folding of all systematic effects on the measurement. The different compositions are denoted by different colors. The dashed line indicates a parameterization for the composition bias. Note that only the parameterization for \xmaxsigma propagates into the uncertainty of the measurement, as for \xmaxmu the method is cross-calibrated using the FD. Note that \epos was used as the hadronic interaction model in the training of the network.}
    \label{fig:comp_bias_parameterizations}
\end{figure*}

\begin{figure*}[t]
    \begin{centering}
        \begin{subfigure}[b]{0.435\textwidth}
            \includegraphics[width=\textwidth]{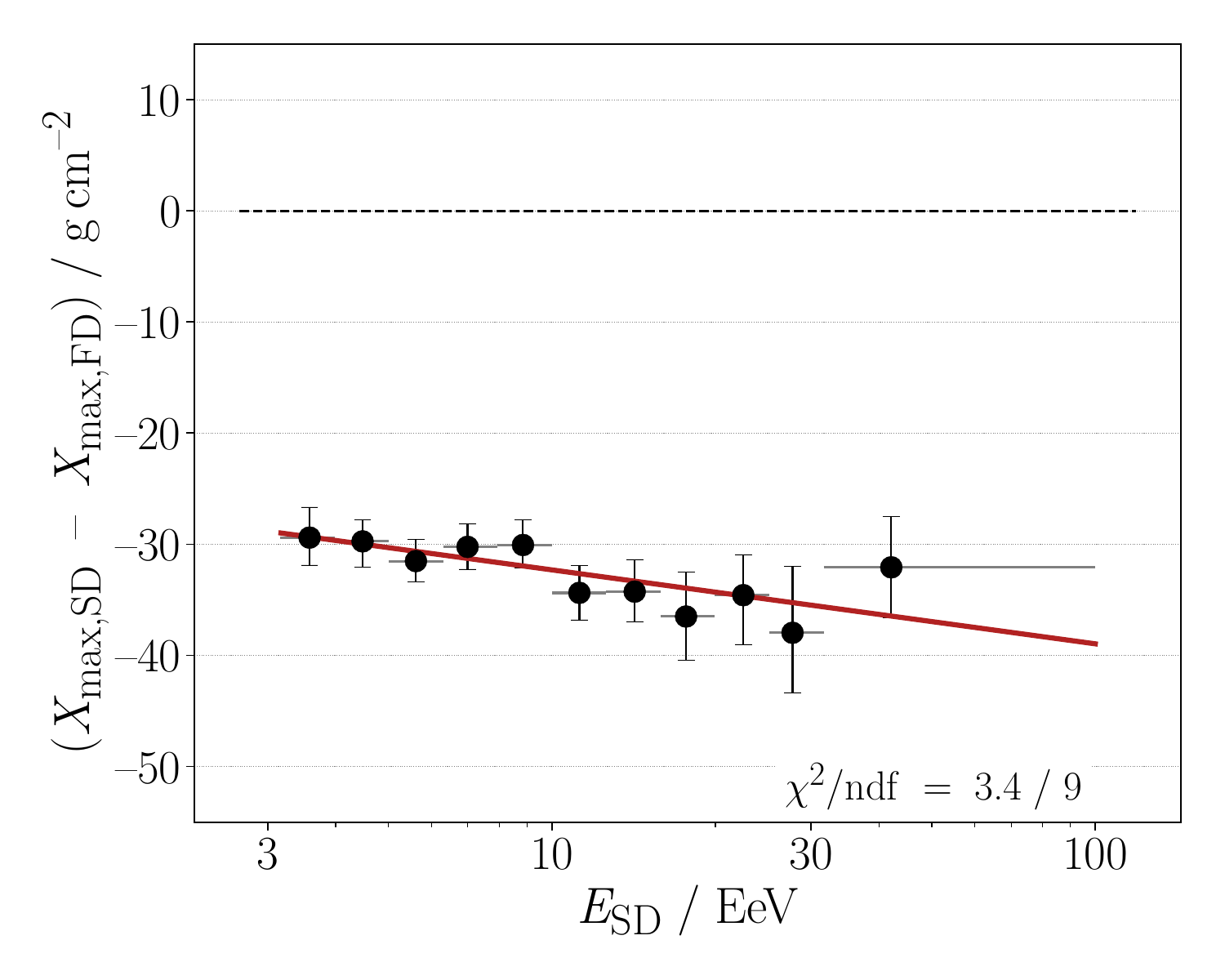}
            \subcaption{}
            \label{fig:dnn_fd_calib_1}          
        \end{subfigure}
         \begin{subfigure}[b]{0.435\textwidth}
            \includegraphics[width=\textwidth]{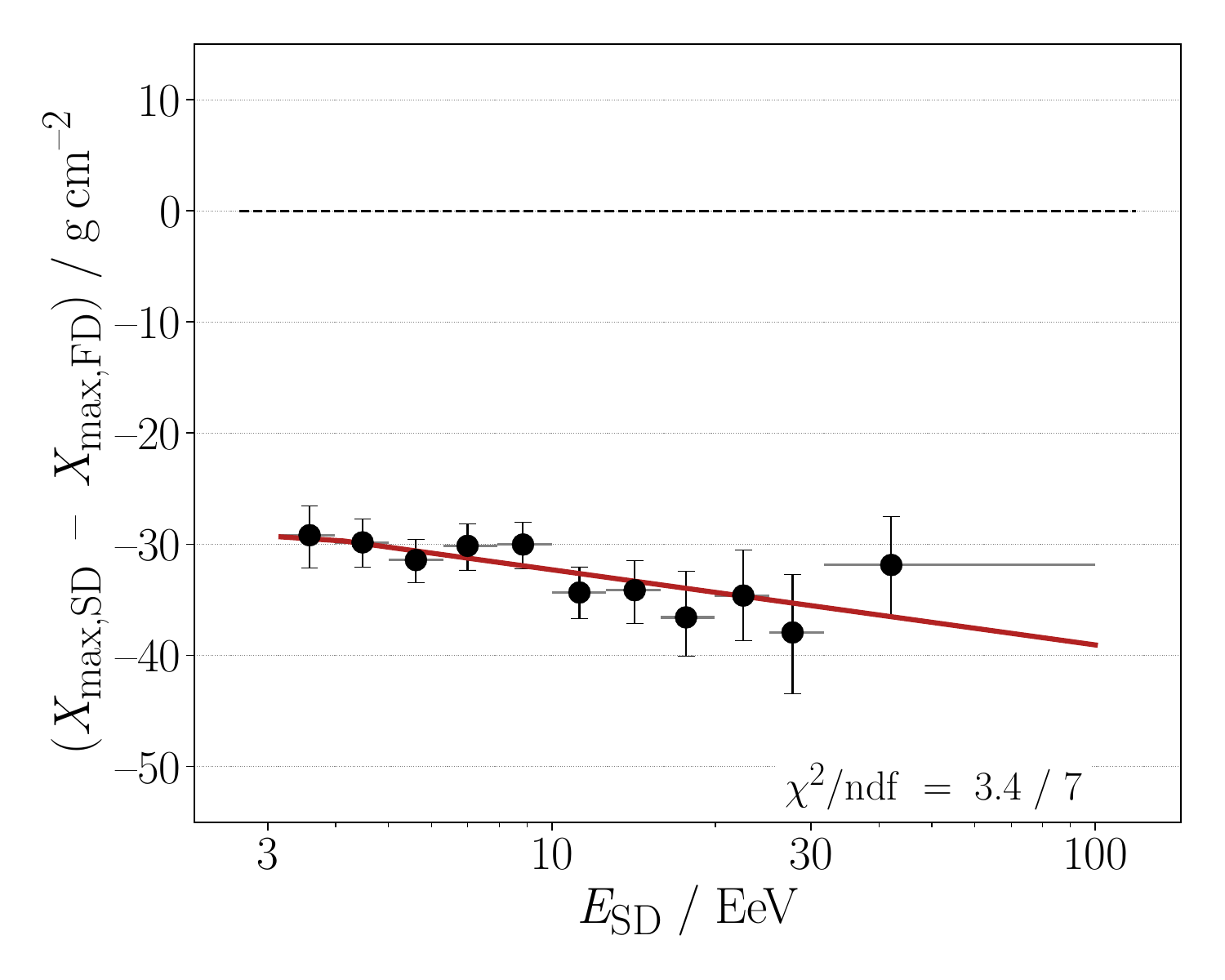}
            \subcaption{}
            \label{fig:dnn_fd_calib_2}
        \end{subfigure}

        \begin{subfigure}[b]{0.435\textwidth}
            \includegraphics[width=\textwidth]{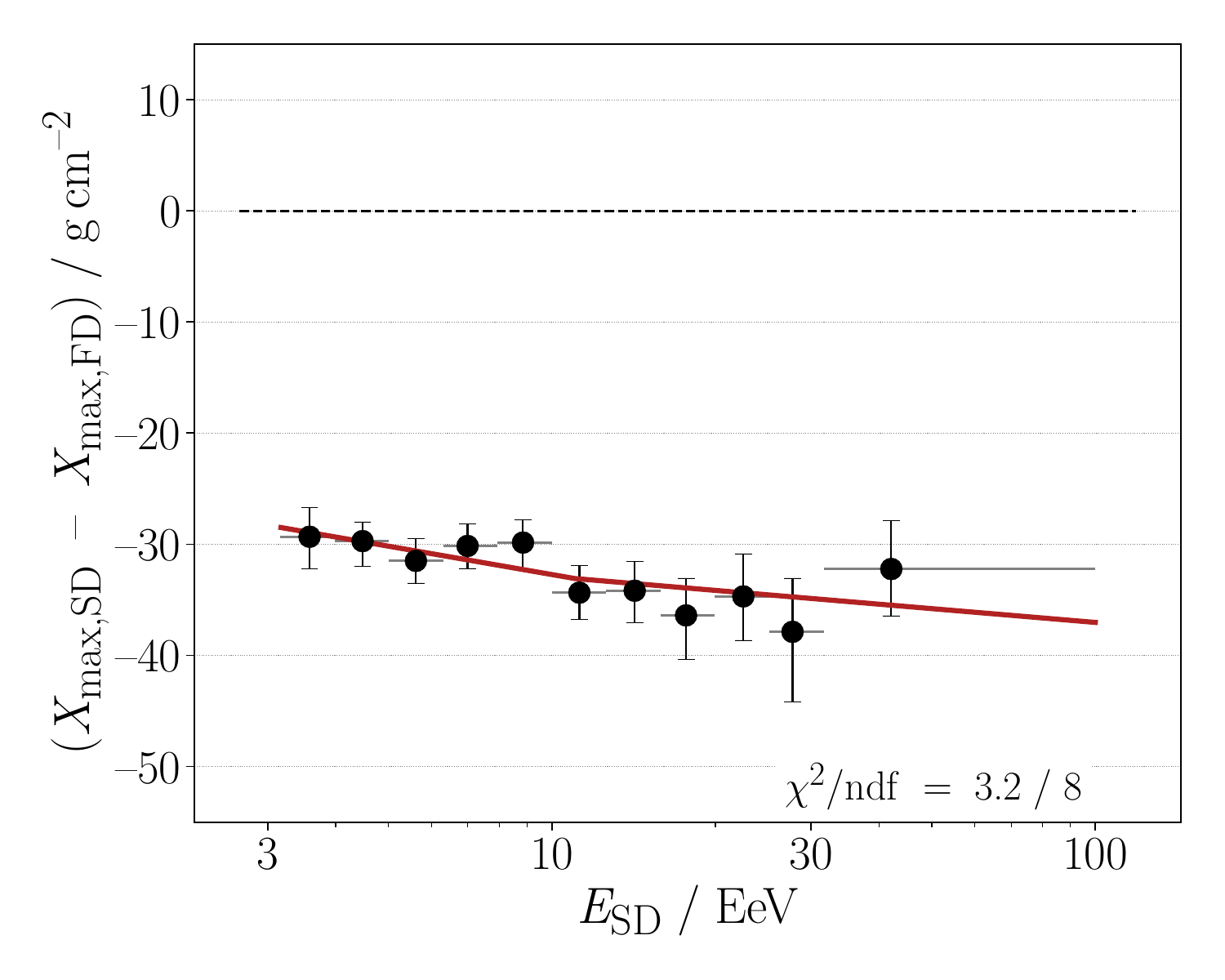}
            \subcaption{}
            \label{fig:dnn_fd_calib_3}          
        \end{subfigure}
         \begin{subfigure}[b]{0.435\textwidth}
            \includegraphics[width=\textwidth]{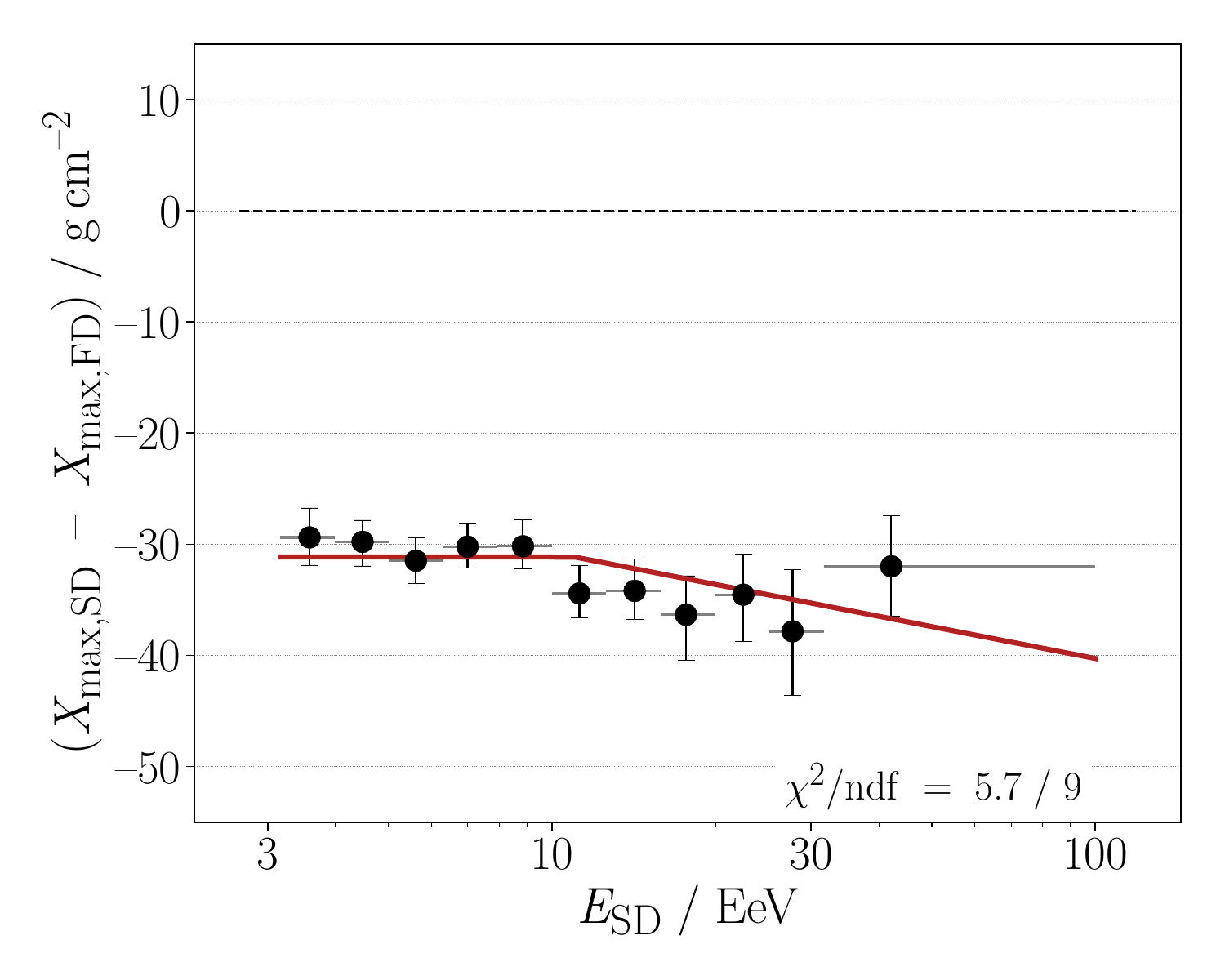}
            \subcaption{}
            \label{fig:dnn_fd_calib_4}       
        \end{subfigure}
        
         \begin{subfigure}[b]{0.435\textwidth}
            \includegraphics[width=\textwidth]{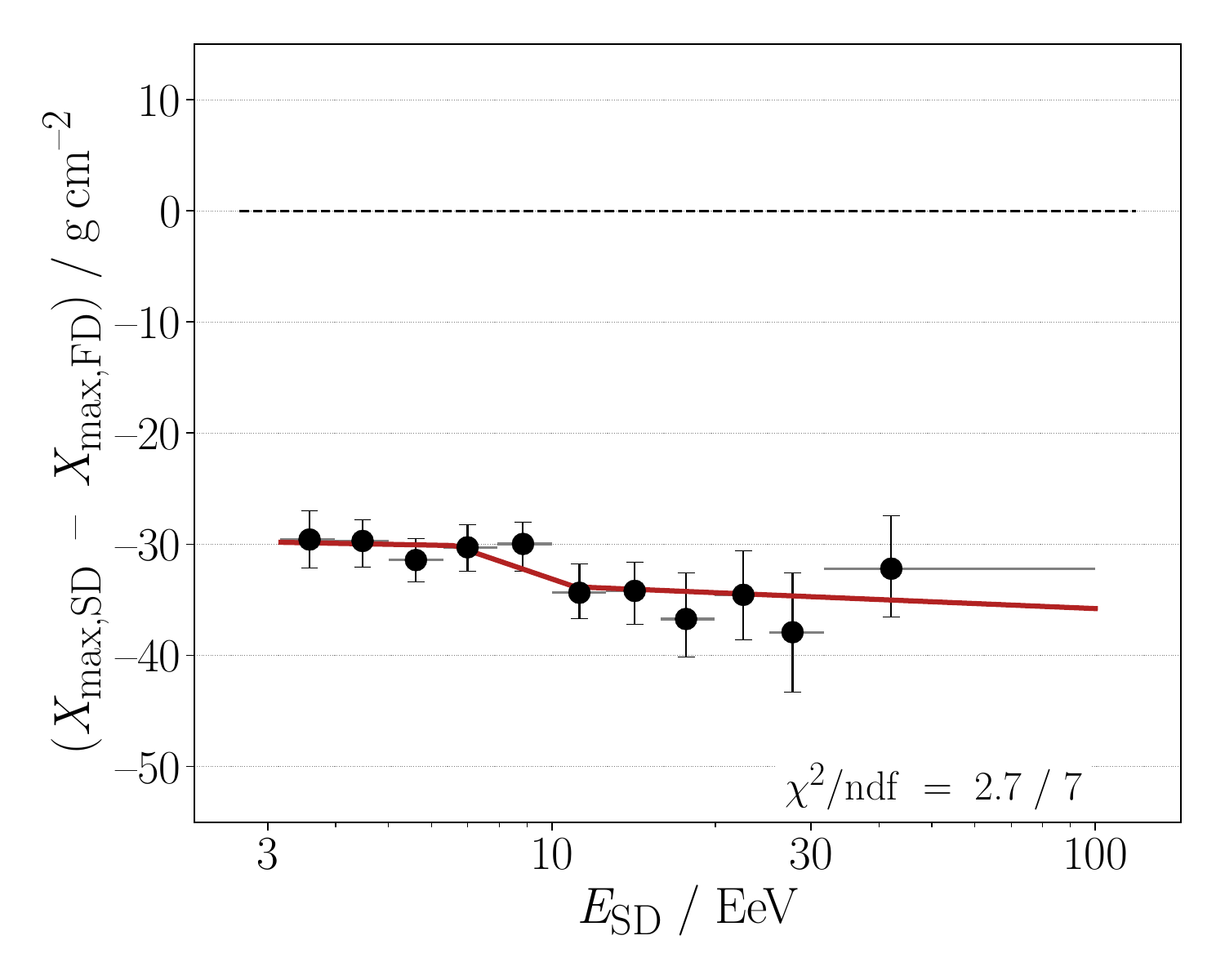}
            \subcaption{}
            \label{fig:dnn_fd_calib_5}       
        \end{subfigure}
         \begin{subfigure}[b]{0.435\textwidth}
            \includegraphics[width=\textwidth]{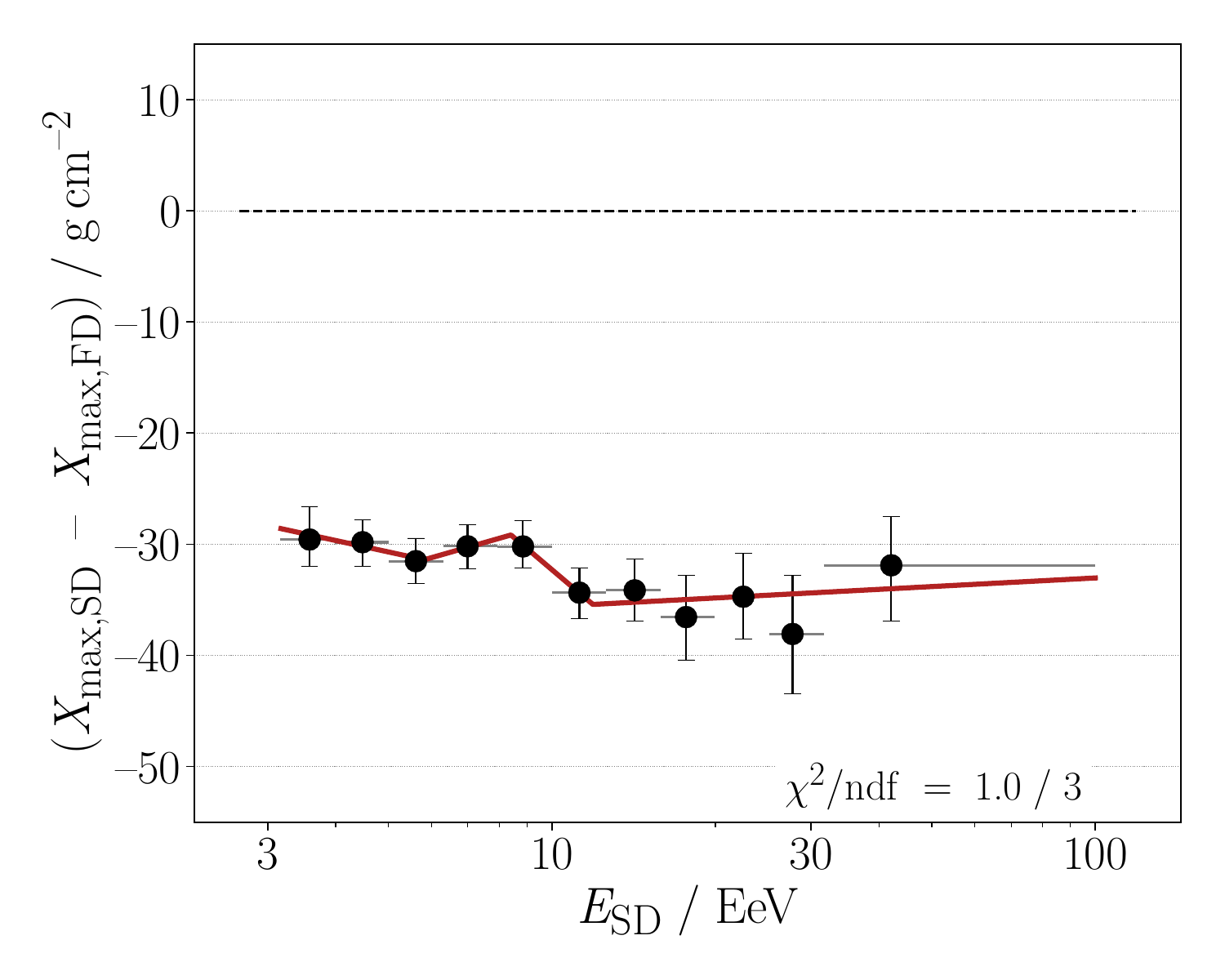}
            \subcaption{}
            \label{fig:dnn_fd_calib_6}       
        \end{subfigure}  

    \end{centering}
    \caption{\label{fig:hybrid_calibration_calib_checks}Models used for studying the effect of energy-dependent calibrations on the measurement of \xmaxmu and the significance of the identified features. (a) Linear function considered for a global energy dependence of the calibration (used for estimating the systematic uncertainty of \xmaxmu). (b) Piecewise-linear function. (c) Piecewise-linear function with the break fixed to the position of the fitted second break. A similar dependence could also be motivated by the composition bias of \epos (used for training) using the Auger mix (cf.~\cref{fig:comp_bias_mean_epos}). (d) Piecewise-linear fit with the first slope fixed to $0$~\gcm and the break fixed to the position of the second break (e) 3-fold piecewise-linear fit with the first and second break fixed to the position of the first and second break. (f) Piecewise-linear function with three adaptive breaks.
    None of the calibrations lowers the significance considerably.}
\end{figure*}

\end{document}